\documentclass[12pt]{article}
\usepackage{amsmath,amssymb,theorem,cite,epsfig,url,psfrag,amsmath,mathtools,mathrsfs,amsbsy,dsfont,esint,braket,cancel,diagbox}
\usepackage{indentfirst}
\usepackage[bookmarks=true,hyperfigures=true,colorlinks=true,linkcolor=blue,citecolor=blue,urlcolor=blue,bookmarksnumbered]{hyperref}
\numberwithin{equation}{section}
\advance \topmargin by -\headheight
\advance \topmargin by -\headsep     
\evensidemargin \oddsidemargin
\usepackage{bm,upgreek}

\makeatletter
\def\Appendix{\appendix
  \def\@seccntformat##1{Appendix~\csname the##1\endcsname.~~}}
\makeatother

\def\XXint#1#2#3{{\setbox0=\hbox{$#1{#2#3}{\int}$}
\vcenter{\hbox{$#2#3$}}\kern-.5\wd0}}

\begin{document}
\title{\textbf{Meson mass spectrum in Ising Field Theory}\vspace*{0.3cm}}
\date{}
\author{Alexey Litvinov$^{1}$, Pavel Meshcheriakov$^{1,2,3}$ and Egor Shestopalov$^{3}$
\\[\medskipamount]
\parbox[t]{0.85\textwidth}{\normalsize\it\centerline{1. Skolkovo Institute of Science and Technology, 121205 Moscow, Russia}}
\\
\parbox[t]{0.85\textwidth}{\normalsize\it\centerline{2. Landau Institute for Theoretical Physics, 142432 Chernogolovka, Russia}}
\\
\parbox[t]{0.85\textwidth}{\normalsize\it\centerline{3. Moscow Institute of Physics and Technology, 141700 Dolgoprudny, Russia}}}
\maketitle
\begin{abstract}
    We study the two-particle approximation of the Ising Field Theory (IFT), formulated in terms of the Bethe-Salpeter (BS) equation. Derived by Fonseca and Zamolodchikov as a systematic realization of the McCoy-Wu approach, this equation captures confinement by modeling ``mesons'' as bound states of two Majorana fermions (``quarks''), in a way analogous to the integral equation in the ’t Hooft model. Despite its approximate nature, the BS equation provides remarkably accurate predictions for the mass spectrum of stable mesons across a wide range of parameters. Motivated by the striking structural similarity between the BS equation in IFT and the 't Hooft equation in two-dimensional QCD, we develop a new non-perturbative analytical framework inspired by the method of Fateev, Lukyanov, and Zamolodchikov (FLZ). Within this approach, we compute spectral sums and systematically derive the large-$n$ WKB expansion for the Bethe-Salpeter equation, which governs the spectrum of an infinite tower of mesons. We further examine how our analytical results capture the known behavior of the spectrum in well-studied asymptotic regimes, such as the $E_8$ limit and the free-fermion point, where exact solutions are available for comparison. Finally, we discuss how the obtained spectral data admit a natural analytic continuation to complex values of the parameters---an extension that was one of the primary motivations for this work.
\end{abstract}
\tableofcontents

\section{Introduction}\label{Introduction}
The two-dimensional Ising model is among the most thoroughly explored systems in statistical mechanics. However, certain aspects of its critical behavior---particularly under a nonzero external magnetic field  $H$---still remain unsolved. The Ising Field Theory (IFT) arises as the scaling limit of the two-dimensional Ising model in a magnetic field $H$, taken in the vicinity of its critical point at $T=T_c$, $H=0$. From a field theory point of view, IFT is the unitary conformal field theory with the central charge $c=\frac{1}{2}$ (also known as minimal model $\mathcal{M}_{3,4}$ \cite{Belavin:1984vu}) perturbed by its two relevant operators: the ``energy density'' $\varepsilon(x)$  and the ``spin density'' $\sigma(x)$. This definition can be expressed via the formal action 
\begin{equation}\label{Ising-action}
    \mathcal{A}_{\text{IFT}}=\mathcal{A}_{c=\frac{1}{2}\text{ CFT}}+\frac{m}{2\pi}\int \varepsilon(x)\;d^2x+h\int\sigma(x)\;d^2x,
\end{equation}
where $\mathcal{A}_{c=\frac{1}{2}\text{ CFT}}$ is the action of the minimal model $\mathcal{M}_{3,4}$ (free massless Majorana fermions), and couplings $m$ and $h$ are related to
the deviations from the critical point in the scaling limit $m\sim T_c-T$, $h\sim H$. It should be noted that the two non-trivial local relevant scalar operators,  $\varepsilon(x)$ and $\sigma(x)$, possess scaling dimensions  $(\Delta_\varepsilon, \bar{\Delta}_\varepsilon) = \left(\tfrac{1}{2}, \tfrac{1}{2}\right)$ and $(\Delta_\sigma, \bar{\Delta}_\sigma)=\left(\tfrac{1}{16}, \tfrac{1}{16}\right)$, respectively. As a result, the associated relevant couplings have mass dimensions $[h]=\frac{15}{8}$ and $[m]=1$. Consequently, apart from an overall mass scale, the theory is effectively characterized by the single scaling parameter
\begin{equation}\label{scaling-parameters}
    \eta=\frac{m}{|h|^{\frac{8}{15}}} \quad \text{ or }\quad \xi=\frac{h}{|m|^{\frac{15}{8}}}.
\end{equation}

IFT exhibits a rich structure as a two-dimensional Quantum Field Theory. As the parameter $\eta$ varies from $-\infty$ to $\infty$, the particle spectrum evolves from a single stable particle into an infinite tower of weakly bound states \cite{McCoy:1978ta}. At zero magnetic field $h=0$ ($\eta=\pm\infty$) it reduces to a free theory of massive Majorana fermions of mass $|m|$ \cite{Onsager:1943jn}. Another exactly solvable case occurs at $m=0$, $h\neq0$ ($\eta=0$), where the model becomes integrable and describes eight stable particles (also known as $E_8$ particles) with factorized $S$-matrix scattering \cite{Zamolodchikov:1989fp,Zamolodchikov:1989zs}. Furthermore, when the parameters are extended into the complex domain, an additional integrable structure emerges near the Yang-Lee edge singularity \cite{Yang:1952be, Lee:1952ig}, located at purely imaginary magnetic field $h=\pm i(0.18935\ldots)m^{15/8}$ \cite{Fonseca:2001dc, Xu:2022mmw,Mangazeev:2023bnt}. In this regime, there is only one stable particle and the model is governed by the integrable Yang-Lee field theory \cite{Fisher:1978pf,Cardy:1985yy,Cardy:1989fw,Xu:2022mmw,Xu:2023nke}. No other real or complex values of the parameters are known to make the IFT integrable.

The physical interpretation of the fermionic excitations changes across the phase transition\footnote{In the high-$T$ regime $T>T_c$, IFT possesses a $\mathbb{Z}_2$ symmetry under $h\mapsto-h$, $\sigma(x)\mapsto-\sigma(x)$. As a result, all scaling functions are even in $h$. In the low-$T$ regime $T<T_c$ and $h=0$, this symmetry is spontaneously broken, leading to two degenerate vacua. A small non-zero $h$ lifts this degeneracy: depending on the sign of $h$, one vacuum becomes energetically favorable, while the other is raised. Without loss of generality, we will assume that $h>0$ in the following.}. In the disordered phase ($T>T_c$) at zero magnetic field $h=0$, they behave as ordinary particles. In contrast, in the ordered ferromagnetic phase ($T<T_c$), they correspond to topological kinks interpolating between two degenerate vacua. When a nonzero magnetic field $h\ne0$ is applied, it induces interaction between fermions and destroys the integrability of the theory away from criticality ($T\ne T_c$). In the ordered phase ($T<T_c$), this field also lifts the degeneracy of the ferromagnetic vacua, resulting in a linear confining potential between the kinks. Consequently, these kinks are no longer free excitations but become confined into two-kink bound states. This confinement scenario was originally proposed by McCoy and Wu \cite{McCoy:1978ta}. Due to the similarity with quark confinement in Quantum Chromodynamics, these topological excitations are often referred to as ``quarks'', and their bound states as ``mesons''\footnote{Although this terminology is introduced for the case $m>0$ ($T<T_c$), we will use it for arbitrary values of $\eta$. It is worth noting that when $h\ne0$, the distinction between the low-temperature and high-temperature regimes becomes blurred and is, to a large extent, artificial. Physical observables such as particle masses vary smoothly -- and indeed analytically—across the point $\eta=0$.}.

Since IFT is not integrable for general couplings $h\ne0$ and $m\ne0$, various approximate methods have been employed to study quark confinement in this regime. These include analytical approaches  \cite{Fonseca:2001dc,Fonseca:2006au,Zamolodchikov:2013ama,Delfino:1996xp,Delfino:2005bh,Rutkevich:2005ai,Rutkevich:2009zz,Rutkevich:2017qdw}, as well as numerical techniques \cite{Fonseca:2001dc,Lencses:2015bpa,Gabai:2019ryw,Fitzpatrick:2023aqm,Jha:2024jan}.
One powerful non-perturbative approach to analyzing the meson mass spectrum in IFT was developed in \cite{Fonseca:2006au} and is based on the Bethe–Salpeter (BS) equation formulated within the two-quark approximation:
\begin{equation}\label{BS-start}
    \left[m^2-\frac{M^2}{4\cosh^2\theta}\right]\psi(\theta)=f_0\fint_{-\infty}^{\infty}\limits\frac{d\theta'}{2\pi}\left[\frac{2\cosh(\theta-\theta')}{\sinh^2(\theta-\theta')}+\frac{\sinh\theta\,\sinh\theta'}{4\cosh^2\theta\,\cosh^2\theta'}\right]\psi(\theta').
\end{equation}
Here $M$ denotes the meson mass, $\psi(\theta)$ is the corresponding wave function, $\theta$ is the rapidity variable and $f_0 \sim |m|^{\frac{1}{8}} h$ represents the bare string tension. This equation, reminiscent of the integral equation in the 't Hooft model \cite{THOOFT1974461}, captures confinement effects by modeling mesons as bound states of two domain-wall fermions.  However, unlike in large $N_c$ two-dimensional QCD, the BS equation in IFT \eqref{BS-start} is only approximate---it neglects contributions from multi-particle sectors and therefore cannot describe effects such as meson decays or inelastic scattering. As a result, the predicted masses $M_n$ remain real for all $\eta>0$, but still, it appears to reproduce most of the real parts of the resonance masses. 

Despite its limitations, the BS equation becomes asymptotically exact in the weak field limit $h\to0$, where confinement is weak and the two-quark approximation becomes valid. In analogy with the 't Hooft model, the parameter $h$ plays a role similar to that of $1/N_c$. In this regime, two asymptotic expansions for the meson masses $M_n(\eta)$ have been developed: a low-energy series in fractional powers of $h$ \cite{McCoy:1978ta,Fonseca:2001dc,Fonseca:2006au,Rutkevich:2009zz} for the lowest-lying states, and a semiclassical expansion in integer powers of $h$ \cite{Fonseca:2006au,Rutkevich:2005ai,Rutkevich:2009zz} describing highly excited mesons with large $n\gg1$. These findings have been validated by numerical simulations which are generalizations of the  Truncated Conformal Space Approach \cite{Fonseca:2006au,Lencses:2015bpa}. The leading terms in both expansions can be derived directly from the BS equation \eqref{BS-start}, confirming the asymptotic validity of the two-quark picture at the lowest order \cite{Fonseca:2006au}. However, starting from the second order in $h$, corrections from four-quark, six-quark, and higher components must be taken into account \cite{Fonseca:2006au}. 

The status of multi-quark corrections is as follows. The first multi-quark correction to the meson masses in IFT, scaling as $h^2$, was derived in \cite{Fonseca:2006au} and stems from the quark mass renormalization. At order $h^3$, additional corrections emerge from three distinct sources: the $h^3$ -- level adjustment of the string tension due to kink interactions \cite{Fonseca:2001dc}; short-distance modifications to the Bethe-Salpeter kernel induced by multi-quark fluctuations \cite{Rutkevich:2009zz}; and third-order radiative corrections to the quark mass itself \cite{Rutkevich:2017qdw}. It turns out that the two-particle approximation describes the mass spectrum of stable mesons $M_n$ remarkably well for all but very small values of $\eta$ ($\eta\lesssim 0.5$), even if all renormalization effects are disregarded \cite{Fonseca:2006au} and for small $\eta$ we should use a non-perturbative definition of the effective string tension $f\overset{\text{def}}{=}\textrm{Re }(\mathcal{F}_{\text{meta}}-\mathcal{F}_{\text{vac}}),$
where $\mathcal{F}$ is the free energy of IFT. 

A central motivation for analyzing the BS equation is the expectation that it may provide insight about the analytic behavior of the mass spectrum $M_n(\eta)$ in the complex $\eta$-plane. Although it is not an exact equation, it provides a simplified framework for probing features that are otherwise difficult to access in the full theory. While the BS masses $M_n$ do not capture the full dynamics of the Ising Field Theory, there are reasons to believe that they reflect certain analytic features of the exact mass functions $M_n(\eta)$ in the complex $\eta$-plane. One especially intriguing line of inquiry concerns the location of zeros of $M_n(\eta)$, as their vanishing could indicate diverging correlation lengths and hint at critical behavior in the full, non-integrable theory. For instance, the zero of the lowest mass $M_1$ at $\eta_{\text{YL}}=(2.429169\ldots)e^{\pm i\frac{11\pi}{15}}$ aligns with the Yang-Lee edge singularity \cite{Fonseca:2001dc,Xu:2022mmw,Mangazeev:2023bnt}, suggesting that this may be one of several critical points accessible through analytic continuation on the $\eta$-plane. 

The similarity between the Bethe-Salpeter equations in IFT and in the 't Hooft model suggests a new approach to studying the analytic structure of IFT using computational tools inspired by integrability. Building on this connection, we adapt a non-perturbative analytical approach originally developed by Fateev, Lukyanov, and Zamolodchikov (FLZ) \cite{Fateev:2009jf}, who demonstrated that in the particular case where the quark masses are equal and fixed to a specific value, the 't Hooft equation reduces to a Baxter-type TQ relation. This observation greatly advanced the understanding of the analytic structure of the ’t Hooft model. In particular, it enabled them to compute the first several spectral sums $G^{(s)}_\pm$ exactly and obtain the large-$n$ WKB expansion. Their method was later successfully extended by us to the case of any equal quark masses in \cite{Litvinov:2024riz}, as well as to the general case in \cite{Artemev:2025cev}. 

In the present work, following \cite{Fateev:2009jf,Litvinov:2024riz, Artemev:2025cev}, we extend the FLZ approach to the Ising Field Theory. We show that the BS equation admits a reformulation reminiscent of the integrable structure observed in the ’t Hooft model. This allows us to develop a new analytical framework for studying the meson spectrum in IFT. With this reformulation in hand, we are able to compute the first few spectral sums and derive the large-$n$ WKB expansion for the eigenvalues of the BS equation. These results yield new non-perturbative insights into the analytic behavior of the meson spectrum as a function of the scaling parameter.

The paper is organized as follows. In Section \ref{BS-and-Integr}, we reformulate the Bethe-Salpeter equation in a form analogous to the ’t Hooft equation (now working in terms of $\alpha=\frac{m^2}{f_0}$ and $\lambda=\frac{M^2}{4\pi f_0}$), discuss its key properties, and outline the limitations of the two-particle approximation. We also introduce two fundamental objects — the $Q$-function and the spectral determinant — which are central to extracting spectral data. Using the $Q$-function, we reformulate the Bethe-Salpeter equation as an algebraic TQ-equation and extend this formulation to arbitrary $\lambda$ by introducing an inhomogeneous term in BS. In Section \ref{TQ-Solutions}, we construct special $Q$-solutions to the TQ-equation in two asymptotic regimes: $\lambda \to 0$ and $\lambda \to -\infty$, represented as series expansions in $\lambda$ and $\lambda^{-1}$, respectively. It is important that this analysis does not rely on any perturbative expansion in the quark mass parameter $m$ of the Bethe-Salpeter equation. In Section \ref{D-Q-relations}, we outline the procedure for extracting spectral data from the TQ-equation, extending the approach previously developed for the ’t Hooft model \cite{Fateev:2009jf,Litvinov:2024riz,Artemev:2025cev} to the case of the Ising Field Theory. The key idea is to establish nontrivial relations that express the spectral determinants through the $Q$-function solutions constructed in Section \ref{TQ-Solutions}. Section \ref{Analytical-results} is devoted to analytical results: the spectral data extracted using the results from the previous two sections. This includes explicit expressions for the spectral sums $\mathcal{G}^{(s)}_{\pm}$ and a systematic large-$n$ (WKB) expansion. In Section \ref{Limiting-cases-section}, we analyze our results in two physically significant limits: near the $E_8$ theory ($\eta \to 0$) and near the free-quark regime ($\eta \to \infty$). This analysis provides a non-trivial consistency check, allowing us to verify our results analytically by comparing them with known exact solutions in these regimes. In Section \ref{BS-analyt-Section}, we analyze the properties of the meson spectrum $\lambda_n(\alpha)$ under analytic continuation of the scaling parameter $\alpha$ to complex values. In particular, based on the analytic results for the spectral sums $\mathcal{G}^{(s)}_\pm$, we demonstrate that the complex scaling parameter plane contains two infinite series of special points $\alpha^*_k$, $\widetilde{\alpha}_k$, where one or two mesons (depending on the series) become massless. Moreover, the series where two mesons simultaneously become massless corresponds to the emergence of four-root branch cuts originating at these points. In this section, we also present new analytic results for the 't Hooft model. Following \cite{Fateev:2009jf}, we conjecture that each of these points corresponds to a ``critical point'' in the infrared, whose dynamics are governed by a non-trivial CFT. Finally, Section \ref{Conclusion} summarizes our conclusions and outlines potential avenues for future research. Proofs of certain statements, along with some technical details, are provided in the appendices.
\section{Bethe-Salpeter Equation and Integrability}\label{BS-and-Integr}
This section aims to present the Bethe-Salpeter equation and outline its key known features. We begin by examining the analytic properties of its wave functions and introducing the central object of our analysis---the $Q$-functions. By investigating their general analytic behavior, we derive the associated TQ equation and establish a connection between its solutions and those of a related inhomogeneous integral equation.
\subsection{Background}
Here are some general comments on the properties of the Bethe-Salpeter equation \eqref{BS-start}.  Firstly, the wave function $\psi(\theta)$ should be considered as a vector in a Hilbert space with the norm
\begin{equation}\label{psi-norm}
    \|\psi\|^2=\int_{-\infty}^{\infty}\limits\frac{d\theta}{2\pi}\;\frac{|\psi(\theta)|^2}{\cosh^2{\theta}}.
\end{equation}
Secondly, equation \eqref{BS-start} is understood as an eigenvalue problem for the parameter $M^2$,
\begin{equation}
    \hat{H}\psi_n=M_n^2\psi_n, \quad n=0,1,2,\ldots,
\end{equation}
where the Hamiltonian is defined as
\begin{equation}\label{Hamiltonian-def}
    \hat{H}\psi(\theta)=4\cosh^2{\theta}\left(m^2\psi(\theta)-f_0\fint_{-\infty}^{\infty}\limits\frac{d\theta'}{2\pi}\left[\frac{2\cosh(\theta-\theta')}{\sinh^2(\theta-\theta')}+\frac{\sinh\theta\,\sinh\theta'}{4\cosh^2\theta\,\cosh^2\theta'}\right]\psi(\theta')\right)
\end{equation}
and is Hermitian with respect to the metric \eqref{psi-norm}. Its eigenfunctions form a complete orthonormal basis, $(\psi_n, \psi_m) = \delta_{nm}$, can be chosen real-valued, $\psi_n^*(\theta) = \psi_n(\theta)$, and have real positive eigenvalues $M_n^2$. We emphasize that the value $M_n$ gives certain approximations to the actual masses of mesons (see Section \ref{Introduction}), and we do not introduce special symbols for these actual masses, since it will be clear from the context which masses are being referred to.

Equation \eqref{BS-start} admits solutions that are symmetric and antisymmetric (even and odd) with respect to the permutation of two quarks 
\begin{equation}
    \psi_n(-\theta)=(-1)^n\psi_n(\theta).
\end{equation}
Since quarks are fermions, we are primarily interested in antisymmetric solutions ($n$ is odd). However, for the sake of completeness in the analysis of the equation, we will also examine symmetric solutions with even $n$, although physically relevant states correspond to antisymmetric solutions, which remain our primary focus. We will also refer to them as ``physical'' and ``non-physical" mesons.

To make the problem more similar to the already studied 't Hooft equation with one flavor \cite{Fateev:2009jf, Litvinov:2024riz}
\begin{equation}\label{BS-tHooft}
    \left(\frac{2\alpha}{\pi}+\nu\coth{\frac{\pi\nu}{2}}\right)\Psi(\nu)=\lambda\int_{-\infty}^{\infty}\limits d\nu'\frac{\pi(\nu-\nu')}{2\sinh{\frac{\pi(\nu-\nu')}{2}}}\Psi(\nu'), \quad \alpha=\frac{\pi m^2}{g^2}-1,\quad M^2=2\pi g^2\lambda, 
\end{equation}
it is useful to apply the Fourier transform to \eqref{BS-start}
\begin{equation}
    \Psi(\nu)\overset{\text{def}}{=}\mathcal{F}[\psi(\theta)]=\int_{-\infty}^{\infty}\limits \frac{d\theta}{2\pi}\;\psi(\theta)e^{i\nu\theta},\quad \psi(\theta)\overset{\text{def}}{=}\mathcal{F}^{-1}[\Psi(\nu)]=\int_{-\infty}^{\infty}\limits d\nu\;\Psi(\nu)e^{-i\nu\theta}
\end{equation}
and introduce new dimensionless notations
\begin{equation}\label{FZ-our-notations}
    \frac{2\alpha}{\pi}=\frac{m^2}{f_0},\quad \lambda=\frac{M^2}{4\pi f_0}.
\end{equation}
Thus, the Fourier representation of the BS equation in IFT, as given in \cite{Fonseca:2001dc}\footnote{For reference, we also present the Bethe-Salpeter equation in the form used in the ’t Hooft model \cite{THOOFT1974461}. To do this, one has to switch from the rapidity variable to the $x$-space, which are related by $\theta=\frac{1}{2}\log\frac{x}{1-x}$
\begin{equation}\label{BS-like-tHooft}
    2\pi^2\lambda\phi(x)=\frac{\alpha}{x(1-x)}\phi(x)-\frac{1}{4}\int_0^1\limits\frac{dy(1-2x)(1-2y)}{2\sqrt{x(1-x)y(1-y)}}\phi(y)+\fint_0^1\limits\frac{dy}{2\sqrt{x(1-x)y(1-y)}} \frac{2xy-x-y}{(x-y)^2}\phi(y).
\end{equation}
}
\begin{equation}\label{BS-eq}
    \left(\frac{2\alpha}{\pi}+\nu\tanh{\frac{\pi\nu}{2}}\right)\Psi(\nu)-\frac{1}{16}\frac{\nu}{\cosh{\frac{\pi\nu}{2}}}\int_{-\infty}^{\infty}\limits d\nu'\frac{\nu'}{\cosh{\frac{\pi\nu'}{2}}}\Psi(\nu')=\lambda\int_{-\infty}^{\infty}\limits d\nu'\frac{\pi(\nu-\nu')}{2\sinh{\frac{\pi(\nu-\nu')}{2}}}\Psi(\nu'),
\end{equation}
serves as the starting point for our analysis and the subsequent elaboration and generalization of the FLZ method \cite{Fateev:2009jf,Litvinov:2024riz,Artemev:2025cev}.

In $\nu$-space, the condition of orthogonality of the basis $\psi_n(\theta)$ can be reformulated as follows:
\begin{multline}
    \lambda_n\delta_{nm}=\lambda_n(\psi_n,\psi_m)=\frac{\lambda_n}{\pi}\int_{-\infty}^{\infty}\limits d\nu\int_{-\infty}^{\infty}\limits d\nu'\;\Psi_n(\nu)\frac{\pi(\nu-\nu')}{2\sinh{\frac{\pi}{2}(\nu-\nu')}}\Psi_m(\nu')\overset{\eqref{BS-eq}}{=}
    \\=\frac{1}{\pi}\int_{-\infty}^{\infty}\limits d\nu\;\left(\frac{2\alpha}{\pi}+\nu\tanh{\frac{\pi\nu}{2}}\right)\Psi_n(\nu)\Psi_m(\nu)-\frac{1}{16\pi}\left(\int_{-\infty}^{\infty}\limits d\nu\;\frac{\nu\cdot\Psi_n(\nu)}{\cosh{\frac{\pi\nu}{2}}}\right)\left(\int_{-\infty}^{\infty}\limits d\nu'\;\frac{\nu'\cdot\Psi_m(\nu')}{\cosh{\frac{\pi\nu'}{2}}}\right).
\end{multline}
To ensure a finite norm and a well-behaved wave function $\psi(\theta)$, the Fourier image $\Psi(\nu)$ must be smooth on the real axis and vanish at least polynomially as $|\nu|\to\infty$\footnote{In fact $\Psi(\nu)$ vanishes exponentially. We will show it in subsection \ref{Fredholm-and-TQ}.}.

A detailed examination of \eqref{BS-eq} reveals that $\Psi(\nu)$ is a meromorphic function in the complex $\nu$-plane, possessing only simple poles located at
\begin{equation}\label{Psi-all-poles}
    \pm i\nu_k^*(\alpha)\pm2iN,\quad N=0,1,2,\ldots,
\end{equation}
where $i\nu^*_k(\alpha)$ is the $k$-th root of the transcendental equation
\begin{equation}\label{main-transcendental-equation}
    \frac{2\alpha}{\pi}+\nu\tanh\left(\frac{\pi\nu}{2}\right)=0.
\end{equation}
The first root of this type $i\nu^*_1(\alpha)$ lies within the interval $[0,1]$ for positive real value of $\alpha$. The corresponding pole of $\Psi(\nu)$ governs the asymptotic behavior of the wave function $\psi(\theta)$ in the rapidity space at large $\theta$
\begin{equation}\label{psi-theta-asympt}
    \psi_n(\theta)\sim\mathcal{F}^{-1}\left[\frac{1}{\nu-i\nu_1^*(\alpha)}-(-1)^n\frac{1}{\nu+i\nu_1^*(\alpha)}\right]\to r_n(\Theta(-\theta)+(-1)^n\Theta(\theta))e^{-\nu^*_1(\alpha)|\theta|}\quad \text{as}\quad \theta\to\pm\infty,
\end{equation}
where $r_n$ is a normalization-dependent constant, $\Theta(\theta)$ is the Heaviside step function. The more distant poles contribute at subleading order. In particular, this allows us to determine the boundary conditions in the $x$-space \eqref{BS-like-tHooft}, since $e^{-\nu^*_1(\alpha)|\theta|}\sim x^{\frac{1}{2}\nu^*_1(\alpha)}$ for $x\sim 0$ and $e^{-\nu^*_1(\alpha)|\theta|}\sim (1-x)^{\frac{1}{2}\nu^*_1(\alpha)}$ for $x\sim1$
\begin{equation}
    \phi(x)\sim
    \begin{cases}
        x^\beta,\quad &x\to0;\\
        (1-x)^\beta,\quad &x\to1,
    \end{cases}\quad \pi\beta\tan{\pi\beta}+\alpha=0, \quad 0\leq\beta<1.
\end{equation}
Since the function $\Psi_n(\nu)$ has no poles at $\nu=\pm i$, this requirement leads to the quantization condition
\begin{equation}\label{Psi_quantization_cond}
    \Psi_n(i)=-\frac{i}{16}\int_{-\infty}^{\infty}\limits d\nu'\frac{\nu'}{\cosh{\frac{\pi\nu'}{2}}}\Psi_n(\nu'),
\end{equation}
moreover, if $\Psi_n(\nu)$ is even, this expression is $0$. We emphasize that an analogous condition does not arise in the 't Hooft model, as it stems from the specific structure of \eqref{BS-eq}---particularly the presence of an additional integral term on the l.h.s. in \eqref{BS-eq}.

Before introducing the FLZ method and extracting the spectral data, following \cite{Fonseca:2006au}, let us briefly discuss the interpretation of the Bethe-Salpeter parameters $f_0$ and $m$ across different positive values of the scaling parameter $\eta$ \eqref{scaling-parameters}.

For sufficiently large values of $\eta$ (weak $h$ limit or equivalently heavy quarks limit), the string tension $f_0$ can be interpreted as the potential energy confining the kinks
\begin{equation}\label{f0-def}
    f_0=2\bar{\sigma}h=2\bar{s}\eta^{\frac{1}{8}}|h|^{\frac{16}{15}},
\end{equation}
where $\bar{\sigma}$ is the spontaneous magnetization \cite{Wu:1975mw}
\begin{equation}\label{spontaneous-magnetiation}
     \bar{\sigma}=|m|^{\frac{1}{8}}\bar{s},\quad \bar{s}=2^{\frac{1}{12}}e^{-\frac{3}{2}\zeta'(-1)}=1.35783834\ldots
\end{equation}
Partial inclusion of multi-quark corrections in the BS equation \eqref{BS-start} modifies the weak-coupling expansions in a straightforward way---through the renormalization of the parameters $m$ and $f_0$, which should be replaced by their ``dressed'' parameters $m_q$ and $f$ 
\begin{equation}
    m^2\longmapsto m^2_q = m^2(1+a_2\lambda^2+a_3\lambda^3+\ldots) ,\quad f_0\longmapsto f=f_0(1+c_2 \lambda^2+c_4\lambda^4+\ldots),
\end{equation}
where $\lambda=\frac{f_0}{m^2}=\frac{2\bar{s}}{\eta^{\frac{15}{8}}}$ is a small parameter, with the first two coefficients $a_k$ determined in \cite{Fonseca:2003ee,Rutkevich:2005ai}, and the coefficients $c_{2k}$ obtained in \cite{Fonseca:2006au}
\begin{equation}
    a_2=0.071010809\ldots,\quad  a_3=0,\quad c_2=-0.003889\ldots, \quad c_4=-0.002499\ldots
\end{equation}
As demonstrated using the Truncated Free Fermion
Space Approach (TFFSA) in \cite{Fonseca:2006au}, the BS equation provides an accurate description of the stable meson mass spectrum $M_n$ across a wide range of $\eta$, except in the regime of very small values ($\eta\lesssim0.5$), even without incorporating all renormalization effects.

For small values of $\eta$, the definition of the string tension should be revised \cite{Voloshin:1985id,Fonseca:2001dc,Fonseca:2006au}. The effective string tension $f$ can be naturally defined as the real part of the difference between the energy densities of the true and metastable vacua, $\mathcal{F}_{\text{meta}}-\mathcal{F}_{\text{vac}}$
\begin{equation}\label{effectife-string-tension-def}
    f_0\quad\longmapsto \quad f\overset{\text{def}}{=}\textrm{Re\;}(\mathcal{F}_{\text{meta}}-\mathcal{F}_{\text{vac}}),
\end{equation}
where $\mathcal{F}_{\text{meta}}$ is obtained by analytically continuing the vacuum energy $\mathcal{F}(h)$ to negative values of the magnetic field $h$. The analogous idea was used for a similar problem in the recent work \cite{Gao:2025mcg}. Although this continuation introduces a small imaginary part---interpreted as a tunneling rate---its real part dominates and provides a meaningful nonperturbative definition of the string tension. Numerical data from \cite{Fonseca:2006au} show that the real part is nearly equal to the bare string tension $f_0$ for most values of the scaling parameter $\eta$, with significant deviations appearing only at small values $\eta\lesssim0.8$. In contrast to the singular behavior $f_0\sim\eta^{\frac{1}{8}}$ as $\eta\to0$, the corrected string tension remains finite and analytic in this regime. Its Taylor expansion in the vicinity of $\eta=0$ is based on the data from \cite{Fonseca:2001dc} and was calculated in \cite{Fonseca:2006au}
\begin{equation}\label{f-series}
    f=|h|^{\frac{16}{15}}\textrm{Re }(\rho_0+\rho_1\eta+\rho_2\eta^2+\rho_3\eta^3+\dots),
\end{equation}
where
\begin{equation}\label{rho-coefs}
    \rho_0=2.3692934\dots,\quad\rho_1=0.3521342\dots,\quad \rho_2=0,\quad \rho_3=-0.0181436\dots,\quad\dots
\end{equation}
At small $\eta$, multi-quark corrections to the quark mass $m_q$ become unreliable, suggesting that the properly defined quark mass remains close to the bare value $m$. According to the arguments presented in \cite{Fonseca:2001dc}, the radius of convergence in \eqref{f-series} can be assumed to be determined by the value $|\eta_{\text{YL}}|=2.42917\dots$ (see also the discussion at the end of Section \ref{BS-analyt-Section}).
\subsection{\texorpdfstring{$Q$}{}-function and TQ equation}\label{Fredholm-and-TQ}
Now we define the $Q$-function:
\begin{equation}\label{Qdef}
    Q(\nu)\overset{\text{def}}{=}\left(\frac{2\alpha}{\pi}\cosh{\frac{\pi\nu}{2}}+\nu\sinh{\frac{\pi\nu}{2}}\right)\Psi(\nu)=\cosh{\frac{\pi\nu}{2}}\left(\frac{2\alpha}{\pi}+\nu\tanh{\frac{\pi\nu}{2}}\right)\Psi(\nu)
\end{equation}
and examine its meromorphic extension to the largest possible domain of analyticity. In terms of $Q(\nu)$, the BS equation \eqref{BS-eq} takes the form
\begin{equation}\label{BS-eq-2}
    Q(\nu)-\nu\int_{-\infty}^{\infty}\limits d\nu'\frac{\nu'}{16\cosh^2{\frac{\pi\nu'}{2}}}\frac{Q(\nu')}{\frac{2\alpha}{\pi}+\nu'\tanh{\frac{\pi\nu'}{2}}}=\lambda\cosh{\frac{\pi\nu}{2}}\int_{-\infty}^{\infty}\limits d\nu'\frac{\pi(\nu-\nu')}{2\sinh{\frac{\pi(\nu-\nu')}{2}}}\frac{Q(\nu')}{\cosh{\frac{\pi\nu'}{2}}\left(\frac{2\alpha}{\pi}+\nu'\tanh{\frac{\pi\nu'}{2}}\right)}.
\end{equation}
Let us note the main properties of the $Q$-function that follow from its definition:
\begin{enumerate}
    \item $Q(\nu)$ is analytic in the strip $\textrm{Im }\nu\in[-2,2]$, since the factor $\frac{2\alpha}{\pi}+\nu\tanh\frac{\pi\nu}{2}$ eliminates the simple poles at $\pm i\nu^*_k(\alpha)$ (when $N=0$) of the function $\Psi(\nu)$ \eqref{Psi-all-poles};
    \item $Q(\nu)$ grows slower than any exponential as $|\textrm{Re}\,\nu| \to \infty$, implying that it stays bounded in this limit as 
    \begin{equation}\label{Q-is-bounded}  
        \forall\epsilon>0\quad Q(\nu)=\mathcal{O}(e^{\epsilon|\nu|}), \quad|\textrm{Re}\,\nu|\to\infty;
    \end{equation}
    \item the quantization condition of the function $\Psi(\nu)$ \eqref{Psi_quantization_cond} extends to $Q(\nu)$:
        \begin{equation}\label{Q_quantization_cond}
            Q(i)=-Q(-i)=\frac{i}{16}\int_{-\infty}^{\infty}\limits d\nu'\frac{\nu'}{\cosh{\frac{\pi\nu'}{2}}}\frac{Q(\nu')}{\frac{2\alpha}{\pi}\cosh{\frac{\pi\nu'}{2}}+\nu'\sinh{\frac{\pi\nu'}{2}}}.
        \end{equation}
\end{enumerate}
A difference equation satisfied by $Q(\nu)$ can be derived by considering the following linear combination:
\begin{equation}\label{Q-linear-comb}
    Q(\nu+2i)+Q(\nu-2i)-2Q(\nu).
\end{equation}
\begin{figure}[h!]
    \centering
    \includegraphics[width=1.\linewidth]{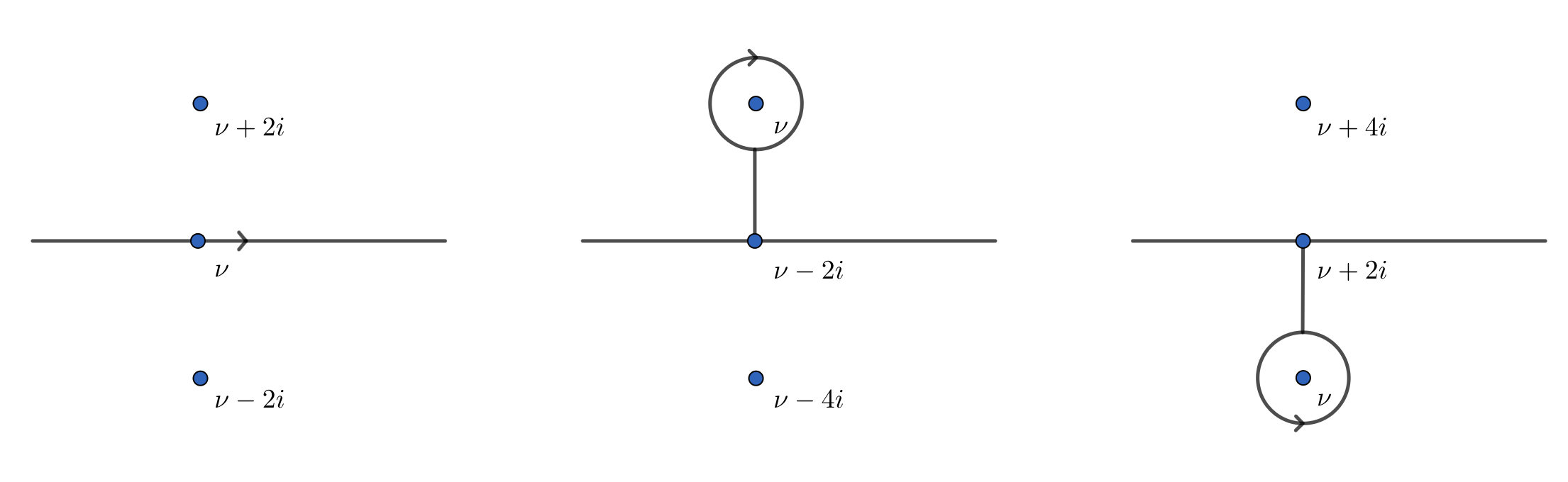}
    \caption{The analytical continuation of $\nu\to\nu\pm2i$ from the real axis leads to the appearance of additional terms, determined by half-residues of poles intersecting the integration contour.}
    \label{fig:TQ}
\end{figure}
Upon analytically continuing equation \eqref{BS-eq-2} to $\nu\pm 2i$ (as illustrated in Fig. \ref{fig:TQ}), additional contributions emerge due to the half-residues of the kernel $\frac{\nu-\nu'}{\sinh\frac{\pi(\nu-\nu')}{2}}$ at $\nu'=\nu\pm2i$, which cross the integration contour. Moreover, the integral terms cancel each other upon substitution into equation \eqref{Q-linear-comb}, yielding the following result:
\begin{equation}\label{TQ-equation}
    Q(\nu+2i)+Q(\nu-2i)-2Q(\nu)=-\frac{4\pi\lambda\cosh{\frac{\pi\nu}{2}}}{\frac{2\alpha}{\pi}\cosh{\frac{\pi\nu}{2}}+\nu\sinh{\frac{\pi\nu}{2}}}Q(\nu)=-\frac{2z}{\nu+\alpha x}Q(\nu),
\end{equation}
where
\begin{equation}\label{x-z-notation}
    z=2\pi\lambda\coth{\frac{\pi\nu}{2}},\quad x=\frac{2}{\pi}\coth{\frac{\pi\nu}{2}}.
\end{equation}
Difference equations of the form \eqref{TQ-equation}, commonly referred to as TQ equations, were first introduced by Baxter in his foundational work on the $8$-vertex model \cite{Baxter:1972hz}. In the context of exactly solvable lattice models, these equations play a central role in determining the eigenvalues of transfer matrices, with the analytic structure of the $Q$-function being essential to the method. In the most elementary cases, the $Q$-function is a polynomial, and its zeros must be chosen such that the associated $T$-function also becomes a polynomial. A similar strategy is employed in the present context, where one seeks for solutions to equation \eqref{TQ-equation} satisfying specific analytic and asymptotic conditions. The TQ equation we obtained is very similar to the equation that arises in the context of the 't Hooft model \cite{Fateev:2009jf,Litvinov:2024riz} with the replacement $\coth\frac{\pi\nu}{2}\to\tanh\frac{\pi\nu}{2}$.

So far, we have not commented on the second condition that the $Q$-function must satisfy \eqref{Q-is-bounded}. When deriving the TQ equation \eqref{TQ-equation}, we did not use this condition; it was sufficient that all integrals converged, which was ensured by the fact already established in the previous subsection that the function $\Psi(\nu)$ decayed at least polynomially. The TQ equation \eqref{TQ-equation} allows us to prove that $\Psi(\nu)$ actually decays exponentially at least as $e^{-\frac{\pi\nu}{2}}$, thereby ensuring that the $Q(\nu)$ function grows slower than any exponential, as required. Indeed, a refined analysis of the asymptotic form of \eqref{TQ-equation} shows that general solutions at $\textrm{Re }\nu\to\infty$ behave as $e^{k\pi\nu}h(\nu)$, where $h(\nu)$ is a periodic function with period $2i$ that grows slower than any exponential: 
\begin{equation}
    Q(\nu+2i)+Q(\nu-2i)-2Q(\nu)+\frac{2z}{\nu+\alpha x}Q(\nu)=e^{k\pi\nu}h(\nu)\left[-4\sin^2\pi k+\mathcal{O}\left(\frac{1}{\nu}\right)\right].
\end{equation}
The vanishing of the r.h.s. implies  $k \in \mathbb{Z}$. Since any positive $k$ would lead to an exponential growth of $Q(\nu)$ and thus violate the decay condition for $\Psi(\nu)$, one must impose $k \leq 0$. This confirms that $\Psi(\nu)$ decays exponentially as $\textrm{Re }\nu \to \infty$, and $Q(\nu)$ satisfies \eqref{Q-is-bounded}.

Let us briefly discuss the significance of the quantization condition \eqref{Q_quantization_cond}. TQ equation \eqref{TQ-equation} admits solutions possessing all three properties described above only for discrete values of $\lambda_n$, which correspond to the eigenvalues of the original spectral problem \eqref{BS-eq}. However, if the quantization condition \eqref{Q_quantization_cond} is omitted, one can construct solutions $Q(\nu|\lambda)$ for arbitrary $\lambda$. In that case, the associated function $Q(\nu|\lambda)$ solves more general inhomogeneous integral equation 
\begin{equation}\label{BS-eq-inhomogeneous}
    \frac{Q(\nu)}{\cosh{\frac{\pi\nu}{2}}}-\lambda\int_{-\infty}^{\infty}\limits d\nu'\frac{\pi(\nu-\nu')}{2\sinh{\frac{\pi(\nu-\nu')}{2}}}\frac{Q(\nu')}{\cosh{\frac{\pi\nu'}{2}}\left(\frac{2\alpha}{\pi}+\nu\tanh{\frac{\pi\nu}{2}}\right)}=F(\nu),
\end{equation}
with the inhomogeneous part 
\begin{equation}
    F(\nu)\overset{\text{def}}{=}\frac{Q(i)+Q(-i)}{2}\frac{1}{\cosh{\frac{\pi\nu}{2}}}+\frac{Q(i)-Q(-i)}{2i}\frac{\nu}{\cosh{\frac{\pi\nu}{2}}}.
\end{equation}
We observe that equation \eqref{BS-eq-inhomogeneous} becomes exactly \eqref{BS-eq-2} when the quantization condition \eqref{Q_quantization_cond} is satisfied. This equation can be derived by applying the operator $\hat{O}$, which defined as
\begin{equation}\label{K-def}
    \hat{O}\cdot f(\nu)\overset{\text{def}}{=}\int_{-\infty}^{\infty}\limits d\nu'\frac{\pi(\nu-\nu')}{2\sinh{\frac{\pi(\nu-\nu')}{2}}}\frac{f(\nu')}{\cosh{\frac{\pi\nu'}{2}}}
\end{equation}
to the left and right sides of the TQ equation \eqref{TQ-equation}. For details of this derivation see \cite{Litvinov:2024riz,Artemev:2025cev}.

It can be shown that, once the values $Q(\pm i)$ are fixed, equation \eqref{BS-eq-inhomogeneous} admits a unique solution. As a result, the space of functions $Q(\nu|\lambda)$ is two-dimensional, and a natural basis can be chosen in terms of symmetric and antisymmetric solutions,
\begin{equation}\label{Qpm-symmetry}
    Q_\pm(-\nu)=\pm Q_{\pm}(\nu).
\end{equation}
Each of these solutions correspond to a specific inhomogeneous term in \eqref{BS-eq-inhomogeneous}
\begin{equation}
    F_+(\nu)=\frac{1}{\cosh\frac{\pi\nu}{2}}\quad \text{and} \quad F_-(\nu) = \frac{\nu}{\cosh\frac{\pi\nu}{2}},
\end{equation}
which are associated with the normalization conditions 
\begin{equation}\label{Qpm-normalization-conditions}
    Q_+(i)=1 \quad \text{and} \quad Q_-(i)=i,    
\end{equation}
respectively.

In the following section, we will present a systematic method for constructing the asymptotic expansion of $Q_\pm(\nu|\lambda)$ in the two distinct regimes: $\lambda \to 0$ and $\lambda \to -\infty$. We then demonstrate how these expansions can be employed to extract non-perturbative information about the spectrum $\lambda_n$, including spectral sums and a systematic large-$n$ WKB expansion.
\subsection{Spectral sums and spectral determinants}
One of the main results of this work (see Section \ref{Analytical-results}) is the derivation of the analytical expressions for the spectral sums of the integral operator appearing in equation \eqref{BS-eq}. The spectral sums (spectral zeta functions) of the operator $\hat{\mathcal{O}}$ are defined in the standard manner as
\begin{equation}
    \zeta_\mathcal{O}(s)\overset{\text{def}}{=}\text{tr }\hat{\mathcal{O}}^{-s}=\sum_{i}\frac{1}{\lambda_i^s}
\end{equation}
initially for those values of $s$ where the series converges, and extended to other values by analytic continuation. In this work, we focus exclusively on integer values of $s$ and are particularly interested in the even and odd spectral sums, defined respectively as the traces of the spectral operator over even and odd states. Namely,
\begin{equation}\label{spectral-sums-def}
    \mathcal{G}_{+}^{(s)}\overset{\text{def}}{=}\text{tr}_+\;\hat{\mathcal{O}}^{-s}=\sum\limits_{n=0}^\infty\left(\frac{1}{\lambda_{2n}^s}-\frac{\delta_{s,1}}{n+1}\right),\quad \mathcal{G}_{-}^{(s)}\overset{\text{def}}{=}\text{tr}_-\;\hat{\mathcal{O}}^{-s}=\sum\limits_{n=0}^\infty\left(\frac{1}{\lambda_{2n+1}^s}-\frac{\delta_{s,1}}{n+1}\right).
\end{equation}
The subtraction at $s=1$ is required to ensure convergence, since, as it will be shown in Section \ref{Analytical-results}, the spectrum exhibits asymptotically linear growth: $\lambda_n \sim n/2$ as $n \to \infty$.

Let us also introduce an important object in the FLZ method---the ``spectral determinants'' (also known as Fredholm determinants)---defined as functions whose zeros correspond to the even/odd eigenvalues of the Bethe–Salpeter equation \eqref{BS-eq}.
\begin{equation}\label{Determinants-def}
    \mathcal{D}_+(\lambda)\overset{\text{def}}{=}\left(\frac{2\pi}{e}\right)^\lambda\prod_{n=0}^\infty\left(1-\frac{\lambda}{\lambda_{2n}}\right)e^{\frac{\lambda}{n+1}},\quad \mathcal{D}_-(\lambda)\overset{\text{def}}{=}\left(\frac{2\pi}{e}\right)^\lambda\prod_{n=0}^\infty\left(1-\frac{\lambda}{\lambda_{2n+1}}\right)e^{\frac{\lambda}{n+1}}.
\end{equation}
This infinite product converges at least in the  small $\lambda$ regime; in particular, it can be represented as follows\footnote{For large $\lambda$, the spectral determinants admit the representation given in \eqref{F_definition}.}
\begin{equation}\label{D-trough-G-def}
    \mathcal{D}_\pm(\lambda)=\left(\frac{2\pi}{e}\right)^\lambda\exp\left[-\sum_{s=1}^\infty s^{-1}\mathcal{G}^{(s)}_{\pm}\lambda^s\right].
\end{equation}
It is important to note that the spectral determinants defined in \eqref{Determinants-def} differ from the conventional notion of a spectral determinant of the operator $\mathcal{O}$, which is typically given by $\text{det }\mathcal{O} = e^{-\zeta'_\mathcal{O}(0)}$. In Section \ref{D-Q-relations}, we will present several key relations linking the solutions of the TQ equation $Q_\pm$ with spectral determinants $\mathcal{D}_\pm$, which will allow us to extract the necessary spectral data for Ising Field Theory.

Let us now clarify which specific operator we are considering in the context of our spectral problem. For brevity, we will adopt the following notation:
\begin{equation}
    t=\frac{\pi\nu}{2},\quad \phi(t)=\sqrt{f(t)}\Psi(t),\quad f(t)=\alpha+t\cdot\tanh t.
\end{equation}
and introduce the integral operators $\hat{A}$ and $\hat{K}$ as follows:
\begin{equation}\label{A-K-def}
    (\hat{A}g)(t)\overset{\text{def}}{=}\frac{1}{4\pi^2}\int_{-\infty}^{\infty}\limits\frac{tt'}{\cosh t\cosh t'}\frac{g(t')}{\sqrt{f(t)f(t')}}dt',\quad (\hat{K}g)(t)\overset{\text{def}}{=}\int_{-\infty}^{\infty}\limits\frac{t-t'}{\sinh{(t-t')}}\frac{g(t')}{\sqrt{f(t)f(t')}}dt'.
\end{equation}
In these notations, the BS equation \eqref{BS-eq} and spectral problem are written in compact forms 
\begin{equation}\label{spectral_problem}
    \phi(t)=(\hat{A}\phi)(t)+\lambda(\hat{K}\phi)(t) \quad \Rightarrow \quad \hat{\mathcal{K}}\phi\overset{\text{def}}{=}(\hat{1}-\hat{A})^{-1}\hat{K}\phi=\frac{1}{\lambda}\phi.
\end{equation}

Let us note that if \eqref{spectral_problem} did not involve the operator $\hat{A}$, the problem of determining the meson mass spectrum (and spectral sums in particular) would be entirely equivalent to the one addressed in \cite{Litvinov:2024riz} for the 't Hooft model with arbitrary equal quark masses. The only difference would lie in the explicit form of the function $f(t)$, which in the 't Hooft model is given by $f(t) = \alpha + t \cdot \coth t$.

Let us discuss the effect of the presence of the operator $\hat{A}$ on the level of spectral sums. First of all, we note that this operator is proportional to the projector $\hat{P}$ onto a certain odd state 
\begin{equation}\label{A-proector}
\begin{gathered}
    \ket{p}=\frac{1}{\sqrt{4\pi^2\mathtt{v}(\alpha)}}\frac{t}{\cosh{t}\sqrt{f(t)}},\\
    \hat{A}=\mathtt{v}(\alpha)\hat{P},\quad \hat{P}^2=\hat{P},\quad \braket{p|p}=1,
\end{gathered}
\end{equation}
where\footnote{The integral $\mathtt{v}(\alpha)$ may be expressed in terms of the integral $\mathtt{u}_3(\alpha)$ \eqref{u-def} as
    \begin{equation}
        \mathtt{v}(\alpha
        )=\frac{1}{16}+\frac{\alpha}{2\pi^2}-\frac{\alpha^2\zeta(3)}{2\pi^4}-\frac{\alpha^3\mathtt{u}_3(\alpha)}{4\pi^2}.
    \end{equation}
}
\begin{equation}\label{Proector-def}
    \mathtt{v}(\alpha)=\frac{1}{4\pi^2}\int_{-\infty}^{\infty}\limits\frac{dt}{f(t)} \frac{t^2}{\cosh^2t},\quad (\hat{P}g)(t)=\frac{1}{4\pi^2\mathtt{v}(\alpha)}\int_{-\infty}^{\infty}\limits dt'\frac{tt'}{\cosh t \cosh t'} \frac{g(t')}{\sqrt{f(t)f(t')}}.
\end{equation}
Using \eqref{A-proector} we may rewrite the spectral operator $\hat{\mathcal{K}}$ in the simpler form 
\begin{equation}
    (\hat{1}-\hat{A})^{-1}=\hat{1}+\sum\limits_{n=1}^\infty\hat{A}^n=\hat{1}+\sum\limits_{n=1}^\infty\mathtt{v}^n(\alpha)\hat{P}^n=\hat{1}+\frac{\mathtt{v}(\alpha)}{1-\mathtt{v}(\alpha)}\hat{P}\quad\Rightarrow\quad \hat{\mathcal{K}}=\left(\hat{1}+\mathcal{V}(\alpha)\hat{P}\right)\hat{K},
\end{equation}
where
\begin{equation}\label{xi-def}
    \mathcal{V}(\alpha)\overset{\text{def}}{=}\frac{\mathtt{v}(\alpha)}{1-\mathtt{v}(\alpha)}.
\end{equation}
We shall henceforth denote the spectral sums of the operator $\hat{K}$ by $G_{\pm}^{(s)}$, while for the operator $\hat{\mathcal{K}}$ we will use the notation $\mathcal{G}_{\pm}^{(s)}$. Since $\hat{P}$ is a projector onto an odd state $\ket{p}$, the even spectral sums of the operators $\hat{K}$ and $\hat{\mathcal{K}}$ coincide: $G_{+}^{(s)}=\mathcal{G}_{+}^{(s)}$, whereas the odd spectral sums differ by terms related to the matrix elements of the operator $\hat{K}$ of the form $\bra{p}\hat{K}^n\ket{p}$. These expressions are obtained by inserting a complete set of states into the explicit formula for the trace of the operator $\hat{\mathcal{K}}$. For example, for the first odd spectral sum $\mathcal{G}_{-}^{(1)}$
\begin{equation}
    \mathcal{G}_{-}^{(1)}=G_{-}^{(1)}+\mathcal{V}(\alpha)\bra{p}\hat{K}\ket{p},\quad \bra{p}\hat{K}\ket{p}=\frac{1}{4\pi^2\mathtt{v}(\alpha)}\int_{-\infty}^{\infty}\limits \frac{dt}{f(t)}\frac{t}{\cosh{t}}\int_{-\infty}^{\infty}\limits \frac{dt'}{f(t')}\frac{t-t'}{\sinh{(t-t')}}\frac{t'}{\cosh{t'}}.
\end{equation}
General formula for $\mathcal{G}_{-}^{(s)}$ looks more complicated
\begin{equation}\label{Gm-with-matr-el}
    \mathcal{G}_{-}^{(s)}=G_{-}^{(s)}+s\cdot\sum\limits_{m=1}^s\frac{\mathcal{V}^m(\alpha)}{m}\sum_{\substack{i_1j_1+\dots+i_mj_m=s
    \\j_1+\dots+j_m=m
    \\i_1\neq i_2\neq\ldots\neq i_m}}\frac{m!}{j_1!\ldots j_m!}\langle p|\hat{K}^{i_1}\ket{p}^{j_1}\cdot\ldots\cdot\bra{p}\hat{K}^{i_m}\ket{p}^{j_m}.
\end{equation}
Thus, to obtain explicit analytical expressions for the spectral sums of the operator $\hat{\mathcal{K}}$, it is necessary to determine the spectral sums of the operator $\hat{K}$ and to develop a method for computing matrix elements of the form $\bra{p}\hat{K}^n\ket{p}$. Various methods for calculating spectral sums of operator $\hat{K}$ are described in Section \ref{D-Q-relations}. The calculation of the matrix elements is presented in Appendix \ref{matrix-el-of-K}.

Finally, let us find the relationship between $D_\pm$ and $\mathcal{D}_\pm$---the spectral determinants of the operators $\hat{K}$ and $\hat{\mathcal{K}}$, respectively. It is clear that the even determinants are equal $\mathcal{D}_+(\lambda)=D_+(\lambda)$, since for even ``non-physical'' wave functions, the operator $\hat{A}$ \eqref{A-K-def} does not contribute to the Bethe-Salpeter equation \eqref{BS-eq}. 

The relationship for odd determinants turns out to be less trivial.  We note that the sum on the right-hand side of \eqref{Gm-with-matr-el} can be represented as the coefficient in the power series expansion in $\lambda$ of the function t $g(\lambda)=\sum_{k=1}^{\infty}\limits g^{(k)}\lambda^k$, with coefficients $g^{(k)}=\mathcal{V}(\alpha)\bra{p}\hat{K}^{k}\ket{p}$:
\begin{equation}
    \frac{1}{m}g^m(\lambda)=\frac{1}{m}\sum_{k_1,\dots,k_m=1}^{\infty}g^{(k_1)}\ldots g^{(k_m)}\lambda^{k_1+\ldots+k_m}=\frac{1}{m}\sum_{n=m}^{\infty}\sum_{\substack{i_1j_1+\ldots+i_mj_m=n\\j_1+\ldots+j_m=m\\i_1\ne\ldots\ne i_m}}^{\infty}\frac{m!}{j_1!\ldots j_m!}(g^{(i_1)})^{j_1}\ldots(g^{(i_m)})^{j_m}\lambda^{n}.
\end{equation}
In Appendix \ref{matrix-el-of-K}, the function $g(\lambda)$ is found explicitly
\begin{equation}\label{g-for-appendix}
    g(\lambda)=-\mathcal{V}(\alpha)-\frac{1}{1-\mathtt{v}(\alpha)}\frac{Q'_-(i)-1}{4\pi^2\lambda}=-\frac{1}{1-\mathtt{v}(\alpha)}\left(\mathtt{v}(\alpha)+\frac{Q'_-(i)-1}{4\pi^2\lambda}\right).
\end{equation}
Thus, using \eqref{D-trough-G-def} we obtain
\begin{equation} \label{newDet-oldDet}
    \mathcal{D}_+(\lambda)=D_+(\lambda),\quad\mathcal{D}_-(\lambda)=\frac{1}{1-\mathtt{v}(\alpha)}\left(1+\frac{Q'_-(i)-1}{4\pi^2\lambda}\right)D_-(\lambda).
\end{equation}

We stress that we obtained the relation between $D_-$ and $\mathcal{D}_-$ as a small $\lambda$ expansion. However, this relation remains valid for all positive values of $\lambda$. By definition, $\mathcal{D}_-(\lambda)$ is a function whose zeros coincide with the spectrum $\lambda_n$ of the Bethe–Salpeter equation \eqref{BS-eq}. In contrast, $D_-(\lambda)$ corresponds to the spectral problem without the operator $\hat{A}$, so its zeros generally differ from the physical meson masses in the two-particle approximation. Therefore, it is expected that the correct masses are determined by the additional factor appearing in \eqref{newDet-oldDet}
\begin{equation}\label{Quant_cond2}
    1+\frac{Q'_-(i)-1}{4\pi^2\lambda} = 0.
\end{equation}
In fact, \eqref{Quant_cond2} coincides with the quantization condition \eqref{Q_quantization_cond} of the $Q$-function, which is a non-perturbative proof (with accuracy to a factor not containing zeros in $\lambda$) of the relations between spectral determinants, see for details Appendix \ref{matrix-el-of-K}.
\section{Solutions of TQ equation}\label{TQ-Solutions}
In this section we adopt the approach proposed in \cite{Fateev:2009jf} and refined in \cite{Litvinov:2024riz,Artemev:2025cev}. Specifically, we look for a solution to the TQ equation \eqref{TQ-equation}  in the two asymptotic regimes---small and large energies---in the form of a formal power series:
\begin{equation}
    \begin{aligned}
        \lambda \to 0:\quad &Q_{\pm}(\nu|\lambda)=\sum_{k=0}^{\infty}q_{\pm}^{(k)}(\nu)\lambda^k,\qquad\\ 
        \lambda \to -\infty:\quad &Q_{\pm}(\nu|\lambda)=\sum_{k=0}^{\infty}\left((-\lambda)^{-\frac{i\nu}{2}}Q_{\pm}^{(k)}(\nu)\pm(-\lambda)^{\frac{i\nu}{2}}Q_{\pm}^{(k)}(-\nu)\right)\lambda^{-k},
    \end{aligned}
\end{equation}
where each of the coefficients $q_{\pm}^{(k)}(\nu)$, $Q_{\pm}^{(k)}(\nu)$ are analytic in the strip $\textrm{Im }\nu\in[-2,2]$ and satisfies the boundedness condition \eqref{Q-is-bounded}. Furthermore, imposing the symmetry condition \eqref{Qpm-symmetry} along with the normalization \eqref{Qpm-normalization-conditions} is expected to ensure the uniqueness of the solution with the desired properties. Importantly, such a construction should be entirely equivalent to the iterative solution of the inhomogeneous integral equation \eqref{BS-eq-inhomogeneous}. Nevertheless, direct evaluation of the arising integrals appears to be technically rather cumbersome. The method outlined below significantly simplifies this process, allowing many of the relevant integrals to be evaluated from the outset.
\subsection{Small \texorpdfstring{$\lambda$}{} expansion}
Following \cite{Fateev:2009jf, Litvinov:2024riz, Artemev:2025cev}, we note that the TQ equation \eqref{TQ-equation} admits solutions in the form of two series associated with the confluent hypergeometric function
\begin{equation}\label{Xi_and_Sigma-def}
    \begin{aligned}
        \Xi(\nu|\lambda)=&\;(\nu+\alpha x)\sum\limits_{k=0}^\infty\frac{\left(1+\frac{i(\nu+\alpha x)}{2}\right)_k}{k!(k+1)!}(-iz)^k=(\nu+\alpha x)F\left(\genfrac{}{}{0pt}{1}{1+\frac{i(\nu+\alpha x)}{2}}{2}\biggl|-iz\right),
        \\
        \Sigma(\nu|\lambda)=&\;1+\sum\limits_{k=1}^\infty\frac{\left(\frac{i(\nu+\alpha x)}{2}\right)_k}{k!(k-1)!}\left(\uppsi_\alpha\left(\nu-2i(k-1)\right)-\psi(k)-\psi(k+1)+\psi\left(\frac{1}{2}\right)+\log16\right)(-iz)^k,
    \end{aligned}
\end{equation}
where $x$, $z$ are defined in \eqref{x-z-notation}, $\psi(k)$ is digamma function, $F\left(\genfrac{}{}{0pt}{1}{a}{b}\big|c\right)$ is the confluent hypergeometric function.

These basic solutions in form completely coincide with those proposed in \cite{Litvinov:2024riz}. The difference here is in the functions $x$ and $z$, which now contain $\coth{\frac{\pi \nu}{2}}$ instead of $\tanh{\frac{\pi \nu}{2}}$. In the definition of $\Sigma(\nu|\lambda)$ we used the function $\uppsi_\alpha(\nu)$, which is defined as (see it's analog for 't Hooft model in \cite{Litvinov:2024riz})
\begin{equation}
    \uppsi_\alpha(\nu+i)=-\gamma_E-\log4+\frac{1}{2}\int_{-\infty}^{\infty}\limits\frac{1}{t+\frac{2\alpha}{\pi}\coth{\frac{\pi t}{2}}}\left(\tanh{\frac{\pi t}{2}}-\tanh{\frac{\pi(t-\nu)}{2}}\right)dt.
\end{equation}
One can check that $\uppsi_\alpha(\nu)$ is a solution of the difference equation
 \begin{equation} \label{dif-eq-for-psi}           \uppsi_\alpha(\nu+2i)=\uppsi_\alpha(\nu)+\frac{2i}{\nu+\frac{2\alpha}{\pi}\coth{\frac{\pi\nu}{2}}}.
\end{equation}
We observe that this relation ensures the analyticity of the function $\uppsi_\alpha(\nu)$ in the strip Im $\nu \in [-2,2]$, except at the single point $\nu = -i \nu_1^*$. Here, $-i \nu_1^*$ denotes the solution to the transcendental equation \eqref{main-transcendental-equation} that lies closest to the origin in the lower half-plane. However, this pole does not affect our construction, since it is precisely canceled by the factor $\left(\frac{i(\nu+\alpha x)}{2}\right)_k$ in \eqref{Xi_and_Sigma-def}.

It will be useful for us to find the expansion of $\uppsi_\alpha(\nu)$ at points $\nu=0$ and $\nu=\pm 2i$ (this is due to the fact that soon we will have to solve the problem of having poles for functions $\Xi(\nu|\lambda)$ and $\Sigma(\nu|\lambda)$ at these points):
\begin{equation}
    \uppsi_\alpha(\nu)=\uppsi_0+\uppsi_1\nu+\uppsi_2\nu^2+...
\end{equation}
We have
\begin{equation}
    \uppsi_1 = \frac{i \pi^2}{4 \alpha},\; \uppsi_3 = -\frac{i \pi^4 (3+\alpha)}{48 \alpha^2},\; \uppsi_5 = \frac{i \pi^6 (15+10 \alpha+2 \alpha^2)}{960 \alpha^3}, \; \uppsi_7 = -\frac{i \pi^8 (315+\alpha(315+17 \alpha(7+\alpha)))}{80640 \alpha^4} 
\end{equation}
and 
\begin{equation}
    \begin{aligned}
        \uppsi_0=&\;-\gamma_E-\log4-\mathtt{u}_0(\alpha), \quad \uppsi_2=\frac{\zeta(3)}{4}+\frac{\pi^2\alpha}{8} \mathtt{u}_3(\alpha), \quad \uppsi_4 = -\frac{\zeta(5)}{16}+\frac{\pi^4 \alpha}{96}(3\mathtt{u}_5(\alpha)+\mathtt{u}_3(\alpha)),
        \\
        \uppsi_6=&\;\frac{\zeta(7)}{64}+\frac{\pi^6\alpha}{5760} (45\mathtt{u}_7(\alpha)+30\mathtt{u}_5(\alpha)+2\mathtt{u}_3(\alpha)),
        \\ 
        \uppsi_8=&\;-\frac{\zeta(9)}{256}+\frac{\pi^8\alpha}{161280}(315\mathtt{u}_9(\alpha)+315\mathtt{u}_7(\alpha)+63\mathtt{u}_5(\alpha)+\mathtt{u}_3(\alpha)),
    \end{aligned}
\end{equation}
where
\begin{equation}\label{u-def}
    \mathtt{u}_0(\alpha)\overset{\text{def}}{=}\int_{-\infty}^\infty\limits dt\frac{1}{2\cosh{t}(\alpha\cosh{t}+t\sinh{t})},\quad \mathtt{u}_{2k-1}(\alpha)\overset{\text{def}}{=}\fint_{-\infty}^{\infty}\limits dt\frac{\cosh^2{t}}{t\sinh^{2k-1}{t}\cdot(\alpha\cosh{t}+t\sinh{t})},
\end{equation}
To find the expansion at $\nu =\pm 2i$ we shall use \eqref{dif-eq-for-psi}.

Let us define two functions, 
\begin{equation}
    M_-(\nu|\lambda)=e^{\frac{iz}{2}}\Xi(\nu|\lambda),\quad M_+(\nu|\lambda)=\frac{1}{2}\left(e^{\frac{iz}{2}} \Sigma(\nu)+e^{-\frac{iz}{2}}\Sigma(-\nu)\right),
\end{equation}
so that 
\begin{equation}
    M_\pm(-\nu|\lambda) = \pm M_\pm(\nu|\lambda).
\end{equation}
The functions $M_\pm(\nu|\lambda)$ solve the TQ-equation, but they do not satisfy the properties we require. Namely, due to the presence of the factors $(-iz)^k$ they have poles at the points $0,\pm 2i$. Thus, similar to \cite{Litvinov:2024riz} we will look for the solution of TQ equation \eqref{TQ-equation} in the following form 
\begin{equation}
    \quad Q_{\pm}(\nu|\lambda) =A_{\pm}(\tau|\lambda)M_{\pm}(\nu|\lambda)+B_{\pm}(\tau|\lambda)zM_{\mp}(\nu|\lambda),\quad \tau\overset{\text{def}}{=}\frac{\pi^2}{4}\coth^2\left(\frac{\pi\nu}{2}\right),
\end{equation}
where $A_{\pm}(\tau|\lambda)$ and $B_{\pm}(\tau|\lambda)$ admit expansions in the parameter $\lambda$ with coefficients that are polynomials in $\tau$
\begin{equation}\label{A-B-def}
    A_\pm(\tau|\lambda)=1+\sum_{s=1}^{\infty}\limits a^{(s)}_\pm(\tau)\lambda^s,\quad B_\pm(\tau|\lambda)=-(1\mp1)\frac{\alpha}{2\pi^2}\lambda^{-1}+\sum_{s=0}^{\infty}\limits b^{(s)}_\pm(\tau)\lambda^s,
\end{equation}
with normalization condition $A_\pm(0|\lambda)=1$. The functions $a^{(s)}_\pm(\tau)$ and $b^{(s)}_\pm(\tau)$ are polynomials in $\tau$ of
degree $s$ and $s+1$ respectively. We determine these expansion coefficients by requiring the Q-function to be pole-free at $\nu = 0$ and $\nu=\pm 2i$. For example
\begin{equation}
\begin{aligned}
    &a^{(1)}_+(\tau)=\frac{8\alpha}{\pi^2}\tau,\quad a^{(2)}_+(\tau)=-\left[2+\frac{8\alpha\left(\pi^2-\alpha\zeta(3)\right)}{\pi^4}-\frac{4\alpha^3\mathtt{u}_3(\alpha)}{\pi^2}\right]\tau+\frac{16\alpha^2}{\pi^4}\tau^2,
    \\
    &b^{(0)}_+(\tau)=\frac{\mathtt{u}_0(\alpha)}{2},\quad b^{(1)}_+(\tau)=2+\frac{\pi^2}{4\alpha}-\alpha^2\mathtt{u}_3(\alpha)-\frac{2\alpha\zeta(3)}{\pi^2}+\frac{4\alpha\left(\mathtt{u}_0(\alpha)+2\right)}{\pi^2}\tau,
\end{aligned}
\end{equation}
and
\begin{equation}
\begin{aligned}
    a^{(1)}_-(\tau)=&- \frac{8 \alpha}{\pi} \left(1+\mathtt{u}_0(\alpha) \right) \tau,\quad a^{(2)}_-(\tau)=-\Biggl[2+\frac{24\alpha}{\pi^2}-\frac{24\alpha^2\zeta(3)}{\pi^4}+\frac{16\alpha}{\pi^4}(\pi^2-\alpha\zeta(3))\mathtt{u}_0(\alpha)-
    \\&-\frac{4\alpha^3}{\pi^2}(3+2\mathtt{u}_0(\alpha))\mathtt{u}_3(\alpha)\Biggl]\tau-\frac{16\alpha^2\left(5+2\mathtt{u}_0(\alpha)\right)}{\pi^4}\tau^2,
    \\
    b^{(0)}_-(\tau)=&-\frac{2\alpha}{\pi^2}+\frac{2\alpha^2\zeta(3)}{\pi^4}+\frac{\alpha^3\mathtt{u}_3(\alpha)}{\pi^2}-\frac{4\alpha^2}{\pi^4}\tau,\quad b^{(1)}_-(\tau)=\frac{2}{3}-\frac{2\alpha}{\pi^2}-\frac{4\alpha^2}{3\pi^2}+\frac{8\alpha^2(3+\alpha)\zeta(3)}{3\pi^4}-
    \\&
     -\frac{4\alpha^3\left(\zeta(3)^2+\zeta(5)\right)}{\pi^6}+\frac{2\alpha^3(\pi^2(2+\alpha)-2\alpha\zeta(3))\mathtt{u}_3(\alpha)}{\pi^4}+\frac{2\alpha^4\mathtt{u}_5(\alpha)}{\pi^2} -\frac{\alpha^5\mathtt{u}^2_3(\alpha)}{\pi^2}+
    \\&
    +\frac{2\alpha}{3\pi^6}\left[\pi^4-12\pi^2\alpha+12\alpha^2\zeta(3)+6\pi^2\alpha^3\mathtt{u}_3(\alpha)\right]\tau-\frac{16\alpha^3}{3\pi^6}\tau^2.
\end{aligned}
\end{equation}
We have calculated explicit expressions for $a^{(s)}_{\pm}(\tau)$ up to $s=6$, and for $b^{(s)}_{\pm}(\tau)$ up to $s=5$. By getting rid of the poles of the Q-function at the desired points, one can obtain analytic expressions for any desired order. 
\subsection{Large \texorpdfstring{$\lambda$}{} expansion}
Here we construct the large $\lambda$ expansion of the functions $Q_\pm(\nu|\lambda)$. For this we note, following \cite{Fateev:2009jf,Litvinov:2024riz,Artemev:2025cev}, that the function 
\begin{equation}\label{S-ansatz}
    S(\nu)=(-\lambda)^{-\frac{i\nu}{2}}S_0(\nu)\sum_{k=0}^\infty\frac{\left(1+\frac{i(\nu+\alpha x)}{2}\right)_k\left(\frac{i(\nu+\alpha x)}{2}\right)_k}{k!}(iz)^{-k},
\end{equation}
solves TQ equation \eqref{TQ-equation} if $S_0(\nu)$ solves the shift equation\footnote{
We note that for $\alpha=0$, $S_0$ may be expressed in terms of the Barnes function 
\begin{equation}
    S^{\alpha=0}_0(\nu)=(2\pi)^{-\frac{1}{2}-\frac{i\nu}{2}}\frac{G\left(\frac{3}{2}+\frac{i\nu}{2}\right) G\left(-\frac{i\nu}{2}\right)}{G\left(\frac{1}{2}-\frac{i\nu}{2}\right)
   G\left(1+\frac{i\nu}{2}\right)},\quad S^{\alpha=0}_0(i)=1.
\end{equation}}
\begin{equation}\label{shift-eq-forS}
    S_0(\nu+2i)=\frac{4\pi\coth{\frac{\pi\nu}{2}}}{\nu+\frac{2\alpha}{\pi}\coth{\frac{\pi\nu}{2}}}S_0(\nu).
\end{equation}
We look for a solution to \eqref{shift-eq-forS} that is analytic in the strip $|\textrm{Im } \nu| \leq 2$, bounded as $|\textrm{Re } \nu| \to \infty$, and has no zeros in the strip $\textrm{Im } \nu \in [0,2]$. The solution method follows Appendix B of \cite{Litvinov:2024riz}, with the modification that the contour $\mathcal{C}$ should now be taken within the strip $\textrm{Im } \nu \in [-1,1]$. The solution with these properties is unique up to a normalization
\begin{equation}\label{S0-integral-representation}
    S_0(\nu+i)=\exp\left[\frac{i}{4}\int_{-\infty}^{\infty}\limits\log\left(\frac{4\pi\coth\left(\frac{\pi t}{2}\right)}{t+\frac{2\alpha}{\pi}\coth\left(\frac{\pi t}{2}\right)}\right)\left(\tanh\frac{\pi(t-\nu)}{2}-\tanh\frac{\pi t}{2}\right)dt\right],\quad
    S_0(i)=1.
\end{equation}
The solution \eqref{S0-integral-representation} is automatically analytic in the strip $\text{Im }\nu \in [0,2]$. Its analytic continuation to the domain $\text{Im }\nu \in [-2,0]$ is achieved through the functional relation \eqref{shift-eq-forS}:
\begin{equation}
    S_0(\nu -2i) = \frac{(\nu -2i) \sinh\left(\frac{\pi\nu}{2}\right) + \frac{2\alpha}{\pi} \cosh\left(\frac{\pi\nu}{2}\right)}{4 \pi \cosh\left(\frac{\pi\nu}{2}\right)} S_0(\nu).
\end{equation}
Thus, $S_0(\nu)$ is analytic in the strip $\textrm{Im }\nu \in [-2,2]$, except for a simple pole at $\nu = -i$. A parallel analysis reveals that $S_0(\nu)$ possesses a meromorphic structure with
\begin{equation}\label{S-zeros-poles}
    \begin{aligned}
        &\text{zeros of order $k$ at:}\quad&&\nu=(2k+1)i,\quad &&&k=1,2,\dots,\\
        &\text{zeros of order $1$ at:}\quad&&\nu=-i\nu_j^{*}-2ki,\quad&&&k=0,1,\dots,\\
        &\text{poles of order $1$ at:}\quad&&\nu=i\nu_j^{*}+2ki,\quad &&&k=1,2,\dots,\\
        &\text{poles of order $k$ at:}\quad&&\nu=-(2k-1)i,\quad &&&k=1,2,\dots,
    \end{aligned}
\end{equation}
where $\pm i\nu_j^*$ are the solutions of the transcendental equation \eqref{main-transcendental-equation}.

It is useful to note that from the integral representation \eqref{S0-integral-representation} it follows that
\begin{equation}
    S_0(\nu)S_0(-\nu)=\frac{\nu+\frac{2\alpha}{\pi}\coth{\frac{\pi\nu}{2}}}{4\pi\coth{\frac{\pi\nu}{2}}}.
\end{equation}

For further analysis, we will need the expansion of $S_0(\nu)$ in the vicinity of the points $\nu=0,\pm 2i, \pm i$. Note that, due to the relation \eqref{shift-eq-forS}, it is sufficient to know the expansion only at the points $\nu=0$ and $\nu=i$
\begin{equation}
    \log S_0(\nu)=\sum_{k=0}^{\infty}s_k(\alpha)\nu^k \quad \text{and} \quad
    \log S_0(\nu)=\sum_{k=0}^{\infty}t_k(\alpha)(\nu-i)^k.
\end{equation}
Similar to \cite{Litvinov:2024riz}, it turns out that the coefficients $s_{2k}(\alpha)$ are elementary functions of $\alpha$
\begin{equation}
    s_0(\alpha)=-\frac{1}{2}\log{\frac{2\pi^2}{\alpha}},\quad s_2(\alpha)=\frac{\pi^2}{8\alpha},\quad s_4(\alpha)=-\frac{\pi^4}{192}\frac{3+2\alpha}{\alpha^2},\quad s_6(\alpha)=\frac{\pi^6}{1920}\frac{5+5\alpha+2\alpha^2}{\alpha^3},
\end{equation}
and $s_{2k-1}(\alpha)$ contain integral parts
\begin{equation}
    \begin{aligned}
        &s_1(\alpha)=-\frac{i}{2}(-1+\gamma_E+\log{2\pi})+\frac{i\alpha}{8}\mathtt{i}_2(\alpha),
        \\ 
        &s_3(\alpha)=-\frac{i}{72}(2\pi^2-3\zeta(3))+\frac{i\pi^2\alpha}{96}\mathtt{i}_4(\alpha),
        \\ 
        &s_5(\alpha)=\frac{i}{7200}(14\pi^4-45\zeta(5))+\frac{i\pi^4\alpha}{1920}(\mathtt{i}_4(\alpha)+3\mathtt{i}_6(\alpha)),\\
        &s_7(\alpha)=-\frac{i}{846720}(124\pi^6-945\zeta(7))+\frac{i\pi^6\alpha}{161280}(2\mathtt{i}_4(\alpha)+30\mathtt{i}_6(\alpha)+45\mathtt{i}_8(\alpha)),
    \end{aligned}
\end{equation}
where\footnote{Note that
\begin{equation}\label{u-i2k-relation}
    \mathtt{i}_{2k}(\alpha)=2\mathtt{u}_{2k-1}(\alpha)-2\alpha\mathtt{u}_{2k+1}(\alpha)+2\mathtt{c}_{2k},\quad \text{with}\quad \mathtt{c}_{2k}=\fint_{-\infty}^{\infty}\limits dt \frac{\cosh{t}}{t\sinh^{2k+1}{t}}.
\end{equation}
The first few $\mathtt{c}_{2k}$
\begin{equation}
    \mathtt{c}_{2}=-\frac{2\zeta(3)}{\pi^2},\quad \mathtt{c}_{4}=\frac{2\zeta(3)}{3\pi^2}+\frac{2\zeta(5)}{\pi^4},\quad \mathtt{c}_{6}=-\frac{16\zeta(3)}{45\pi^2}-\frac{4\zeta(5)}{3\pi^4}-\frac{2\zeta(7)}{\pi^6}.
\end{equation}
}
\begin{equation}\label{i2k_def}
    \mathtt{i}_{2k}(\alpha)\overset{\text{def}}{=}\fint_{-\infty}^{\infty}\limits\frac{\cosh{t}(\sinh2t+2t)}{t\sinh^{2k}{t}(\alpha\cosh t+t\sinh t)}dt.
\end{equation}
Also, one can obtain that all $t_{2k}(\alpha)=0$ and
\begin{equation}
    \begin{aligned}
        &t_1(\alpha)=-\frac{i}{2}(1+\gamma_E+\log{8\pi})+\frac{i\alpha}{8}\mathtt{i}_{1}(\alpha),
        \\ 
        &t_3(\alpha)=\frac{i}{72}(2\pi^2+21\zeta(3))-\frac{i\pi^2\alpha}{96}\mathtt{i}_3(\alpha),
        \\
        &t_5(\alpha)=-\frac{i}{7200}(14\pi^4+1395\zeta(5))+\frac{i\pi^4\alpha}{1920}(3\mathtt{i}_5(\alpha)-\mathtt{i}_3(\alpha)),
        \\
        &t_7(\alpha)=\frac{i}{846720}(124\pi^6+120015\zeta(7))-\frac{i\pi^6\alpha}{161280}(45\mathtt{i}_7(\alpha)-30\mathtt{i}_5(\alpha)+2\mathtt{i}_3(\alpha)),
    \end{aligned}
    \end{equation}
where 
\begin{equation}\label{i2k-1_def}
    \mathtt{i}_{2k-1}(\alpha)\overset{\text{def}}{=}\int_{-\infty}^{\infty}\limits\frac{\sinh 2t+2t}{t\cosh^{2k-1}t(\alpha\cosh t+t\sinh t)}dt.
\end{equation}

The ansatz \eqref{S-ansatz} cannot satisfy the analyticity conditions for the $Q$-functions due to the presence of poles originating from the terms $(iz)^{-k}$ (for the same reason, we searched for the coefficients of the $S_0(\nu)$ decomposition above). However, this ansatz will serve as a fundamental building block for constructing the asymptotic expansion of $Q_\pm(\nu|\lambda)$ in the limit $\lambda \to -\infty$. We note that for $\lambda \to +\infty$, the function defined by \eqref{S-ansatz} exhibits exponential growth.
Following \cite{Fateev:2009jf,Litvinov:2024riz} we represent the solution of TQ equation \eqref{TQ-equation} in the form
\begin{equation}\label{Q-ansatz}
    Q_{\pm}(\nu|\lambda)=T(c^{-1}|\lambda)R_{\pm}(c|\lambda)S(\nu)\pm
    T(-c^{-1}|\lambda)R_{\pm}(-c|\lambda)S(-\nu),
\end{equation}
where $S(\nu)$ is given by \eqref{S-ansatz} and we use the notation
\begin{equation}
    c=i\pi \tanh{\frac{\pi\nu}{2}}.
\end{equation}
Since $c(\nu)$ is a $2i$-periodic function, the ansatz \eqref{Q-ansatz} satisfies the TQ-equation for arbitrary $T(c^{-1}|\lambda)$ and $R_{\pm}(c|\lambda)$. These functions play an analogous role to \eqref{A-B-def} in the small-$\lambda$ expansion: they systematically cancel the poles at $\nu = 0, \pm i, \pm 2i$ order-by-order in $\lambda^{-1}$. We seek $T(c^{-1}|\lambda)$ and $R_{\pm}(c|\lambda)$ in the form of large-$\lambda$ asymptotic expansions:
\begin{equation}
    T(c^{-1}|\lambda)=1+\sum_{k=1}^{\infty}T^{(k)}(c^{-1})\lambda^{-k},\quad
    R_{\pm}(c|\lambda)=1+\sum_{k=1}^{\infty}
    R_{\pm}^{(k)}\big(c|\log(-\lambda)\big)\lambda^{-k},
\end{equation}
where $T^{(k)}(c^{-1})$ and $R_{\pm}^{(k)}\big(c|\log(-\lambda)\big)$ are polynomials in their variables. More precisely, the function $R_{\pm}(c|\lambda)$ is responsible for canceling the poles at $\nu=\pm i$, while $T(c^{-1}|\lambda)$ is responsible for canceling the poles at $\nu = 0$ and $\nu = \pm 2i$.
We have found the closed form expression for the function $T(y|\lambda)$ (the same as in the 't Hooft model \cite{Litvinov:2024riz} for one flavor)
\begin{equation}
    T(y|\lambda)=\exp{\left[\alpha f\left(\frac{\alpha}{\pi^2\lambda}\right)y\right]}.
\end{equation}
where
\begin{equation*}
    f(t)=\frac{\sqrt{1-2t}-1+t(1+\log 4)-2t\log(1+\sqrt{1-2t})}{t}=\frac{t}{2}+
    \frac{t^2}{4}+\frac{5t^3}{24}+\frac{7t^4}{32}+\dots
\end{equation*} 
While we have not obtained closed-form expressions for $R_\pm(c|\lambda)$, the pole cancellation condition at each order of $\lambda^{-1}$ uniquely determines the coefficients $R_{\pm}^{(k)}\big(c|\log(-\lambda)\big)$, given a fixed normalization of $Q_{\pm}(\nu|\lambda)$. The most natural and convenient normalization choice is $R_{\pm}(0|\lambda) = 1$, which leads to 
\begin{equation}
    \begin{aligned}
        R_{-}(c|\lambda)&=1-\frac{3c}{8\pi^2}\lambda^{-1}-\frac{c}{128\pi^4}\left(15c+8q^+_1\right)\lambda^{-2}+\dots,\\
        q^-_1&=11(1+\alpha)-6\log{(-\lambda)}-12it_1(\alpha),
    \end{aligned}
\end{equation} 
\begin{equation}
    \begin{aligned}
        R_{+}(c|\lambda)&=1+\frac{c}{8\pi^2}\lambda^{-1}+\frac{c}{128\pi^4}\left(9c+8q^-_1\right)\lambda^{-2}+\dots,\\
        q^+_1&=5(1+\alpha)-2\log{(-\lambda)}-4it_1(\alpha).
    \end{aligned}
\end{equation} 
We have computed the coefficients $R^{(k)}_\pm (c|\log(-\lambda))$ for $k\leq6$.
\section{Relation between \texorpdfstring{$Q$}{}-functions and spectral determinants}\label{D-Q-relations}
In this section, we explore how the solutions of TQ equation \eqref{TQ-equation} derived in Section \ref{TQ-Solutions} can be applied to analyze the meson mass spectrum in the Ising Field Theory. Specifically, we present and explain three non-perturbative relations -- or at least ones valid in both asymptotic regimes -- linking the spectral determinants to the constructed functions $Q_\pm(\nu)$.
\subsection{Resolvent and integral relations}
The operator $\hat{K}$ \eqref{A-K-def} with the kernel
\begin{equation}
    K(\nu,\nu')=\frac{1}{\sqrt{f(\nu)f(\nu')}}\frac{\pi(\nu-\nu')}{2\sinh\frac{\pi(\nu-\nu')}{2}},\quad f(\nu)=\frac{2\alpha}{\pi}+\nu\tanh{\frac{\pi\nu}{2}},
\end{equation}
belongs to the class of completely integrable operators \cite{Its:1980,Its:1990MPhysB}. As it is well established in the literature \cite{Its:1980,Its:1990MPhysB,zbMATH01284258}, the resolvent kernel of such operators, namely the kernel of $\frac{\hat{K}}{1 - \lambda \hat{K}}$, can be expressed in the form
\begin{equation}\label{Resolvent}
    R(\nu,\nu'|\lambda)=\frac{\cosh\frac{\pi\nu}{2}\cosh\frac{\pi\nu'}{2}}
    {\sinh\frac{\pi(\nu'-\nu)}{2}}\sqrt{f(\nu)f(\nu')}\left[\Psi_+(\nu|\lambda)\Psi_-(\nu'|\lambda)-
    \Psi_-(\nu|\lambda)\Psi_+(\nu'|\lambda)\right],
\end{equation}
where $\Psi_{\pm}(\nu|\lambda)$ are solutions of the inhomogeneous equation
\begin{equation}
    f(\nu)\Psi_{\pm}(\nu|\lambda)-\lambda\int_{-\infty}^{\infty}\limits
    \frac{\pi(\nu-\nu')}{2\sinh\frac{\pi(\nu-\nu')}{2}}\Psi_{\pm}(\nu'|\lambda)d\nu'=
    e_{\pm}(\nu),
\end{equation}
with $e_{\pm}(\nu)$ being
\begin{equation}\label{epm-choice}
    e_{+}(\nu)=\frac{1}{\cosh\frac{\pi\nu}{2}}\quad\text{and}\quad
    e_{-}(\nu)=\frac{\nu}{\cosh\frac{\pi\nu}{2}}.
\end{equation}
Formally, the functions $e_{\pm}(\nu)$ can be chosen arbitrarily, provided there exists a function $M(\nu)$ satisfying the condition
\begin{equation}
    \frac{\pi(e_+(\nu)e_-(\nu')-e_+(\nu')e_-(\nu))}{2(M(\nu)-M(\nu'))}=\frac{\pi(\nu-\nu')}{2\sinh\frac{\pi(\nu-\nu')}{2}}.
\end{equation}
The choice \eqref{epm-choice} corresponds to $M(\nu)=-\tanh\frac{\pi\nu}{2}$.

Since this expression coincides with the right-hand side of \eqref{BS-eq-inhomogeneous}, the resolvent $R(\nu,\nu'|\lambda)$ can be written in terms of the functions $Q_{\pm}(\nu)$.  The spectral sums $G^{(s)}_\pm$ of the spectral operator $\hat{K}$ are related to the resolvent by the trace identities
\begin{equation} \label{sum/differ_of_spec_sums}
    \begin{aligned}
        \sum_{s=1}^\infty\limits \left[G_+^{(s)}+G_-^{(s)}\right] \lambda^{s-1}&=C+\fint_{-\infty}^\infty\limits d\nu\left[R(\nu,\nu|\lambda)-R^{(0)}(\nu)\right],
        \\
        \sum_{s=1}^\infty\limits\left[G_+^{(s)}-G_-^{(s)}\right]\lambda^{s-1}&=\int_{-\infty}^\infty\limits d\nu\; R(\nu,-\nu|\lambda).
    \end{aligned}
\end{equation}
The constant $C$ in \eqref{sum/differ_of_spec_sums} is determined by the specific choice of the subtraction term $R^{(0)}(\nu)$, which is introduced to ensure the convergence of the integral. We take $R^{(0)}(\nu)=\frac{1}{\nu\tanh{\frac{\pi\nu}{2}}}$ and with this choice the constant can be shown to be exactly $C=2\log{2\pi}-4$.

Consequently, from \eqref{D-trough-G-def}, both the ratio $D_+(\lambda)/D_-(\lambda)$ and the product $D_+(\lambda)D_-(\lambda)$ of the spectral determinants associated with the operator $\hat{K}$ admit representations via the resolvent, leading to the integral identities that relate $D_{\pm}(\lambda)$ and $Q_{\pm}(\nu)$
\begin{equation}\label{DD-integral}
    \begin{aligned}
        \partial_\lambda\log(D_+D_-)&=2-\fint_{-\infty}^\infty\limits d\nu\left[\frac{1}{f(\nu)}\left(Q_+(\nu)\partial_\nu Q_-(\nu)-Q_-(\nu)\partial_\nu Q_+(\nu)\right)-\frac{1}{\nu\tanh\frac{\pi\nu}{2}}\right],
        \\
        \partial_\lambda\log\left(\frac{D_+}{D_-}\right)&=-\int_{-\infty}^{\infty}\limits\frac{\pi}{f(\nu)}\frac{Q_+(\nu)Q_-(\nu)}{\sinh{\pi\nu}}d\nu.
    \end{aligned}
\end{equation}
By substituting the explicitly constructed in Section \ref{TQ-Solutions} solutions $Q_\pm(\nu)$ of the TQ equation \eqref{TQ-equation} into the right-hand side, the established integral relations \eqref{DD-integral} allow for the analytical computation of the first spectral sums $G^{(1)}_\pm$, as well as the numerical evaluation of higher-order spectral sums $G^{(s)}_\pm$. However, in practice, this approach turns out to be rather inefficient. Furthermore, in analogy with Appendix A of \cite{Litvinov:2024riz}, these integral relations serve as an additional non-trivial consistency check for the more powerful relations presented in the next two subsections.
\subsection{Quantum Wronskian and rational relations}
The so-called ``quantum Wronskian'' plays a central role in the analysis of second-order difference equations of the type \eqref{TQ-equation}. It is constructed from a pair of linearly independent solutions---namely, $Q_+$ and $Q_-$ in our case, as follows:
\begin{equation}
    W(\nu)\overset{\text{def}}{=}Q_-(\nu+i)Q_+(\nu-i)-Q_+(\nu+i)Q_-(\nu-i).
\end{equation}
By definition, this object is $2i$-periodic, a property that follows directly from TQ equation \eqref{TQ-equation}
\begin{equation}
    W(\nu+2i)=W(\nu).
\end{equation}
Given that the functions $Q_\pm(\nu)$ are analytic in the strip $\textrm{Im }\nu\in[-2,2]$ and grow slower than any exponential (as stated in \eqref{Q-is-bounded}), the Wronskian is in fact an entire and bounded function. It must therefore be constant, and its value can be fixed by evaluating it at a single point, such as $\nu=0$
\begin{equation}
    W(\nu)=W(0)\overset{\eqref{Qpm-symmetry}}{=}2Q_+(i)Q_-(i)\overset{\eqref{Qpm-normalization-conditions}}{=}2i.
\end{equation}
The validity of this equation can be directly verified for any $\nu$ by employing the previously constructed expansions of $Q_\pm(\nu)$ in powers of $\lambda$ or $1/\lambda$.

By evaluating the Wronskian at $\nu=i$,
\begin{equation}
 Q_+(0)Q_-(2i)=i,
\end{equation}
one finds that the product $Q_+(0)$ and $Q_-(2i)$ is a regular function of $\lambda$. Furthermore, if one of these functions develops a pole at some point, the other must simultaneously vanish at that point. In fact, a more concrete statement can be deduced \cite{Fateev:2009jf,Litvinov:2024riz} 
\begin{equation}\label{DD-QQ-relation}
    \frac{Q_-(i)}{Q_-(2i)} = \frac{1}{2} \frac{D_-(\lambda)}{D_+(\lambda)} \quad \text{and} \quad \frac{Q_+(i)}{Q_+(0)}= \frac{D_+(\lambda)}{D_-(\lambda)},
\end{equation}
where $D_{\pm}(\lambda)$ are ``non-physical'' spectral determinants for the operator $\hat{K}$.
Using \eqref{newDet-oldDet}, one can express the ratio of the spectral determinants $\mathcal{D}_{\pm}(\lambda)$ 
\begin{equation}\label{DD-QQ-relation-new}
    \frac{Q_-(i)}{Q_-(2i)}=\frac{1-\mathtt{v}(\alpha)}{2}\left(1+\frac{Q'_-(i)-1}{4\pi^2\lambda}\right)^{-1}\frac{\mathcal{D}_-(\lambda)}{\mathcal{D}_+(\lambda)} \quad \text{and} \quad \frac{Q_+(i)}{Q_+(0)}= \frac{1}{1-\mathtt{v}(\alpha)}\left(1+\frac{Q'_-(i)-1}{4\pi^2\lambda}\right)\frac{\mathcal{D}_+(\lambda)}{\mathcal{D}_-(\lambda)}.
\end{equation}

The advantage of \eqref{DD-QQ-relation-new} is that it allows to find the ratio of normalizations of spectral determinants \eqref{dm/dp} and explicit expressions for the difference of the spectral sums.
\subsection{Log-derivative relations}
The expressions in \eqref{DD-QQ-relation} are insufficient for analyzing $D_+(\lambda)$ and $D_-(\lambda)$ separately and must be complemented by additional relations. In the case of the 't Hooft model studied in \cite{Fateev:2009jf,Litvinov:2024riz,Artemev:2025cev}, a different class of identities was discovered. They establish a relation between the logarithmic derivative of the solution to the TQ equation at the point $\nu=i$ and the logarithmic derivatives of the spectral determinants\footnote{For example for the case of equal quark masses in 't Hooft model \cite{Litvinov:2024riz} they are
\begin{equation}\label{2.17-tHooft}
\begin{aligned} 
    &\partial_\lambda\log{D_-(\lambda)}-\frac{\alpha\mathtt{i}_2(\alpha)}{4}=2i\partial_\nu\log{Q_+(\nu)}\Big|_{i},\\
    &\partial_\lambda\log{D_+(\lambda)}-\frac{\alpha\mathtt{i}_2(\alpha)}{4}=2i\left(1-\frac{2\alpha}{\pi^2}\lambda^{-1}\right)\partial_\nu\log{Q_-(\nu)}\Big|_{i},
\end{aligned}
\end{equation}
where $\mathtt{i}_2(\alpha)$ is an integral defined in \cite{Litvinov:2024riz}. While a formal proof of \eqref{2.17-tHooft} is lacking, these identities are highly practical and have successfully passed nontrivial analytical and numerical validations.}.

In the case of the Ising Field Theory, we were able to establish the following relations of this kind for the spectral determinants of the operator $\hat{K}$ \eqref{A-K-def}
\begin{equation}\label{2.17-old}
\begin{aligned} 
    &\partial_\lambda\log{D_-(\lambda)}-\frac{\alpha\mathtt{i}_2(\alpha)}{4}=2i\left(1-\frac{\alpha}{\pi^2}\lambda^{-1}\right)\partial_\nu\log{Q_+(\nu)}\Big|_{2i},\\
    &\partial_\lambda\log{D_+(\lambda)}-\frac{\alpha\mathtt{i}_2(\alpha)}{4}=2i\left(1+\frac{1}{1-\frac{\pi^2\lambda}{\alpha}}\right)\partial_\nu\log{Q_-(\nu)}\Big|_{2i},
\end{aligned}
\end{equation}
where $\mathtt{i}_2(\alpha)$ defined in \eqref{i2k_def}. These identities are compatible with those derived from the quantum Wronskian \eqref{DD-QQ-relation} in the sense that any poles in $\lambda$ on the right-hand side arise only from the zeros of $Q_+(0)\overset{\eqref{TQ-equation}}{=}(1-\frac{\pi^2\lambda}{\alpha})Q_+(2i)$, $Q_-(2i)$.

Using the relations between the spectral determinants $D_\pm$ and $\mathcal{D}_\pm$ \eqref{newDet-oldDet}, these formulas can be rewritten for the spectral determinants of the operator $\hat{\mathcal{K}}$ \eqref{spectral_problem}
\begin{equation}\label{2.17-new}
\begin{aligned} 
    &\partial_\lambda\log{\mathcal{D}_-(\lambda)}-\frac{\alpha\mathtt{i}_2(\alpha)}{4}=2i\left(1-\frac{\alpha}{\pi^2}\lambda^{-1}\right)\partial_\nu\log{Q_+(\nu)}\Big|_{2i}+\partial_\lambda\log{\left(1+\frac{Q'_-(i)-1}{4\pi^2\lambda}\right)},\\
    &\partial_\lambda\log{\mathcal{D}_+(\lambda)}-\frac{\alpha\mathtt{i}_2(\alpha)}{4}=2i\left(1+\frac{1}{1-\frac{\pi^2\lambda}{\alpha}}\right)\partial_\nu\log{Q_-(\nu)}\Big|_{2i}.
\end{aligned}
\end{equation}
Again, the status of equation \eqref{2.17-new} remains unchanged: although it does not have a rigorous proof, it has successfully passed all the tests that we have conducted.
\section{Analytical results}\label{Analytical-results}
\subsection{Spectral sums}
As noted in \eqref{D-trough-G-def}, the coefficients in the small-$\lambda$ expansion of $\log\mathcal{D}_\pm(\lambda)$ correspond to the spectral sums. While these sums do not have direct physical interpretation, they serve as useful benchmark “observables” for validating our analytic results. Moreover, when combined with the WKB expansion discussed in the following subsection, they provide a tool to compute the spectrum numerically with improved accuracy---particularly for the low-lying states, where WKB methods are least reliable.

Our conjectured formulas provide non-perturbative predictions for the values of the spectral sums $G^{(s)}_\pm$ of ``non-physical'' spectral operator $\hat{K}$ \eqref{A-K-def} as well the values $\mathcal{G}^{(s)}_\pm$ of ``physical'' operator $\hat{\mathcal{K}}$ \eqref{spectral_problem}. By applying the logarithmic derivative relations \eqref{2.17-new}, we are able to compute these sums exactly, one by one. The first few results are listed below
\begin{equation}\label{G1pm-analytical}
    \begin{aligned}
        \mathcal{G}^{(1)}_+=&\;\log(2\pi)-3-\frac{\alpha}{4}\mathtt{i}_2(\alpha),\\
        \mathcal{G}^{(1)}_-=&\;\log(2\pi)-1-\frac{2\alpha\zeta(3)}{\pi^2}-\frac{\alpha}{4}\mathtt{i}_2(\alpha)-\alpha^2\mathtt{u}_3(\alpha)+\mathcal{V}(\alpha)\bra{p}\hat{K}\ket{p},
    \end{aligned}
\end{equation}
\begin{equation}\label{G2pm-analytical}
    \begin{aligned}
        \mathcal{G}^{(2)}_+=&\;2+\frac{\pi^2}{\alpha}-\frac{4\alpha}{3}-\frac{4\alpha^2\zeta(5)}{\pi^4}+\frac{8\alpha^2\zeta(3)}{3\pi^2}+2\alpha^3\left(\mathtt{u}_3(\alpha)+\mathtt{u}_5(\alpha)\right),
        \\
        \mathcal{G}^{(2)}_-=&\;-\frac{\pi^2}{\alpha}+\frac{8\alpha}{3}-\frac{8\alpha}{3\pi^2}(3+2\alpha)\zeta(3)+\frac{4\alpha^2}{\pi^4}\left(\zeta^2(3)+2\zeta(5)\right)-
        \\
        &-4\alpha^2\mathtt{u}_3(\alpha)\left(1+\alpha-\frac{\alpha\zeta(3)}{\pi^2}\right)+\alpha^4\mathtt{u}^2_3(\alpha)-4\alpha^3\mathtt{u}_5(\alpha)+ 2\mathcal{V}(\alpha)\bra{p}\hat{K}^2\ket{p}+\mathcal{V}^2(\alpha)\bra{p}\hat{K}\ket{p}^2,
    \end{aligned}
\end{equation}
\begin{equation}\label{G3pm-analytical}
    \begin{aligned}
        \mathcal{G}^{(3)}_+=&\;\frac{10\pi^2}{9}-\frac{3\pi^2}{\alpha}-\frac{4}{15}\left(10+6\alpha^2-5\alpha(2-\zeta(3))\right)+\frac{184\alpha^3\zeta(3)}{45\pi^2}+\frac{2\alpha}{\pi^2}\left(1-\frac{16\alpha^2}{3\pi^2}\right)\zeta(5)+\frac{8\alpha^3\zeta(7)}{\pi^6}-
        \\
        &-\alpha^2\left(\left(\pi^2-4\alpha^2\right)\mathtt{u}_3(\alpha)+\left(\pi^2-8\alpha^2\right)\mathtt{u}_5(\alpha)-4\alpha^2\mathtt{u}_7(\alpha)\right),
        \\
        \mathcal{G}^{(3)}_-=&\;-\frac{4\pi^2}{3}+\frac{4\alpha(5+3\alpha)}{5}+2\alpha^2(3+2\alpha)\left(\frac{2\zeta(3)}{\pi^2}+\alpha\mathtt{u}_3(\alpha)\right)^2-\alpha^3\left(\frac{2\zeta(3)}{\pi^2}+\alpha\mathtt{u}_3(\alpha)\right)^3-
        \\&
        -\frac{\alpha\left(\pi^2-4\alpha(3+2\alpha)\right)}{3}\left(\frac{6\zeta(5)}{\pi^4}-\alpha\left(\mathtt{u}_3(\alpha)+3\mathtt{u}_5(\alpha)\right)\right)+
        \\&+2\left(\frac{2\zeta(3)}{\pi^2}+\alpha\mathtt{u}_3(\alpha)\right)\left(\pi^2+\frac{(\pi^2-9)\alpha}{3}-\frac{(90+23\alpha)\alpha^2}{15}-\frac{6\alpha^3\zeta(5)}{\pi^4}+\alpha^4\left(\mathtt{u}_3(\alpha)+3\mathtt{u}_5(\alpha)\right)\right)+
        \\&-2\alpha^3\left(\frac{\alpha}{15}\left(\frac{6\zeta(7)}{\pi^6}+2\mathtt{u}_3(\alpha)+30\mathtt{u}_5(\alpha)+45\mathtt{u}_7(\alpha)\right)\right)+
        \\&+3\mathcal{V}(\alpha)\bra{p} \hat{K}^3\ket{p}+3\mathcal{V}^2(\alpha)\bra{p}\hat{K}^2\ket{p}\bra{p} \hat{K}\ket{p}+\mathcal{V}^3(\alpha)\bra{p}\hat{K}\ket{p}^3.
    \end{aligned}
\end{equation}
For clarity, we omit the explicit expressions for the matrix elements, as including them would make the formulas unwieldy. The first two values of these matrix elements can be found in Appendix \ref{matrix-el-of-K}. The first few spectral sums, as well as matrix elements (up to $s=5$), are compiled in the supplementary file \texttt{Spectral-sums-Ising.nb}.

Let us emphasize that the first spectral sums $G^{(1)}_\pm$ can be obtained analytically from the integral relations \eqref{DD-integral} by substituting the leading terms in the asymptotic expansions of the $Q$-functions
\begin{equation}
    Q_+(\nu)=1+\mathcal{O}(\lambda),\quad Q_-(\nu)=\nu+\mathcal{O}(\lambda).
\end{equation}
Moreover, expressions for the differences of spectral sums can be derived from the rational Wronskian identities \eqref{DD-QQ-relation-new}. We have also performed a numerical verification of the resulting spectral sums using a standard discretization method, and our results agree with the numerical values to high precision.
\subsection{WKB expansion}\label{WKB-section}
For large negative values of $\lambda$, the spectral determinants $\mathcal{D}_\pm(\lambda)$ admit the following asymptotic expansion (it follows directly from its definition \eqref{Determinants-def})
\begin{equation}\label{F_definition}
    \mathcal{D}_{\pm}(\lambda)=d_{\pm}\left(2\pi e^{-2+\gamma_E}\right)^{\lambda}(-\lambda)^{\lambda+\frac{1}{8}\pm\frac{1}{4}}
    \exp\Bigl(F^{(0)}_{\pm}(L)+F^{(1)}_{\pm}(L)\lambda^{-1}+F^{(2)}_{\pm}(L)\lambda^{-2}+\dots\Bigr),
\end{equation}
with $F^{(k)}_{\pm}(L)$ being the polynomials in $L=\log(-2\pi\lambda)+\gamma_E$. The factor $(2\pi e^{-2+\gamma_E})^\lambda$ in \eqref{F_definition} precisely determines the form of the additional term $-\frac{\alpha\mathtt{i}_2(\alpha)}{4}$ on the left-hand side of the logarithmic derivative identities \eqref{2.17-new}. It is fixed by matching the $\mathcal{O}(\lambda^0)$ terms in the limit 
$\lambda \to -\infty$ on both sides of the relation. By employing the asymptotic expansion of the $Q$-functions and integrating the log-derivative identities with respect to $\lambda$, one can explicitly determine the coefficients $F^{(k)}_{\pm}(L)$ order by order in $1/\lambda$; for example:
\begin{equation}
    \begin{aligned}
        F^{(0)}_{\pm}(L)=&\;-\frac{\alpha L^2}{2\pi^2}+\frac{\alpha\mathtt{i}_2(\alpha)L}{4\pi^2},
        \\
        F^{(1)}_{+}(L)=&\;\frac{\left(3\pi^2+4\alpha^2\right)L}{8\pi^4}-\frac{8\alpha^2-2\pi^2(1-9\alpha+12\log2)+4\alpha^3\mathtt{i}_2(\alpha)+3\pi^2\alpha\mathtt{i}_1(\alpha)}{32\pi^4},
        \\
        F^{(1)}_{-}(L)=&\;\frac{\alpha^2L}{2\pi^4}-\frac{\pi^2+4\alpha^2-7\pi^2\alpha+2\alpha^3\mathtt{i}_2(\alpha)}{16\pi^4},
    \end{aligned}
\end{equation}
where $\mathtt{i}_1(\alpha)$ and $\mathtt{i}_2(\alpha)$ are defined in \eqref{i2k-1_def},\eqref{i2k_def}. The relation \eqref{DD-QQ-relation-new} enables us to derive expressions for the differences $F^{(k)}_+(L)-F^{(k)}_-(L)$, offering a valuable non-trivial consistency check. In addition, it provides a tool to compute the ratio\footnote{For the spectral determinants $D_\pm(\lambda)$ of the operator $\hat{K}$, this ratio is $\frac{\sqrt{2\alpha}}{\pi}$.}
\begin{equation}\label{dm/dp}
    \frac{d_-}{d_+}=\frac{1}{1-\mathtt{v}(\alpha)}\frac{\sqrt{2\alpha}}{\pi}
\end{equation}
although we have no way to derive $d_+$ and $d_-$ separately. The ratio given in \eqref{dm/dp} can be rewritten as a rapidly converging infinite product (it follows from the definition \eqref{Determinants-def})
\begin{equation}\label{dp/dm-prod}
    \frac{d_-}{d_+}=\frac{\Gamma({\frac{5}{8}})}{\Gamma({\frac{1}{8}})}\left(\prod_{m=0}^{\infty}\limits\,\frac{m+\frac{5}{8}}{\lambda_{2m+1}}\right)\cdot\left(\prod_{m=0}^{\infty}\limits\,\frac{m+\frac{1}{8}}{\lambda_{2m}}\right)^{-1}.
\end{equation}
We have verified through numerical computation that this product agrees with equation \eqref{dm/dp} to a high degree of accuracy. Equation \eqref{dm/dp} represents a significant and nontrivial prediction of our theoretical framework. Fortunately, the constants $d_{\pm}$ do not influence the derivation of the large-$\lambda$ expansion of the spectrum discussed below.

The expressions in \eqref{F_definition} are asymptotically valid only for large negative values of $\lambda$, and therefore cannot be applied near the positive spectrum points $\lambda_n>0$, which correspond to the meson masses. Nevertheless, for large positive $\lambda$, the spectral determinants $\mathfrak{D}_\pm(\lambda)$ can still be obtained via the following representation:
\begin{equation}
    \mathfrak{D}_{\pm}(\lambda)=\frac{1}{2}
    \left(\mathcal{D}_{\pm}(-e^{-i\pi}\lambda)+\mathcal{D}_{\pm}(-e^{+i\pi}\lambda)\right).
\end{equation}
In this expression, the two terms represent analytic continuations of equation \eqref{F_definition} through the upper and lower half-planes, respectively. This continuation procedure was originally proposed for the 't Hooft model in \cite{Fateev:2009jf}, based on the requirement that the associated $Q$-function decays as $\nu\to\infty$. The method was later confirmed and effectively applied in the subsequent analysis in \cite{Litvinov:2024riz,Artemev:2025cev}. In Ising Field Theory it leads to the following result:
\begin{equation}\label{D-analytical-continuation}
    \mathfrak{D}_{\pm}(\lambda)=2d_{\pm}
    \left(2\pi e^{-2+\gamma_E}\right)^{\lambda}
    \lambda^{\lambda+\frac{1}{8}\pm\frac{1}{4}}\exp{\left(\sum_{k=0}^{\infty}\limits\Xi^{(k)}_{\pm}(l)\lambda^{-k}\right)}\cos\left(
    \frac{\pi}{2}\left[2\lambda+\frac{1}{4}\pm\frac{1}{2}+\sum_{k=0}^{\infty}\Phi^{(k)}_{\pm}(l)\lambda^{-k}\right]\right),
\end{equation}
where $\Xi^{(k)}_{\pm}(l)$ and $\Phi^{(k)}_{\pm}(l)$ are again polynomials of $l$
\begin{equation}
    l=\log{(2\pi\lambda)}+\gamma_E,
\end{equation}
which are symmetrized/antisymmetrized versions of $F^{(k)}_{\pm}(L)$
\begin{equation}
    \Xi^{(k)}_{\pm}(l)\overset{\text{def}}{=}\frac{1}{2}\left(F^{(k)}_{\pm}(l+i\pi)+F^{(k)}_{\pm}(l-i\pi)\right),\quad \Phi^{(k)}_{\pm}(l)\overset{\text{def}}{=}\frac{i}{\pi}\left(F^{(k)}_{\pm}(l-i\pi)-F^{(k)}_{\pm}(l+i\pi)\right).
\end{equation}

The zeros of $\mathfrak{D}_{\pm}(\lambda)$ are governed by the last factor $\cos(\dots)$, which effectively imposes the “quantization conditions” on $\lambda$
\begin{equation}\label{quantisation-condition-lambda}
    2\lambda+\frac{1}{4}\pm\frac{1}{2}+\Phi^{(0)}_{\pm}(l)+\Phi^{(1)}_{\pm}(l)\lambda^{-1}+\dots=2m+1,\quad m=0,1,2,\dots
\end{equation}
The first few phases $\Phi_\pm^{(k)}(l)$ are explicitly given by the following expressions
\begin{equation}\label{Phi-pm-phases}
    \resizebox{\textwidth}{!}{$
    \begin{aligned}
        \Phi^{(0)}_{\pm}(l)=&\;\frac{\alpha^2\mathtt{i}_2(\alpha)}{2\pi^2}-\frac{2\alpha l}{\pi^2},\quad \Phi^{(1)}_{\pm}(l)=\frac{\alpha^2}{\pi^4}+\frac{3(1\pm1)}{8\pi^2},
        \\
        \Phi^{(2)}_+(l)=&\;\frac{8\alpha^3+2\pi^2(11\alpha+5-12\log2)+3\pi^2\alpha\mathtt{i}_1(\alpha)}{16\pi^6}-\frac{3l}{4\pi^4},
        \\
        \Phi^{(3)}_+(l)=&\;\frac{80\alpha^4+8\pi^2\alpha(71\alpha+76-68\log2)-63\pi^4}{192\pi^8}+\frac{(6\log2-7)(6\log2-1)}{12\pi^6}+
        \\&+\frac{\alpha\mathtt{i}_1(\alpha)\left(8(7\alpha+4-6\log2)+3\alpha\mathtt{i}_1(\alpha)\right)}{64\pi^6}-\frac{\left(4(7\alpha+4-6\log2)+3\alpha\mathtt{i}_1(\alpha)\right)l}{8\pi^6}+\frac{3l^2}{4\pi^6},
        \\
        \Phi^{(2)}_{-}(l)=&\;\frac{64\alpha^3-4\pi^2(11+28\alpha-30\log2)-15\pi^2\alpha\mathtt{i}_1(\alpha)}{128\pi^6}+\frac{15l}{32\pi^4}
        \\
        \Phi^{(3)}_{-}(l)=&\;\frac{320\alpha^4-\pi^2\left(1280\alpha^2+8\alpha(197-174\log4)+293+3\log4(75\log4-326)\right)+135\pi^4}{768\pi^8}-
        \\&\;-\frac{\alpha\mathtt{i}_1(\alpha)\left(8(232\alpha+163-75\log4)+75\alpha\mathtt{i}_1(\alpha)\right)}{4096\pi^6}+\frac{\left(4(232\alpha+163-75\log4)+75\alpha\mathtt{i}_1(\alpha)\right)l}{512\pi^6}-\frac{75l^2}{256\pi^6}.
    \end{aligned}
    $}
\end{equation}
More values $\Phi^{(k)}_\pm(l)$ for $k\leq5$ can be found in \texttt{Phi.nb}.

For large values of $\lambda$ (or $m$), equation \eqref{quantisation-condition-lambda} can be inverted to yield the following asymptotic expansion for $\lambda_n$. Since the phases $\Phi^{(k)}_\pm(l)$ begin to differ significantly beyond a certain point, we present the result as two separate series
\begin{equation}\label{WKB-normal-alpha-odd}
    \resizebox{\textwidth}{!}{$
    \begin{aligned}
        \lambda^{\text{odd}}_{n}=&\;\frac{1}{2}\mathfrak{n}+\frac{\alpha}{\pi^2}\log\rho+\frac{\alpha^2}{\pi^4}\frac{2\log\rho-1}{\mathfrak{n}}-\frac{1}{\mathfrak{n}^2}\Biggl[\frac{2\alpha^3\log^2\rho}{\pi^6}-\frac{\left(96\alpha^3-15\pi^2\right)\log\rho}{16\pi^6}+\frac{3\alpha^3}{\pi^6}-
        \\
        &-\frac{28\alpha+11-30\log2}{16\pi^4}-\frac{15\alpha\mathtt{i}_1(\alpha)}{64\pi^4}\Biggl]+\frac{1}{\mathfrak{n}^3}\Biggl[\frac{8\alpha^4\log^3\rho}{3\pi^8}+\frac{\left(15\pi^2(16\alpha+5)-1024\alpha^4\right)\log^2\rho}{64\pi^8}+
        \\
        &+\frac{\left(3072\alpha^4-4\pi^2(8\alpha(28\alpha+55-5\log64)+163-75\log4)-15\pi^2(8\alpha+5)\alpha\mathtt{i}_1(\alpha)\right)\log\rho}{128\pi^8}+
        \\
        &+\frac{-1856\alpha^4+\pi^2\left(1952\alpha^2+1840\alpha-12(292\alpha+163)\log2+293+900\log^22\right)-135\pi^4}{192\pi^8}+
        \\
        &+\frac{\alpha\mathtt{i}_1(\alpha)\left(8(292\alpha+163-150\log2)+75\alpha\mathtt{i}_1(\alpha)\right)}{1024\pi^6}\Biggl]+\mathcal{O}\left(\frac{\log^4{\mathfrak{n}}}{\mathfrak{n}^4}\right),
    \end{aligned}
    $}
\end{equation}
\begin{equation}\label{WKB-normal-alpha-even}
    \resizebox{\textwidth}{!}{$
    \begin{aligned}
        \lambda^{\text{even}}_{n}=&\;\frac{1}{2}\mathfrak{n}+\frac{\alpha}{\pi^2}\log\rho+\frac{1}{\mathfrak{n}}\left[\frac{\alpha^2}{\pi^4}(2\log\rho-1)-\frac{3}{4\pi^2}\right]-\frac{1}{\mathfrak{n}^2}\Biggl[\frac{2\alpha^3\log^2\rho}{\pi^6}-\frac{3\left(4\alpha^3+\pi^2(1+\alpha)\right)\log\rho}{2\pi^6}-
        \\&-\frac{-24\alpha^3+2\pi^2(-17\alpha-5+12\log2)-3\pi^2\alpha\mathtt{i}_1(\alpha)}{8\pi^6}\Biggl]+\frac{1}{\mathfrak{n}^3}\Biggl[\frac{8\alpha^4\log^3\rho}{3\pi^8}-\frac{\left(16\alpha^4+3\pi^2(1+\alpha)^2\right)\log^2\rho}{\pi^8}+
        \\&+\frac{\left(48\alpha^4+2\pi^2\left(20\alpha^2+25\alpha-12(1+\alpha)\log2+8\right)+3\pi^2(1+\alpha)\alpha\mathtt{i}_1(\alpha)\right)\log\rho}{2\pi^8}
        \\&-\frac{464\alpha^4+8\pi^2(\alpha(140\alpha+91-204\log2)+2(6\log2-7)(6\log2-1))-9\pi^4}{48\pi^8}-
        \\&-\frac{\alpha\mathtt{i}_1(\alpha)\left(68\alpha+3\alpha\mathtt{i}_1(\alpha)+32-48\log2\right)}{16\pi^6}\Biggl]+\mathcal{O}\left(\frac{\log^4{\mathfrak{n}}}{\mathfrak{n}^4}\right),
    \end{aligned}
    $}
\end{equation}
where we have introduced
\begin{equation}
    \mathfrak{n}=n+\frac{1}{4}-\frac{\alpha^2\mathtt{i}_2(\alpha)}{2\pi^2},\quad \rho=\pi e^{\gamma_E}\mathfrak{n}.
\end{equation}
The odd values of $n=0,1,2\dots$ correspond to physical mesons, while the even values correspond to non-physical ones (see the explanation in Section \ref{BS-and-Integr}). We provide only the first few leading terms here, although our method allows for the computation of an arbitrary number of terms in the expansion. For the values of $\alpha$ of order one, equations \eqref{WKB-normal-alpha-odd}-\eqref{WKB-normal-alpha-even} remain highly accurate even for relatively small values of $n$. We have collected more terms (including $\mathcal{O}\left(\frac{1}{\mathfrak{n}^5}\right)$) of this expansions in Wolfram Mathematica \texttt{WKB.nb} 
 
As previously noted, for large $n$ (i.e., for heavy mesons satisfying the Bethe-Salpeter equation \eqref{BS-eq}), equation \eqref{WKB-small-alpha} should be understood as providing a two-particle approximation to the real part of the resonance masses in the Ising field theory.
\section{Analysis of results in limiting cases}\label{Limiting-cases-section}
In this section, we will demonstrate how our exact formulas can be applied and simplified in several physically relevant limiting regimes. These limits allow us to establish relations between our results and previously known findings in the literature.
\subsection{Near \texorpdfstring{$E_8$}{} theory: \texorpdfstring{$\alpha\to0$}{}}
For pure magnetic deformation (corresponding to $m=0$ and therefore $\eta=0$), Alexander Zamolodchikov \cite{Zamolodchikov:1989fp,Zamolodchikov:1989zs} demonstrated that the Ising Field Theory becomes an integrable model admitting eight stable particle excitations. The masses of these particles are proportional to the components of the Perron-Frobenius eigenvector of the Cartan matrix associated with the exceptional Lie algebra $E_8$
\begin{equation}
\begin{aligned}
    &M_1,\quad M_2=2\cos{\frac{\pi}{5}}M_1,\quad M_3=2\cos{\frac{\pi}{30}}M_1,\quad M_4=4\cos{\frac{\pi}{5}}\cos{\frac{7\pi}{30}}M_1,\quad M_5=4\cos{\frac{\pi}{5}}\cos{\frac{2\pi}{15}}M_1,
    \\
    &M_6=4\cos{\frac{\pi}{5}}\cos{\frac{\pi}{30}}M_1,\quad M_7=8\cos^2{\frac{\pi}{5}}\cos{\frac{7\pi}{30}}M_1,\quad M_8=8\cos^2{\frac{\pi}{5}}\cos{\frac{2\pi}{15}}M_1.
\end{aligned}
\end{equation}
The lightest mass $M_1$ has been calculated exactly in terms of $|h|^{\frac{8}{15}}$ in \cite{Fateev:1993av}
\begin{equation}
    M_1=\mathcal{C}\;|h|^{\frac{8}{15}},\quad \mathcal{C}=\frac{4\sin\frac{\pi}{5}\Gamma\left(\frac{1}{5}\right)}{\Gamma\left(\frac{2}{3}\right)\Gamma\left(\frac{8}{15}\right)}\left(\frac{4\pi^2\Gamma\left(\frac{3}{4}\right)\Gamma^2\left(\frac{13}{16}\right)}{\Gamma\left(\frac{1}{4}\right)\Gamma^2\left(\frac{3}{16}\right)}\right)^{\frac{4}{15}}=4.40490858...
\end{equation}
Owing to this correspondence, the integrable limit is commonly referred to as the $E_8$ field theory. Furthermore, the structure of the exact $S$-matrices reflects the underlying $E_8$ symmetry. Notably, the five heaviest particles in this spectrum lie above the two-particle decay threshold and remain stable solely due to the constraints imposed by integrability.

Near the integrable $E_8$ point ($|m|\ll1\Rightarrow|\eta|\ll1$; following the analogy with the ’t Hooft model, we will also refer to this as the chiral limit), only the three lightest mesons remain stable. The other five, which lie above the two-particle threshold, become resonances once integrability is broken. Despite this, the integrable point is believed to be analytic: the meson masses $M_n(\eta)$ can be expanded in a convergent power series around $\eta=0$,
\begin{equation}\label{M-chiral-expansion}
    M_n(\eta)=M^{(0)}_n+\eta M^{(1)}_n+\eta^2 M^{(2)}_n+\eta^3 M^{(3)}_n+\dots
\end{equation}
The leading corrections $M^{(1)}_n$ for small $n$ are known exactly from form factor perturbation theory \cite{Delfino:1996xp}, and numerical data from TFFSA \cite{Fonseca:2006au} agrees precisely with these results. Table \ref{tab:mesons-nearE8} presents these values, along with numerical estimates of the next-order terms for the lightest mesons. For higher-order coefficients, only numerical results are available. For $n>3$, the coefficients $M^{(2)}_n$, $M^{(3)}_n$, etc., typically develop nonzero imaginary parts, indicating that the corresponding particles become unstable \cite{Delfino:1996xp,Delfino:2005bh}.

\begin{table}[h!]
    \centering
    \resizebox{\textwidth}{!}{$
    \begin{tabular}{|c||c|c|c||c|c|c||c|c|c||}
    \hline
     & \multicolumn{3}{c||}{IFT} & \multicolumn{6}{c||}{2 particle approximation of IFT}\\
    \hline
    $n$ & $M_n^{(0)}$ & $M_n^{(1)}$ & $M_n^{(2)}$ & $M_n^{(0)}$ & $M_n^{(1)}$ & $M_n^{(2)}$ & $M_n^{(0)}$ & $M_n^{(1)}$ & $M_n^{(2)}$
    \\
    \hline
    1 & 4.404908579981 & 1.295045 & 0.2003 & 4.248274145 & 1.316490 & 0.2815 & 4.248274116 & 1.316490 & 0.2984\\
    \hline
    2 & 7.127291799746 & 1.115886 & 0.2072 & 6.948600321 & 1.077762 & 0.3726 &  
    6.948600276 & 1.077763 & 0.3754
    \\
    \hline
    3 & 8.761556059774 & 1.953268 & $-1.5110$ & 8.838163544 & 1.095065 & 0.3471 & 
    8.838163436 & 1.095066 & 0.3473
    \\
    \hline
    4 & 10.59322004129 & 1.484334 & *.***** & 10.387731645 & 1.144113 & 0.3212 & 
    10.387731843 & 1.144112 & 0.3212
    \\
    \hline
    5 & 13.02221009790 & *.****** & *.***** & 11.733999773 & 1.253651 & *.**** & 
    11.733999280 & 1.201195 & 0.2994
    \\
    \hline
    \end{tabular}
    $}
    \caption{First three rows: $M_n^{(0)}$ \cite{Zamolodchikov:1989fp,Zamolodchikov:1989zs,Fateev:1993av} and $M_n^{(1)}$ \cite{Delfino:1996xp} are exact,  the coefficients $M^{(2)}_n$ are estimated from the TFFSA data \cite{Fonseca:2006au}. Second three rows: expansion for BS mesons from numerical diagonalization in \cite{Fonseca:2006au}. Last three rows: $M^{(0)}_n$, $M^{(1)}_n$ are calculated from the system \eqref{chiral-system}, $M^{(2)}_n$ are obtained from WKB in the chiral limit \eqref{WKB-small-alpha} (all available expansion terms were used).}
    \label{tab:mesons-nearE8}
\end{table}

Within the two-particle approximation studied in this work, the analytic results for the spectral sums \eqref{G1pm-analytical}-\eqref{G3pm-analytical} and the large-$n$ WKB expansion \eqref{WKB-normal-alpha-odd} obtained in Section \ref{Analytical-results} enable us to construct analogous small-$\eta$ expansions for the mesons of the Bethe–Salpeter equation \eqref{BS-eq}. Since our analysis was conducted in terms of the parameter $\alpha$ \eqref{FZ-our-notations}, we now make explicit the relation between $\alpha$ and $\eta$ in the small-$\eta$ limit. As discussed at the end of Section \ref{BS-and-Integr}, in this limit it is appropriate to define the effective string tension $f$ as \eqref{effectife-string-tension-def}. Then, according to the expansion \eqref{f-series}, the parameter $\alpha$ is related to $\eta$ as follows (we assume that $\eta\in\mathbb{R}$)
\begin{multline}
    \alpha=\frac{\pi\eta^2}{2\rho_0}\left(1-\frac{\rho_1}{\rho_0}\eta+\frac{\rho_1^2-\rho_0\rho_2}{\rho_0^2}\eta^2-\frac{\rho_1^3-2\rho_0\rho_2\rho_1+\rho_0^2\rho_3}{\rho_0^3}\eta^3+\mathcal{O}\left(\eta^4\right)\right)\overset{\eqref{rho-coefs}}{\approx} 0.662981\eta^2-0.098535\eta^3+
    \\+0.014645\eta^4+0.002901\eta^5+\mathcal{O}\left(\eta^6\right).
\end{multline}

To derive the asymptotic behavior of the spectral sums, we rely on the known asymptotics of the integrals $\mathtt{u}_{2k-1}(\alpha)$ and $\mathtt{i}_{2k}(\alpha)$ (see Appendix \ref{u-asymptotics-chiral}). In the limit $\alpha\to0$, this leads to the expansion of the spectral sums in powers of $\alpha$ (here we present the first three)
\begin{equation}\label{G1pm-chiral}
    \begin{aligned}
        \mathcal{G}^{(1)}_+(\alpha)\Big|_{\alpha \rightarrow0}=&\;\frac{\pi}{\sqrt{\alpha}}+\log2\pi-3-\frac{\pi\sqrt{\alpha}}{6}+\mathcal{O}(\alpha),
        \\
        \mathcal{G}^{(1)}_-(\alpha)\Big|_{\alpha \rightarrow0}=&\;\log2\pi-\frac{41}{45}-\left(\frac{\pi}{3}+\frac{376}{675\pi}\right)\sqrt{\alpha}+\mathcal{O}(\alpha),
    \end{aligned}
    \end{equation}
\begin{equation}\label{G2pm-chiral}
    \begin{aligned}
        \mathcal{G}^{(2)}_+(\alpha)\Big|_{\alpha\rightarrow0}=&\;\frac{\pi^2}{\alpha} - \frac{2 \pi}{\sqrt{\alpha}}+2+\mathcal{O}(\sqrt{\alpha}),
        \\
        \mathcal{G}^{(2)}_-(\alpha)\Big|_{\alpha\rightarrow0}=&\;\frac{556}{2025}+\frac{\pi^2}{3}-\frac{4\left(8042+11025\pi^2-30375\zeta(3)\right)\sqrt{\alpha}}{30375\pi}+
        \mathcal{O}\left(\alpha\right),
    \end{aligned}
\end{equation}
\begin{equation}\label{G3pm-chiral}
    \begin{aligned}
        \mathcal{G}^{(3)}_+(\alpha)\Big|_{\alpha \rightarrow0}=&\;\frac{\pi ^3}{\alpha^{\frac{3}{2}}}-\frac{3\pi^2}{\alpha}+\frac{4\pi-\frac{\pi^3}{6}}{\sqrt{\alpha}}+\frac{10\pi^2}{9}-\frac{8}{3}+\mathcal{O}\left(\sqrt{\alpha}\right),
        \\
        \mathcal{G}^{(3)}_-(\alpha)\Big|_{\alpha \rightarrow0}=&\;\frac{27604}{91125}+\frac{158\pi^2}{225}-2\zeta(3)-\frac{\left(481756+45\pi^2\left(47356-675\pi^2\right)-5953500\zeta(3)\right)\sqrt{\alpha}}{455625\pi}+\mathcal{O}(\alpha).
    \end{aligned}
\end{equation}
A notable feature of the even spectral sums is their divergent behavior in the limit $\alpha\to0$, as evident from their asymptotic expansions. Based on the structure of the lowest even sums (the more we include, the more expansion terms we can identify), one can infer that the mass of the lightest even meson vanishes in this limit as
\begin{equation}\label{even-meson-alpha=0}
    \lambda_{0}(\alpha)\Big|_{\alpha\to0}=\frac{\sqrt{\alpha}}{\pi}+\frac{\alpha}{\pi^2}+\frac{\left(12+\pi^2\right)\alpha^{3/2}}{18 \pi ^3}-\frac{\left(2\pi^2-3\right)\alpha^2}{18\pi^4}+\mathcal{O}\left(\alpha^{\frac{5}{2}}\right).
\end{equation}
Of course, in the context of the Ising Field Theory, even mesons arise as an artifact of the Bethe–Salpeter equation, and the above result is of limited physical significance. However, in the ’t Hooft model, the divergence of even spectral sums in the chiral limit provides an independent verification of the Gell-Mann–Oakes–Renner relation \cite{Litvinov:2024riz,Artemev:2025cev}.

Using the same integral expansions from Appendix \ref{u-asymptotics-chiral}, we find that the large‑$n$ WKB expansion in the chiral limit takes the following form (here, we present the result only for the odd physical mesons)
\begin{equation}\label{WKB-small-alpha}
\resizebox{\textwidth}{!}{$
\begin{aligned}
    \lambda^{\text{odd}}_{n}(\alpha)\Big|_{\alpha\to0}=&\;\left[n-\frac{3}{8}+\frac{11-15\log\left(8\pi e^{\gamma_E}\left(n-\frac{3}{8}\right)\right)}{64\pi^4\left(n-\frac{3}{8}\right)^2}+\mathcal{O}\left(\frac{\log^2 n}{n^3}\right)\right]+\left[1+\frac{15}{64\pi^2\left(n-\frac{3}{8}\right)^2}+\mathcal{O}\left(\frac{\log n}{n^3}\right)\right]\frac{\sqrt{\alpha}}{\pi}+
    \\&+\left[\log\left(2\pi e^{\gamma_E}\left(n-\frac{3}{8}\right)\right)+\frac{112+15C^{(0)}_{1}}{256\pi^2\left(n-\frac{3}{8}\right)^2}+\mathcal{O}\left(\frac{\log^2n}{n^3}\right)\right]\frac{\alpha}{\pi^2}-
    \\&-\left[\frac{1}{6}-\frac{1}{\pi^2\left(n-\frac{3}{8}\right)}-\frac{5}{128\pi^2(n-\frac{3}{8})^2}+\mathcal{O}\left(\frac{\log n}{n^3}\right)\right]\frac{\alpha^{3/2}}{\pi}+\dots+\#\alpha^{\frac{1}{2}+k}+\dots
\end{aligned}
$}
\end{equation}
Equation \eqref{WKB-small-alpha} represents a direct generalization of the results obtained in \cite{Fonseca:2006au}. In that work, based on numerical estimates of the eigenvalues of the spectral problem, an approximation was proposed for the first few coefficients in the expansion \eqref{WKB-small-alpha}, neglecting $1/n$ corrections. In contrast, our formula provides exact analytic expressions for these expansion coefficients. Expansion \eqref{WKB-small-alpha} indicates that the point $\alpha=0$ corresponds to a square-root branch point (see discussion in Section \ref{BS-analyt-Section}). This behavior arises from the fact that, at $\alpha=0$, each eigenfunction $\Psi_n(\nu)$ develops a simple pole at $\nu=0$. As a result, the associated rapidity-space wavefunction $\Psi_n(\theta)$ approaches a constant $r_n$ at large $|\theta|$.

At the same time, it is evident that for sufficiently small $n$, the accuracy of formula \eqref{WKB-small-alpha} is limited. To obtain more precise values for the low-energy mesons of the Bethe-Salpeter equation \eqref{BS-eq} in the chiral limit, one must supplement the WKB expansion \eqref{WKB-small-alpha} with the asymptotic expansions for the spectral sums given in equations \eqref{G1pm-chiral}-\eqref{G3pm-chiral} (the more spectral sums are included, the higher the resulting accuracy)
\begin{equation}\label{chiral-system}
    \begin{cases}
        \left(\frac{1}{\lambda^{\text{odd}}_1}-1\right)+\ldots+\left(\frac{1}{\lambda^{\text{odd}}_s}-\frac{1}{s}\right)\approx\mathcal{G}^{(1)}_-(\alpha)\Big|_{\alpha\to0}-\sum_{k=s+1}^{\infty}\limits\left[\frac{1}{\lambda^{\text{odd}}_{k}}\Big|_{\text{WKB}}-\frac{1}{k}\right],
        \\
        \vdots
        \\
        \frac{1}{(\lambda^{\text{odd}}_1)^s}+\ldots+\frac{1}{(\lambda^{\text{odd}}_s)^s}\approx\mathcal{G}^{(s)}_-(\alpha)\Big|_{\alpha\to0}-\sum_{k=s+1}^{\infty}\limits\frac{1}{(\lambda^{\text{odd}}_{k})^s}\Big|_{\text{WKB}}.
    \end{cases}
\end{equation}

By matching the coefficients of like powers of $\alpha$ on both sides of the equation, one obtains a linear system that determines the expansion coefficients $M^{(k)}_n$ \eqref{M-chiral-expansion} for the low-lying Bethe-Salpeter mesons (recall that $\lambda_n\sim M_n^2$ \eqref{FZ-our-notations}). In our calculations, we employed all five available spectral sums and first $6$ available terms in WKB expansion (including $\mathcal{O}\left(\frac{1}{n^4}\right)$ terms). The results are presented in Table \ref{tab:mesons-nearE8}. The leading two coefficients agree remarkably well with the numerical values reported in \cite{Fonseca:2006au}, with the exception of the subleading term for the fifth meson. Higher-order coefficients $M^{(k)}_n$, $k>1$ are highly sensitive to the accuracy of the WKB expansion \eqref{WKB-small-alpha} and require the inclusion of additional spectral sums for reliable determination. We also note that the precision of the coefficients $\rho_k$ in \eqref{f-series} plays an important role: starting from $k=3$, only numerical estimates from TFFSA are available \cite{Fonseca:2006au,Fonseca:2001dc}. The entries listed in the table as coefficients of $\eta^2$ correspond to the values obtained from the WKB expansion \eqref{WKB-small-alpha}. Note that their accuracy increases with the order number.
\subsection{Expansion around the free fermion theory: \texorpdfstring{$\alpha\to\infty$}{}}
For purely thermal deformation at low-$T$ regime (corresponding to $h=0$, $m>0$ and therefore $\eta=\infty$), the Ising Field Theory action \eqref{Ising-action} can be reduced to the free Majorana fermion action 
\begin{equation}\label{free-fermion-action}
    \mathcal{A}_{FF}=\frac{1}{2\pi}\int \left[\psi\bar{\partial}\psi+\bar{\psi}\partial\bar{\psi}+m\bar{\psi}\psi \right]\;d^2x.
\end{equation}
The presence of the weak magnetic field $h\ne0$ generates interactions between fermions, thereby breaking the integrability of the theory (we will also refer to this case as the heavy quark mass limit). As a consequence, the spectrum of the system forms a tower of mesonic states whose masses $M_n$ asymptotically approach $2m$ from above.

While all the spectral sums except $\mathcal{G}^{(1)}_\pm$ are expected to vanish as $\alpha\to\infty$, this behavior is not immediately evident from the explicit expressions. However, by applying the large-$\alpha$ asymptotics of the integrals $\mathtt{u}_{2k-1}(\alpha)$, $\mathtt{i}_{k}(\alpha)$ (see Appendix \ref{u-asymptotics-heavy}), one can systematically derive the corresponding expansions of the spectral sums in the limit $\alpha\to\infty$
\begin{equation}\label{G1pm-heavy}
    \resizebox{\textwidth}{!}{$
    \begin{aligned}
        \mathcal{G}^{(1)}_+(\alpha)\Big|_{\alpha\to\infty}=&\;\log\left(\frac{2 \pi^2e^{-\gamma_E-2}}{\alpha }\right)+\frac{\pi^2}{8\alpha}-\frac{\pi^2}{24\alpha^2}+\frac{5\pi^2-16}{24\alpha^3}-\frac{\pi^2(120+7\pi^2)}{2880\alpha^4}+\frac{\pi^2(30+7\pi^2)}{720\alpha^5}+\mathcal{O}\left(\frac{1}{\alpha^6}\right),
        \\
        \mathcal{G}^{(1)}_-(\alpha)\Big|_{\alpha\to\infty}=&\;\log\left(\frac{2\pi^2e^{-\gamma_E-2}}{\alpha}\right)-\frac{\pi^2}{8\alpha}+\frac{2\pi^2}{15\alpha^2}-\frac{640+73\pi^2}{960\alpha^3}+\frac{\pi^2(60165+3592\pi^2)}{161280\alpha^4}-
        \\&-\frac{\pi^2(3409385+821152\pi^2)}{6451200\alpha^5}+\mathcal{O}\left(\frac{1}{\alpha^6}\right),
    \end{aligned}
    $}
\end{equation}
\begin{equation}\label{G2pm-heavy}
    \begin{aligned} 
        &\mathcal{G}^{(2)}_+(\alpha)\Big|_{\alpha\to\infty}=\frac{\pi^2}{3\alpha}+\frac{\pi^4}{16\alpha^2}-\frac{7\pi^4}{60\alpha^3}+\frac{7\pi^4}{24\alpha^4}-\frac{\pi^4\left(588+31\pi^2\right)}{1008\alpha^5}+\mathcal{O}\left(\frac{1}{\alpha^6}\right),
        \\
        &\mathcal{G}^{(2)}_-(\alpha)\Big|_{\alpha\to\infty}=\frac{\pi^2}{3\alpha}-\frac{\pi^4}{16\alpha^2}+\frac{13\pi^4}{84\alpha^3}-\frac{15671}{33600\alpha^4}+\frac{\pi^4\left(1263857+73200\pi^2\right)}{1209600\alpha^5}+\mathcal{O}\left(\frac{1}{\alpha^6}\right),
    \end{aligned}
\end{equation}
\begin{equation}\label{G3pm-heavy}
    \begin{aligned}
        &\mathcal{G}^{(3)}_+(\alpha)\Big|_{\alpha\to\infty}=\frac{\pi^4}{15\alpha^2}+\frac{\pi^6}{32\alpha^3}-\frac{53\pi^6}{420\alpha^4}+\frac{\pi^6}{2\alpha^5}-\frac{\pi^6\left(7210+367\pi^2\right)}{5040\alpha^6}+\mathcal{O}\left(\frac{1}{\alpha^7}\right),\\
        &\mathcal{G}^{(3)}_-(\alpha)\Big|_{\alpha\to\infty}=\frac{\pi^4}{15\alpha^2}-\frac{\pi^6}{32\alpha^3}+\frac{2\pi^6}{15\alpha^4}-\frac{3743\pi^6}{6300\alpha^5}+\frac{\pi^6(34654301+1965600\pi^2)}{19008000\alpha^6}+\mathcal{O}\left(\frac{1}{\alpha^7}\right).     
    \end{aligned}
\end{equation}
These asymptotics confirm that $\mathcal{G}_+^{(s)}>\mathcal{G}_-^{(s)}$, as anticipated from the definition of the spectral sums (since $\lambda_{2n}<\lambda_{2n+1}$), although this inequality is not immediately evident from equations \eqref{G1pm-analytical}-\eqref{G3pm-analytical}. It is also important to highlight that the lowest energy levels yield only a subleading contribution to the spectral sums. The dominant contribution arises from relativistic mesons whose masses deviate substantially from twice the quark mass.  This is true since
\begin{equation}\label{heavy-lambda-estimate}
    \lambda_n \sim M_{n}^2\sim m^2 \sim \alpha\quad \Rightarrow\quad (\lambda_n)^{-s}\sim\alpha^{-s}.   
\end{equation}

Remarkably, the two leading terms in the spectral sums \eqref{G1pm-heavy}–\eqref{G3pm-heavy} coincide with those previously obtained for the ’t Hooft model in \cite{Litvinov:2024riz}, although the parameter $\alpha$ has a different physical meaning in that context. As it was shown in \cite{Litvinov:2024riz}, these contributions can be systematically extracted using a simplified form of the quasi-classical expression (5.26) from \cite{Fonseca:2006au}\footnote{In \cite{Litvinov:2024riz} as a start semiclassical equation we take expressions (3.29), (3.30) from \cite{ZIYATDINOV:2010ModPhysA} which have exactly the same form.}, which describes the meson spectrum in the heavy quark limit, along with its analogue for non-physical ``even'' mesons
\begin{equation}\label{FZ-semiclassics}
    \frac{\sinh{2\theta_n}-2\theta_n}{2}=\pi\lambda_{\textrm{FZ}}\left(n-\left[\frac{1}{2}\pm\frac{1}{4}\right]\right)+\mathcal{O}(\lambda^2_{\textrm{FZ}}),\quad n=1,2,3\dots,
\end{equation}
where $\lambda_{\textrm{FZ}}=\frac{f_0}{m^2}\overset{\eqref{FZ-our-notations}}{=}\frac{\pi}{2\alpha}$, $\theta$ is the rapidity parameter related with the meson masses $\frac{M^2}{4m^2}=\cosh^2{\theta}$. The ``+'' sign corresponds to even energy levels, and ``-'' to odd energy levels.

As it was shown in Section \ref{WKB-section}, the spectrum of highly excited states in the two-particle approximation of the Ising Field Theory exhibits a linear behavior, as described by equation \eqref{WKB-normal-alpha-odd}-\eqref{WKB-normal-alpha-even}. This result applies in the limit $n\to\infty$ for fixed mass $m$. We now turn our attention to a different regime, in which the mass $m$ tends to infinity first, followed by the large excitation number limit $n\gg1$. It is important to emphasize that these two limiting procedures do not commute.

We follow the procedure described for the 't Hooft model in \cite{Artemev:2025cev}. To analyze the limit $\alpha \to \infty$ more thoroughly, it is important to understand how the WKB expansion behaves for levels with $n \lesssim \alpha$. This situation is not straightforward: for large $\alpha$, the expansion parameter $\mathfrak{n}$ can become negative for sufficiently small $n$, rendering the expressions \eqref{WKB-normal-alpha-odd}-\eqref{WKB-normal-alpha-even} inapplicable. From \eqref{heavy-lambda-estimate}, one expects that for such values of $n$, the corresponding eigenvalue remains of order $\alpha$.

Moreover, the phase functions $\Phi_{\pm}^{(k)}$ contain terms that scale with increasing powers of $\alpha$, and as a result, contributions from higher-order phases are no longer suppressed in this regime. A possible resolution is to perform a resummation of the leading contributions in the series \eqref{quantisation-condition-lambda}. Examining the explicit expressions, we observe that the dominant terms scale as $\frac{\alpha^{k+1}}{\lambda^k}$ for $k \geq 1$, while the $k = 0$ term contributes as $\alpha\log\frac{\alpha}{\lambda}$. Subleading corrections are suppressed by additional factors of $\alpha^{-2}$ or smaller.

The phases \eqref{Phi-pm-phases} together with the asymptotic expansions for the integrals $\mathtt{u}_{2k-1}(\alpha),\mathtt{i}_k(\alpha)$ (see Appendix \ref{u-asymptotics-heavy}) allow us to write \eqref{quantisation-condition-lambda} as follows
\begin{equation}
   \frac{2\alpha}{\pi^2}\left(\frac{2}{A}+\log{A}-1-\log{4}+\frac{A}{4}\left(1+\frac{A}{4}+\frac{5A^2}{48}+\frac{7A^3}{128}+\frac{21A^4}{640}+...\right)\right)+\mathcal{O}\left(\frac{1}{\alpha}\right)=n+\frac{1}{2},
\end{equation}
where $A=\frac{2\alpha}{\pi^2\lambda}$, $n=0,1,2,\dots$ The expression in square brackets can be identified as the first terms in the Maclaurin expansion of the hypergeometric series ${}_3F_2\left(1, 1, \frac{3}{2}; 2, 3; A\right).$
It turns out that the coefficient at $\alpha$ vanishes precisely at $A=1$. For $n$ of order one, this condition determines the leading term in the $1/\alpha$ expansion of $\lambda_n$ to be $\lambda_n = \frac{2\alpha}{\pi^2}$. This corresponds to $M_n^2=4m^2+\dots$, as expected in the heavy quark limit. Moreover, the expansion around $A=1$ yields for the left-hand side:
\begin{equation}\label{A-to-1-quantization-cond}
    \frac{2\alpha}{\pi^2}\left(A^{-1}-1\right)^{3/2}\left(\frac{4}{3}-\frac{2(A^{-1}-1)}{5}+\frac{3}{14}(A^{-1}-1)^2+\dots\right)+\mathcal{O}\left(\frac{1}{\alpha} \right)=n+\frac{1}{2},
\end{equation}
which leads to
\begin{equation} \label{large-alpha-wkb}
    \frac{\lambda_n}{2\alpha/\pi^2}=1+\left(\frac{3\pi^2}{8\alpha}(n+1/2)\right)^{2/3}+\dots
\end{equation}
This coincides with the subleading term of the ``relativistic'' expansion in \cite{McCoy:1978ta}, applied for $1\ll n\ll\alpha$.

In the strict $\alpha \to \infty$ limit, the meson mass spectrum is more accurately captured within the non-relativistic approximation, where the corresponding coefficients are proportional to the zeros $\mu_k$ of the Airy function $\mathrm{Ai}(z)$ or its derivative $\mu'_k$
\begin{equation} \label{wkb-nonrel-asympt}
    \lim \limits_{\alpha \to \infty} \frac{\frac{\lambda_n}{2\alpha/\pi^2}-1}{\alpha^{-2/3}} = -\left(\frac{\pi}{2} \right)^{2/3}\cdot 
    \begin{cases}
        \mathrm{\mu}_{k+1}',\quad&n =2k; \\
        \mathrm{\mu}_{k+1},\quad&n=2k+1.
    \end{cases}
\end{equation}
While the approximation given in \eqref{large-alpha-wkb} becomes increasingly accurate as $n$ grows, one might anticipate that it should reproduce the exact result in the large-$\alpha$ limit, since the omitted terms in the quantization condition are formally suppressed by powers of $1/\alpha$. However, this expectation is misleading, as some of these subleading terms may yield leading-order contributions if they possess singular behavior near the point $A \to 1$. A representative example is a correction term of the form $\frac{a/\alpha}{(A^{-1} - 1)^{3/2}}$, which modifies the leading-order solution to the quantization condition \eqref{A-to-1-quantization-cond}. In this case, one finds
\begin{equation}
    A^{-1}-1\approx\left(\frac{3\pi^2}{8\alpha}\right)^{2/3}\left(n+\frac{1}{2}+\frac{8a}{3\pi^2}\frac{1}{n+1/2}\right)^{2/3}\approx\left(\frac{3\pi^2}{8\alpha}(n+1/2)\right)^{2/3}\left(1+\frac{16a}{9\pi^2(n+1/2)^2}+\dots\right).
\end{equation}

The resulting $n$-dependence aligns with the first correction term in the asymptotic expansion for the zeros of the Airy function. Therefore, a systematic resummation of $\mathcal{O}(1/\alpha)$ corrections to the phase functions $\Phi^{(k)}_{\pm}$ would be desirable to validate the asymptotic formula \eqref{wkb-nonrel-asympt} within this approach.
\section{Bethe-Salpeter equation in the complex \texorpdfstring{$\alpha$}{}-plane}\label{BS-analyt-Section}
In this section, we take initial steps toward analyzing the behavior of the Bethe-Salpeter meson masses $\lambda_n(\alpha)$ under analytic continuation of the scaling parameter $\alpha$ into the complex plane, $\alpha\in\mathbb{C}$, a direction first proposed and briefly explored in \cite{Fonseca:2006au}\footnote{For complex values of $\alpha$, the Hamiltonian \eqref{Hamiltonian-def} ceases to be Hermitian, and the spectrum of the theory is no longer guaranteed to remain real; it may instead shift into the complex plane, away from the real axis. There are, however, notable exceptions where a non-Hermitian Hamiltonian can still possess a purely real spectrum. This occurs, for instance, when the Hamiltonian commutes with the $\mathcal{PT}$-symmetry operator. In the $\mathcal{PT}$-symmetric ordered phase, the spectrum remains entirely real despite the non-Hermitian nature of the Hamiltonian. In contrast, when the system enters the phase of spontaneously broken $\mathcal{PT}$ symmetry, the spectrum contains both real eigenvalues and pairs of complex-conjugate levels.}. Building on the results of \cite{Litvinov:2024riz}, we also present previously unknown behavior of meson masses in the ’t Hooft model in the vicinity of complex critical points.

\subsection{Analytical continuation}
As mentioned in the Introduction, the main motivation for studying this question lies in exploring the critical behavior of meson masses in the Ising Field Theory near complex critical points. The hope is that the Bethe-Salpeter equation will capture the key qualitative features of the full theory.

\begin{figure}[h!]
    \centering
    \includegraphics[width=0.49\linewidth]{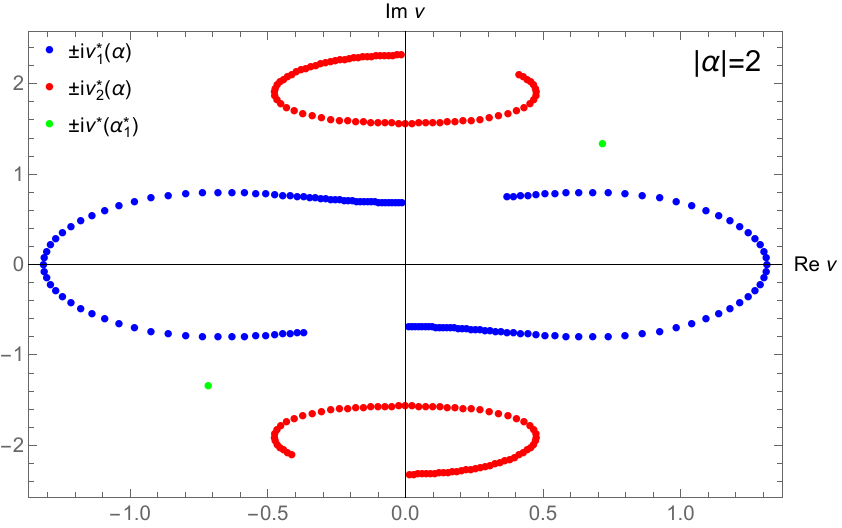}
    \includegraphics[width=0.49\linewidth]{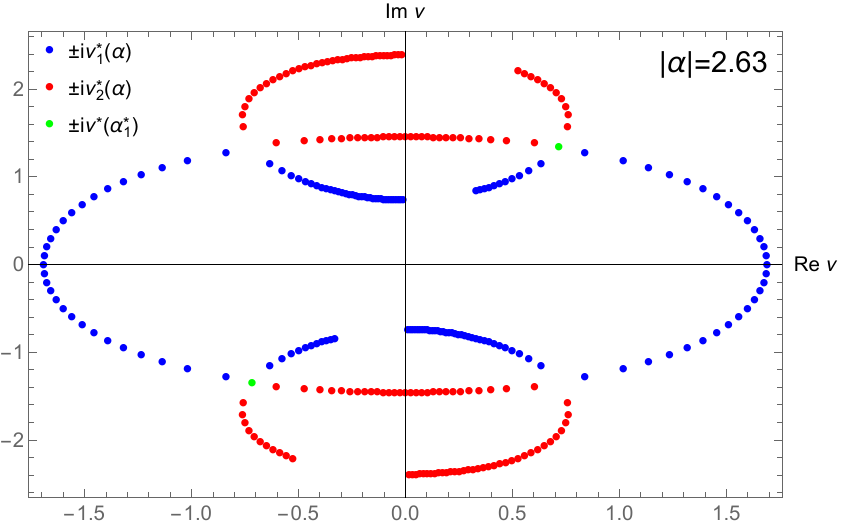}
    \includegraphics[width=0.49\linewidth]{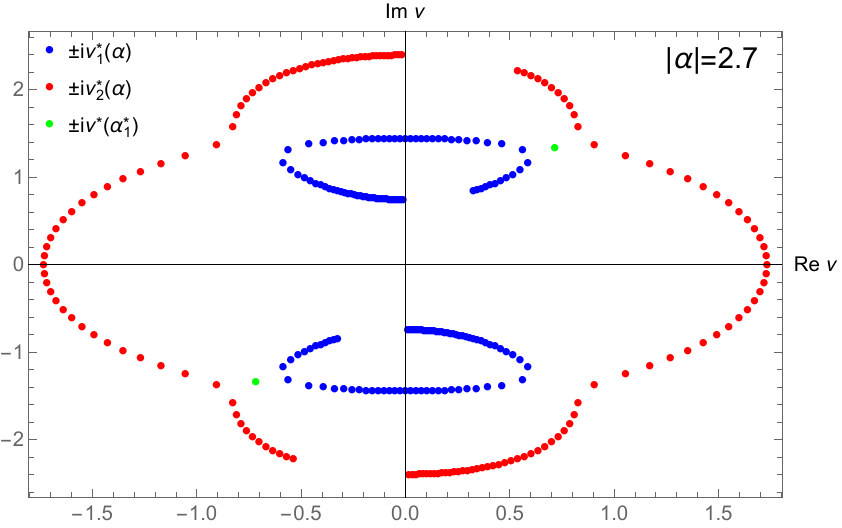}
    \includegraphics[width=0.49\linewidth]{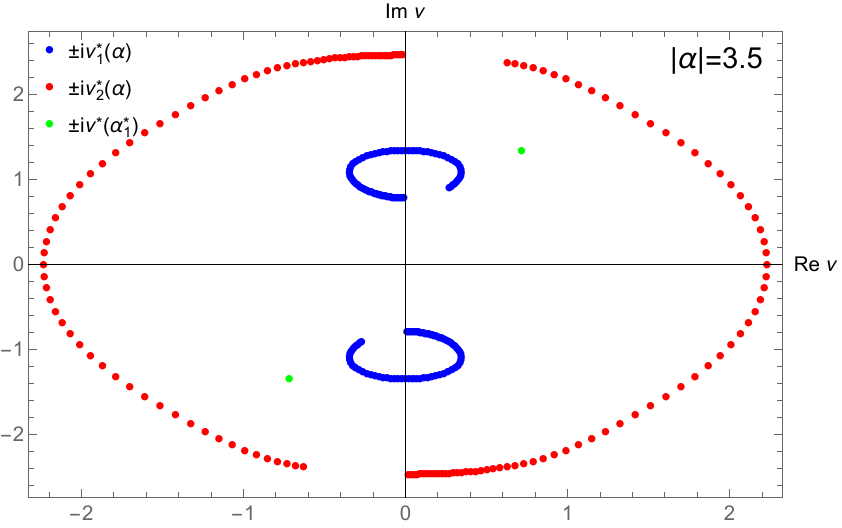}
    \caption{Trajectories of the first two pairs of poles of $\Psi(\nu)$ under the analytic continuation \eqref{analyt-cont}, shown for several values of $|\alpha| = \{2,\ 2.63,\ 2.7,\ 3.5\}$. Each blue and red point corresponds to the positions $\pm i\nu^*_1(\alpha)$ and $\pm i\nu^*_2(\alpha)$ at varying values of $\arg \alpha\in\{0,\frac{3\pi}{2}\}$. The green points mark the location of the double zero of the transcendental equation \eqref{main-transcendental-equation}, given by $\pm i\nu^*(\alpha^*_1)=\pm\frac{2}{\pi}\sqrt{\alpha^*_1(1+\alpha^*_1)}$, which corresponds to the critical value $\alpha^*_1\approx-1.65061-2.05998i$.}
    \label{fig:poles-evolution}
\end{figure}

To investigate the analytic structure of the eigenvalues, we analytically continue the Bethe-Salpeter equation \eqref{BS-eq}, originally defined for real nonnegative $\alpha\geq0$, into the complex scaling parameter plane. For any real $\alpha>0$, this continuation is performed along a circular path in the complex $\alpha$-plane in the counterclockwise direction\footnote{Of course, the continuation can also be performed in the clockwise direction. This only affects the direction of pole evolution and, consequently, the signs of the residue contributions that arise when poles cross the real axis. It also reflects the positions of the critical points: the points in Fig. \ref{fig:crit-points} should be replaced by their complex conjugates.}
\begin{equation}\label{analyt-cont}
    \mathbb{R}\ni\alpha\quad\longmapsto\quad \alpha=e^{i\phi}|\alpha|\in\mathbb{C}.
\end{equation}
The procedure remains straightforward while $|\arg\alpha|<\pi$. However, as $\arg\alpha$ approaches $\pm\pi$, the poles of $\Psi(\nu)$ at $\nu=\pm i\nu^*(\alpha)$ reach the real $\nu$-axis (see Fig. \ref{fig:poles-evolution} and more explanations in the next subsection). 
At this moment, we do not specify which poles of $\Psi(\nu)$ do this, except that there are two of them and they differ in sign. Upon continuing to the second Riemann sheet, where $|\arg\alpha|>\pi$, these poles cross the real axis and contribute additional residue terms. As a result, on the second sheet, the Bethe-Salpeter equation acquires an extra term due to these residues, modifying \eqref{BS-eq} 
\begin{multline}\label{BS_second_worldsheet}
    \left(\frac{2\alpha}{\pi}+\nu\tanh{\frac{\pi\nu}{2}}\right)\Psi(\nu)-\frac{1}{16}\frac{\nu}{\cosh{\frac{\pi\nu}{2}}}\int_{-\infty}^{\infty}\limits\limits d\nu'\frac{\nu'}{\cosh{\frac{\pi\nu'}{2}}}\Psi(\nu')-\lambda(\alpha)\int_{-\infty}^{\infty}\limits d\nu'\frac{\pi(\nu-\nu')}{2\sinh{\frac{\pi(\nu-\nu')}{2}}}\Psi(\nu')-
    \\-2\pi i\left(\underset{\nu'=i\nu^*}{\textrm{Res }}-\underset{\nu'=-i\nu^*}{\textrm{Res }}\right)\left[\frac{1}{16}\frac{\nu}{\cosh{\frac{\pi\nu}{2}}}\frac{\nu'}{\cosh{\frac{\pi\nu'}{2}}}\Psi(\nu')-\lambda(\alpha)\frac{\pi(\nu-\nu')}{2\sinh{\frac{\pi(\nu-\nu')}{2}}}\Psi(\nu')\right]=0.
\end{multline}
Equation \eqref{BS_second_worldsheet} clearly remains valid at $\arg\alpha=2\pi$, that is, after analytic continuation along a closed contour centered at $\alpha=0$. A second continuation along the same closed path generates an additional residue contribution, which provided $|\alpha|$ is sufficiently small---exactly cancels the previous one. This cancellation supports the conclusion that $\alpha=0$ is a square-root branch point of the solution, as discussed below equation \eqref{WKB-small-alpha}.
\subsection{Critical points on the second sheet}
It can be argued that the Bethe-Salpeter masses $M_n$ (recall that $\lambda_n\sim M_n^2$), considered as functions of complex $\alpha$, exhibit an infinite number of  singularities on the second Riemann sheet of the $\alpha$-plane. These singularities can be shown to accumulate near $\alpha=\infty$, making it an essential singularity for all BS masses. They arise in two distinct series, both of which converge toward infinity.

\paragraph{Evolution of the poles of $\Psi$: Part I.}
As discussed above, upon analytic continuation to the second sheet with $|\arg\alpha|>\pi$, the pole at $\nu=i\nu^*(\alpha)$ moves into the lower half-plane, where it can collide with other poles of $\Psi(\nu)$ at specific values of $\alpha=\alpha^*$. Note that only the poles $\pm i\nu^*_k(\alpha)$ of $\Psi(\nu)$   corresponding to $N=0$ in \eqref{Psi-all-poles} are permitted to cross the real axis, and collisions occur exclusively with poles of this type. It should be noted right away that the collision of the poles of $\Psi(\nu)$ is not the only source of singular points, which will be discussed in the third paragraph of this subsection. The trajectories of the first two pairs of poles, $\pm i\nu^*_1(\alpha)$ and $\pm i\nu^*_2(\alpha)$, as functions of $\alpha$, are illustrated in Fig. \ref{fig:poles-evolution}.
 
First, we observe that the pole trajectories are symmetric under the transformation $\nu\to-\nu$ and move counterclockwise when $\arg\alpha>0$. Therefore, it suffices to track the evolution of only the upper pole from the first pair and the lower pole from the second pair. Second, for sufficiently small $|\alpha|<|\alpha^*_1|$, analytic continuation causes the first pole to cross the real axis. As $|\arg\alpha|=2\pi$, the two first poles effectively exchange positions. During this process, the second pole traces out a closed contour and returns to its original location without crossing real axis. Higher-order poles exhibit analogous behavior.

However, as the modulus of $\alpha$ increases, we observe that the trajectory of the first upper pole gradually approaches that of the second lower pole. At a certain critical value $\alpha=\alpha^*_1$, the contours of these poles intersect. Since the poles $i\nu^*_1(\alpha)$, $-i\nu^*_2(\alpha)$ are solutions of the same transcendental equation \eqref{main-transcendental-equation}, this intersection corresponds to a second-order zero of this function. For $|\alpha|>|\alpha_1^*|$, the evolution scenario changes: now the first pole traces a closed path and no longer crosses the real axis, while the second pole begins to cross the real axis and exchanges positions with its partner at $\arg\alpha=2\pi$.

As $\alpha$ increases further, this pattern repeats an infinite number of times at a discrete set of critical points $\alpha=\alpha_k^*$. At each such point, the $k$-th upper pole collides with the $(k+1)$-th lower pole during its evolution. For $|\alpha_{k+1}^*|>|\alpha|>|\alpha_k^*|$, it is the $(k+1)$-th pole $i\nu^*_{k+1}(\alpha)$ that crosses the real axis and undergoes the exchange. And it is precisely at this point that the residue in \eqref{BS_second_worldsheet} should be taken.

\paragraph{Location of critical points from pole collisions of $\Psi$.} The detailed pole collision scenario described above is governed by a system of equations -- namely, the transcendental equation \eqref{main-transcendental-equation} (we also multiplied it by $\cosh{t}$, $t=\frac{\pi\nu}{2}$) together with its derivative with respect to $t$
\begin{equation}\label{crit-points-system}
    \begin{cases} 
      \alpha\cosh{t}+t\sinh{t}&=0;\\
      (1+\alpha)\sinh{t}+t\cosh{t}&=0,
    \end{cases}\quad\Rightarrow\quad
    \begin{cases}
        it^*_k&=\pm\sqrt{\alpha^*_k(1+\alpha^*_k)};\\
        2it^*_k+\sinh{(2it^*_k)}&=0.
    \end{cases}
\end{equation}

\begin{figure}[h!]
    \centering
    \includegraphics[width=0.65\linewidth]{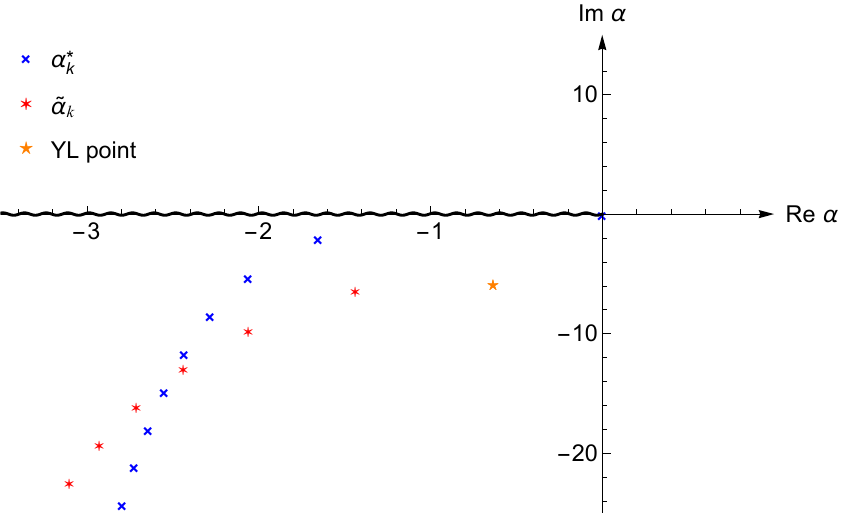}
    \caption{Critical points of $\lambda$ on the second sheet of the $\alpha$-plane correspond to the values $\alpha^*_k$ \eqref{criical-points-approx} (marked in blue) and $\widetilde{\alpha}_k$ \eqref{v=1-eq} (marked in red). The point $\alpha=\alpha^*_0=0$ is a square root branching point. The remaining points $\alpha^*_k$ in accordance with \eqref{lambda-even-critical} are fourth-order branch points on the second sheet. The data for points $\alpha^*_k$, $\widetilde{\alpha}_k$ are given in Tables \ref{tab:crit-points} and \ref{tab:crit-points-2}, respectively. The orange star corresponds to the approximate location of Yang-Lee point $\alpha_{\text{YL}}\approx-0.63-5.95i$ in the full Ising Field Theory.} 
    \label{fig:crit-points}
\end{figure}

This yields an infinite set of solutions to \eqref{crit-points-system}, corresponding to an infinite sequence of values $\alpha^*_k$ located on the second sheet of the $\alpha$-plane ($\arg\alpha>\pi$), beyond the branch cut $(-\infty,0)$. We collect the first several points in Table \ref{tab:crit-points}, see also Fig. \ref{fig:crit-points}. 
\begin{table}[h!]
    \begin{center}
    \resizebox{\textwidth}{!}{$
        \begin{tabular}{| c | c | c | c | c |}
        \hline 
        $k$ & $-it^*_k$ & $-\alpha^*_k$ & $|\alpha^*_k|$ & $\phi^*_k$ \\
        \hline
        0 & 0 & 0 & 0 & $-$\\
        \hline
        1 & $1.12536430580 + 2.10619611525i$ & $1.6506112935 + 2.0599814572i$ & 2.6397047650 & 0.28497600762 \\
        \hline
        2 & $1.55157437291 + 5.35626869864i$ & $2.0578451096 + 5.3347083072i$ & 5.7178526754 & 0.38281140736 \\
        \hline 
        3 & $1.77554367351 + 8.53668242658i$ & $2.2784697373 + 8.5226372750i$ & 8.8219482239 & 0.41684664443 \\
        \hline
        4 & $1.92940449655 + 11.69917761283i$ & $2.4311221188 + 11.6887718663i$ & 11.9389171410 & 0.43472603131 \\
        \hline
        5 & $2.04685246238 + 14.85405991264i$ & $2.5479913751 + 14.8457993910i$ & 15.0628689035 & 0.44589547061 \\
        \hline
        \dots & \dots & \dots & \dots & \dots \\
        \hline
        $k\gg1$ & $\frac{1}{2}\log{\left(4\pi(k-\frac{1}{4})\right)}+\pi(k-\frac{1}{4})i$ & $\frac{1}{2}\log{\left(4\pi e(k-\frac{1}{4})\right)}+\pi(k-\frac{1}{4})i$ & $\pi\left(k-\frac{1}{4}\right)+\frac{\log^2\left(4\pi\left(k-\frac{1}{4}\right)\right)}{8\pi\left(k-\frac{1}{4}\right)}$ & $\frac{1}{\pi}\left(\frac{\pi}{2}-\frac{\log{\left(4\pi e(k-\frac{1}{4})\right)}}{2\pi(k-\frac{1}{4})}\right)$\\
        \hline
        \end{tabular}
        $}
    \end{center}
    \caption{Numerical values for critical points $\alpha^*_k=|\alpha^*_k|e^{i\pi(1+\phi^*_k)}$ and corresponding points $it^*_k$ in the lower-left quadrant of the complex plane.
    \label{tab:crit-points}}
\end{table}
Each critical value of $\alpha^*_k$ (except for the point $\alpha=\alpha^*_0=0$, which was previously discussed in Section \ref{Limiting-cases-section}) corresponds to two values of $i\nu^*_k=\frac{2}{\pi}it^*_k$, one in the upper and one in the lower half-plane. This reflects the fact that one pole evolves downward from the upper half-plane and collides with another pole of $\Psi(\nu)$, while the other evolves upward from the lower half-plane and undergoes a similar collision.
Using the iteration method, we can obtain an approximate formula for the roots $it^*_k$, $\alpha^*_k$ of the system \eqref{crit-points-system}
\begin{equation}\label{criical-points-approx}
    \resizebox{\textwidth}{!}{$
    \begin{aligned}
        -it^*_k&=\left[\frac{1}{2}\log\kappa+\frac{\log^2\kappa-2\log\kappa-\frac{1}{2}}{\kappa^2}+\mathcal{O}\left(\frac{\log^4\kappa}{\kappa^4}\right)\right]+i\left[\frac{\kappa}{4}-\frac{\log \kappa}{\kappa}+\frac{2\log^3\kappa-6\log^2\kappa+2\log\kappa+1}{\kappa^3}+\mathcal{O}\left(\frac{\log^5\kappa}{\kappa^5}\right)\right],\\
        -\alpha^*_k&=\left[\frac{1}{2}(\log\kappa+1)+\frac{2\log^2\kappa-2\log\kappa-1}{2\kappa^2}+\mathcal{O}\left(\frac{\log^4\kappa}{\kappa^4}\right)\right]+i\left[\frac{\kappa}{4}-\frac{2\log\kappa+1}{2\kappa}+\frac{4\log^3\kappa-8\log^2\kappa+1}{2\kappa^3}+\mathcal{O}\left(\frac{\log^5\kappa}{\kappa^5}\right)\right],
    \end{aligned}
    $}
\end{equation}
where $\kappa=4\pi\left(k-\frac{1}{4}\right),$ $k=1,2,3,\ldots$. Even for the first non-zero root ($k=1$), the relative error is less than $1$ percent. 

\paragraph{Critical point analysis via the spectral sums.}
Let us now return to the spectral sums $\mathcal{G}^{(s)}_\pm$ computed in Section \ref{Analytical-results}. When referring to spectral sums $\mathcal{G}^{(s)}_\pm$ at complex values of the parameter $\alpha$, we mean that they are computed using the analytic continuations of the eigenvalues $\lambda_n(\alpha)$. By analyzing their structure, we conclude that these sums may, in principle, exhibit singular behavior at two types of points: the singularities of the integrals $\mathtt{u}_{2k-1}(\alpha)$ \eqref{u-def} (affecting both even and odd sums), and, in the case of odd sums, additional singularities may arise due to the matrix elements when the integral $\mathtt{v}(\alpha)=1$ \eqref{Proector-def}.

We begin our discussion with the first type of potential singularities. By definition, the singular points of the integrals $\mathtt{u}_{2k-1}(\alpha)$ coincide with the solutions $\alpha^*_k\ne0$ (we have already discussed the case $\alpha=0$ in detail in Section \ref{Limiting-cases-section}) of the transcendental equation \eqref{main-transcendental-equation}, whose behavior we have already analyzed in detail above. Using the expansion of the integrals $\mathtt{u}_{2k-1}(\alpha)$ near these points $\alpha^*_k$ (see Appendix \ref{u-asymptotics-critical-points}), we derive the corresponding expansions of the spectral sums in the vicinity of these potential singularities. Here we present only the essential part of the expansions, which allows us to draw conclusions about the behavior of the eigenvalues; the explicit expressions for first three spectral sums are provided in Appendix \ref{u-asymptotics-critical-points}
\begin{equation}
    \resizebox{\textwidth}{!}{$
    \begin{aligned}
        \mathcal{G}^{(s)}_{-}(\alpha)\Big|_{\alpha\to\alpha^*_k}&=\text{const}+\mathcal{O}\left(\sqrt{\alpha-\alpha^*_k}\right),\quad &&\mathcal{G}^{(1)}_{+}(\alpha)\Big|_{\alpha\to\alpha^*_k}=\text{const}+\mathcal{O}\left(\sqrt{\alpha-\alpha^*_k}\right),\\
        \mathcal{G}^{(2)}_{+}(\alpha)\Big|_{\alpha\to\alpha^*_k}&=-\frac{4\pi(1+\alpha^*_k)}{\sqrt{\alpha-\alpha^*_k}}+\mathcal{O}\left(1\right),\quad &&\mathcal{G}^{(3)}_{+}(\alpha)\Big|_{\alpha\to\alpha^*_k}=\frac{2\pi(1+\alpha^*_k)\left(\pi^2-4\alpha^*_k(1+\alpha^*_k)\right)}{\alpha^*_k\sqrt{\alpha-\alpha^*_k}}+\mathcal{O}\left(1\right),\\
        \mathcal{G}^{(4)}_{+}(\alpha)\Big|_{\alpha\to\alpha^*_k}&=\frac{8\pi^2(1+\alpha^*_k)^2}{\alpha-\alpha^*_k}+\mathcal{O}\left((\alpha-\alpha^*_k)^{-\frac{1}{2}}\right),\quad  &&\mathcal{G}^{(5)}_{+}(\alpha)\Big|_{\alpha\to\alpha^*_k}=-\frac{20\pi^2(1+\alpha^*_k)^2\left(\pi^2-4\alpha^*_k(1+\alpha^*_k)\right)}{3\alpha^*_k(\alpha-\alpha^*_k)}+\mathcal{O}\left((\alpha-\alpha^*_k)^{-\frac{1}{2}}\right).
    \end{aligned}  
    $}
\end{equation}
From the expansions presented, it is evident that the odd spectral sums remain finite, while the even ones diverge, with the rate of divergence gradually increasing with the index of the spectral sum. This divergence behavior leads us to conjecture that, under the analytic continuation $\alpha \to \alpha_k^*$, the spectrum contains a pair of mesons $\lambda_{2k_1}, \lambda_{2k_2}$ (precise indices $2k_1$ and $2k_2$ of these even mesons cannot be determined from the form of the spectral sum expansion) whose masses vanish according to the following law:
\begin{equation}\label{lambda-even-critical}
    \lambda^{\pm}_{2k}(\alpha)\Big|_{\alpha\to\alpha^*_k}=\frac{\pm i}{\sqrt{2\pi(1+\alpha^*_k)}}(\alpha-\alpha^*_k)^{\frac{1}{4}}-\frac{\pi^2-4\alpha_k^*(1+\alpha^*_k)}{12\pi\alpha^*_k(1+\alpha^*_k)}\sqrt{\alpha-\alpha^*_k}+\mathcal{O}\left((\alpha-\alpha^*_k)^{\frac{3}{4}}\right).
\end{equation}
In this notation, the subscript $2k$ refers to an even-parity meson pair associated with the critical point $\alpha_k^*$, while the $\pm$ index distinguishes the two mesons within the pair. The existence of these mesons indicates that the points $\alpha_k^*$ are fourth-root branch points. This suggests that for $\alpha>|\alpha_1^*|$, under analytic continuation of the BS equation \eqref{BS_second_worldsheet} by an angle of $4\pi$, i.e., $\alpha \to e^{4\pi i}\alpha$, the pole cancellation observed for small $|\alpha|<|\alpha_1^*|$ no longer occurs. Instead, such cancellation appears to take place only after a full $8\pi$ rotation. 

At first glance, this unusual behavior might appear to be an artifact of the two-particle approximation in the full Ising Field Theory. As we have emphasized multiple times, the even mesons are non-physical: their wavefunctions are even for $\alpha > 0$, which contradicts the nature of mesons composed of two Majorana fermions, whose wavefunction must be odd. However, it is surprising that a similar phenomenon also occurs in the 't Hooft model \eqref{BS-tHooft}. Using the first $7$ spectral sums computed in \cite{Litvinov:2024riz}, we obtain their expansion near the critical points $\alpha_k^*$ (we only list the leading terms)\footnote{In the ’t Hooft model, critical points are defined by a system analogous to \eqref{crit-points-system}, as first noted in \cite{Zamolodchikov:2009pres}
\begin{equation*}
    \begin{cases}
        \alpha\sinh{t}+t\cosh{t}&=0;\\
        (1+\alpha)\cosh{t}+t\sinh{t}&=0,
    \end{cases}
\end{equation*}
}
\begin{equation}
    \resizebox{\textwidth}{!}{$
    \begin{aligned}
        &G^{(s)}_+\approx\left(\frac{2\pi}{\sqrt{\alpha-\alpha^*_k}}\right)^s,\quad G^{(1)}_-\approx\text{const},\quad G^{(2)}_-\approx-\frac{4\pi(1+\alpha^*_k)}{\sqrt{\alpha-\alpha^*_k}}, \quad G^{(3)}_-\approx\frac{2\pi(1+\alpha^*_k)\left(\pi^2-4\alpha^*_k(1+\alpha^*_k)\right)}{\alpha^*_k\sqrt{\alpha-\alpha^*_k}},\quad G^{(4)}_-\approx\frac{8 \pi ^2 (1+\alpha^*_k)^2}{\alpha-\alpha^*_k},
        \\
        &G^{(5)}_-\approx-\frac{20\pi^2(1+\alpha_k^*)^2\left(\pi^2-4\alpha_k^*(1+\alpha_k^*)\right)}{3\alpha_k^*(\alpha-\alpha_k^*)},\quad G^{(6)}_-\approx-\frac{16\pi^3(1+\alpha_k^*)^3}{(\alpha-\alpha_k^*)^{\frac{3}{2}}},\quad G^{(7)}_-\approx\frac{56\pi^3(1+\alpha_k^*)^3\left(\pi^2-4\alpha_k^*(1+\alpha_k^*)\right)}{3\alpha_k^*(\alpha-\alpha_k^*)^{\frac{3}{2}}}.
    \end{aligned}
    $}
\end{equation}
Near each critical point $\alpha_k^*$, the following asymptotic behavior of eigenvalues is observed:
\begin{equation}\label{lambda-tHooft-crit}
    \begin{aligned}
        ^{\text{'t Hooft}}\lambda_{2k}(\alpha)\Big|_{\alpha\to\alpha^*_k}&=\frac{1}{2\pi}\sqrt{\alpha-\alpha_k^*}+\mathcal{O}\left(\alpha-\alpha_k^*\right),\\
        ^{\text{'t Hooft}}\lambda^{\pm}_{2k-1}(\alpha)\Big|_{\alpha\to\alpha^*_k}&=\frac{\pm i}{\sqrt{2\pi(1+\alpha^*_k)}}(\alpha-\alpha^*_k)^{\frac{1}{4}}-\frac{\pi^2-4\alpha_k^*(1+\alpha^*_k)}{12\pi\alpha^*_k(1+\alpha^*_k)}\sqrt{\alpha-\alpha^*_k}+\mathcal{O}\left((\alpha-\alpha^*_k)^{\frac{3}{4}}\right).
    \end{aligned}
\end{equation}
These expressions describe the vanishing of one even and two odd meson masses at $\alpha=\alpha_k^*$. The notation for indices is the same as for IFT: we do not know what the order of these mesons was before the analytical continuation; the index $k$ corresponds to the number of the critical point $\alpha_k^*$. The next term in the expansion of the even eigenvalue is also known but not shown here, as it follows from subleading contributions to the spectral sums. Interestingly, the asymptotic formula for the pair of odd 't Hooft mesons is structurally identical to that of the two even BS mesons \eqref{lambda-even-critical}. The square-root behavior (with a previously unknown coefficient) of even mesons in the 't Hooft model was previously conjectured by Alexander Zamolodchikov in \cite{Zamolodchikov:2009pres}. The presence of two odd mesons with quartic-root behavior is a new result.

Let us also note that at the points $\alpha_k^*$, in addition to the vanishing of the even meson masses \eqref{lambda-even-critical}, all other masses—both even and odd—become multivalued functions, admitting expansions in powers of $(\alpha - \alpha_k^*)^{1/2}$. This behavior can be seen---at least for mesons with sufficiently large excitation numbers---by substituting the expansions of the integrals $\mathtt{i}_k(\alpha)$ \eqref{i2k_def}, \eqref{i2k-1_def} near $\alpha_k^*$ (see Appendix \ref{u-asymptotics-critical-points}) into the WKB formulas \eqref{WKB-normal-alpha-odd}–\eqref{WKB-normal-alpha-even}
\begin{equation}
    \resizebox{\textwidth}{!}{$
    \begin{gathered}
        \lambda_n(\alpha)\Big|_{\alpha\to\alpha_k^*}=\left[\frac{N_n}{2}+\frac{\alpha_k^*\log\left(\pi e^{\gamma_E}N_n\right)}{\pi^2}+\mathcal{O}\left(\frac{\log N_n}{N_n}\right)\right]+\frac{2\sqrt{\alpha-\alpha_k^*}}{\pi}\left[1+\frac{2\alpha_k^*}{\pi^2N_n}+\mathcal{O}\left(\frac{\log N_n}{N_n^2}\right)\right]+\mathcal{O}(\alpha-\alpha_k^*),\\ N_n=n+\frac{1}{4}+\frac{(\alpha_k^*)^2}{\pi^2}\left(\frac{2\zeta(3)}{\pi^2}+\alpha_k^*c^{(0)}_{3}-c^{(0)}_{1}-\frac{4\pi i}{\alpha_k^*}\sqrt{\frac{1+\alpha_k^*}{\alpha_k^*}}\right).
    \end{gathered}
    $}
\end{equation}
A similar statement and reasoning also apply to the 't Hooft model.

Let us now return to the second source of potential singularities. As mentioned earlier, this can occur in the case of odd spectral sums when the integral $\mathtt{v}(\widetilde{\alpha})=1$ \eqref{Proector-def}. In fact, this equation arises from the original Bethe-Salpeter equation \eqref{BS-eq} by setting $\lambda=0$ and integrating it against the kernel $\frac{\nu}{\cosh{\frac{\pi\nu}{2}}}$. Analysis of this integral shows that if $\alpha$ lies on the first sheet, its maximum value is reached at $\alpha=0$, where it equals $\frac{1}{16}$. Therefore, the solutions to this equation must be sought only on the second sheet. It is also evident that, since the integral diverges at infinitely many points $\alpha_k^*$ (as the denominator of the integrand contains as the factor the same transcendental function as in equation \eqref{main-transcendental-equation}), this implies that the equation $\mathtt{v}(\widetilde{\alpha})=1$ must also have infinitely many solutions, which are likely to be found in the vicinity of the critical points $\alpha^*_k$.

Analogously to the reasoning in Appendix \ref{u-asymptotics-critical-points} for the integrals $\mathtt{u}_{2k-1}(\alpha)$, the analytic continuation of the integral $\mathtt{v}(\alpha)$ to the second sheet contains contributions from residues, and then the equation that defines the critical points $\widetilde{\alpha}_k$ looks as follows
\begin{equation}\label{v=1-eq}
    \mathtt{v}(\widetilde{\alpha}_k)=\frac{2i}{\pi}\frac{\widetilde{t}^2_k}{\sinh{2\widetilde{t}_k}+2\widetilde{t}_k}+\frac{1}{4\pi^2}\int^{\infty}_{-\infty}\limits dt\frac{t^2}{\cosh^{2}{t}\cdot(\widetilde{\alpha}_k+t\tanh{t})}=1,\quad \widetilde{\alpha}_k+\widetilde{t}_k\tanh{\widetilde{t}_k}=0,
\end{equation}
where the first term comes from the residues at the poles $\pm \widetilde{t}_k$ ($\widetilde{t}_k$ is located in the lower-left quadrant of the complex plane), which cross the real axis. Even the numerical search for solutions to this equation is significantly more involved compared to \eqref{crit-points-system}.  As discussed in paragraph $1$ of this subsection, it is important to carefully track which specific pole of the wavefunction $\Psi$ (equivalently, which solution of the transcendental equation $\alpha+t\tanh t=0$) is being substituted into the residue term. This is because, under the analytic continuation \eqref{analyt-cont}, only a specific pair of poles $\pm\widetilde{t}_k(\widetilde{\alpha})$ crosses the real axis. Table \ref{tab:crit-points-2} lists the first few roots in the lower-left quadrant of the complex plane, see also Fig. \ref{fig:crit-points}.
\begin{table}[h!]
    \begin{center}
    \resizebox{\textwidth}{!}{$
        \begin{tabular}{| c || c | c | c | c | c |c |}
        \hline 
        $k$ & $-\widetilde{t}_k$ & $-\widetilde{\alpha}_k$ & $|\widetilde{\alpha}_k|$ & $\phi_k$ & $\mathtt{v}(\widetilde{\alpha}_k)$ & error $=|\mathtt{v}-1|$ \\
        \hline
        1 & $1.780584964+6.712607883i$ & $1.43597629+6.53591945i$ & $6.691806258$ & $0.4311593711$ & $0.9999999986 - 2\cdot10^{-10}i$ & $2.46\cdot10^{-10}$ \\
        \hline
        2 & $2.265940096+9.940137453i$ & $2.05917956+9.87044083i$ & 10.082947125 & 0.4345329158 & $1.0000000003-3\cdot10^{-10}i$ & $3.61\cdot10^{-10}$ \\
        \hline
        3 & $2.584371022+13.130022399i$ & $2.43729765+13.09179770i$ & 13.316740849 & 0.4414110424 & $1.0000000020-4\cdot10^{-10}i$ & $2.07\cdot10^{-9}$ \\
        \hline
        4 & $2.823589387+16.303504158i$ & $2.70960814+16.27903131i$ & $16.502994778$ & $0.4474993456$ & $0.9999999994+1\cdot10^{-10}i$ & $6.00\cdot10^{-10}$ \\
        \hline
        5 & $3.015746504+19.468038844i$ & $2.92275384+19.45088519i$ & $19.669250739$ & $0.4525248998$ & $1.0000000006+2\cdot10^{-10}i$ & $ 2.07 \cdot10^{-9}$ \\
        \hline
        6 & $3.176526430+22.627066254i$ & $3.09801852+22.61430407i$ & $22.825522275$ & $0.4566632927$ & $1.00000000005+7\cdot10^{-10}i$ & $3.33\cdot10^{-9}$\\
        \hline
        \end{tabular}
     $}
     \end{center}
     \caption{Numerical values for critical points $\widetilde{\alpha}_k=|\widetilde{\alpha}_k|e^{i\pi(1+\phi_k)}$ and corresponding points $\widetilde{t}_k$ in the lower-left quadrant of the complex plane.}
     \label{tab:crit-points-2}
\end{table}
For higher-order roots, the integral term in \eqref{v=1-eq} becomes suppressed relative to the first term, allowing for a very rough approximation given by
\begin{equation}\label{odd-crit-approx}
    \widetilde{\alpha}_k\approx \pi\left(k+\frac{1}{4}\right)\exp\left\{{i\frac{3\pi}{2}-i\frac{\log{\left(4\pi(k-1+\frac{1}{4})^2\right)}}{2\pi(k+\frac{1}{4})}}\right\},\quad k\gg1.
\end{equation}

Since the roots $\widetilde{\alpha}_k$ of \eqref{v=1-eq} are simple zeros, we have $1-\mathtt{v}(\alpha)\sim\alpha-\widetilde{\alpha}_k$ in their vicinity. Moreover, since the maximal power of $\mathcal{V}(\alpha)\sim\frac{1}{1-\mathtt{v}(\alpha)}$ \eqref{xi-def} in the odd spectral sums \eqref{Gm-with-matr-el} matches the index of the sum, this implies that the spectrum in this limit contains a single meson whose mass vanishes linearly as
\begin{equation}\label{lambda-odd-critical}
    \widetilde{\lambda}_{2k-1}(\alpha)\Big|_{\alpha\to\widetilde{\alpha}_k}\sim\alpha-\widetilde{\alpha}_k.
\end{equation}
Since these points arise from the Bethe-Salpeter equation \eqref{BS-eq} in the context of odd-parity physical mesons, they are expected to provide certain predictions for the full theory. In particular, the physical interpretation of the first critical point will be discussed in the concluding paragraph below.

We note that the behavior of vanishing mesons \eqref{even-meson-alpha=0}, \eqref{lambda-even-critical}, \eqref{lambda-odd-critical} near the critical points $\alpha^*_k$, $\widetilde{\alpha}_k$, extracted from the spectral sums, agrees with the analytic formula for the ratio of spectral determinant constants $d_-/d_+$ \eqref{dm/dp}. In particular, near the critical points $\alpha^*_k$ (including $\alpha^*_0=0$), this ratio vanishes as $\sqrt{\alpha-\alpha^*_k}$, while near $\widetilde{\alpha}_k$ it diverges as $1/(\alpha-\widetilde{\alpha}_k)$. This agreement further confirms the consistency of our results and indicates that the formulas \eqref{DD-integral} (but modified in accordance with \eqref{newDet-oldDet}), \eqref{DD-QQ-relation-new}, \eqref{2.17-new} capture genuinely non-perturbative physics, being valid simultaneously in both the large- and small-$\lambda$ limits. A similar agreement near the critical points between the behavior of vanishing mesons \eqref{lambda-tHooft-crit} and the ratio $d_-/d_+$ (formula 4.25 in \cite{Litvinov:2024riz}) also holds in the ’t Hooft model.

\paragraph{Evolution of the poles of $\Psi$: Part II.}
Thus far, we have performed the analytic continuation of the parameter $\alpha$ into the complex plane along a circular path centered at $0$ with radius $|\alpha|$. As a result, after completing a full rotation, the Bethe-Salpeter equation acquires contributions from two residues at the poles $\pm i\nu_k^*(\alpha)$ of the meson wavefunction $\Psi(\nu)$ on the imaginary axis \eqref{BS_second_worldsheet}, where the order number $k$ is determined by the modulus of $\alpha$. However, the final outcome of the analytic continuation must be invariant under homotopic deformations of the contour. What matters is only the starting point of the continuation -- namely, the initial value of $\alpha$ -- since it determines the location of the poles of $\Psi(\nu)$. To illustrate this, we consider the specific example $\alpha=7$. We show that the result of analytic continuation along a closed contour encircling the critical points $\alpha^*_0$, $\alpha^*_1$, and $\alpha^*_2$ -- that is, along a path homotopic to a circle of radius $7$ centered at $0$ -- leads to the emergence of residues at the poles $\pm i\nu^*_3(7)$. Along the way, we also formulate several general patterns observed in this behavior.
\begin{figure}[h!]
    \centering
    \includegraphics[width=0.45\linewidth]{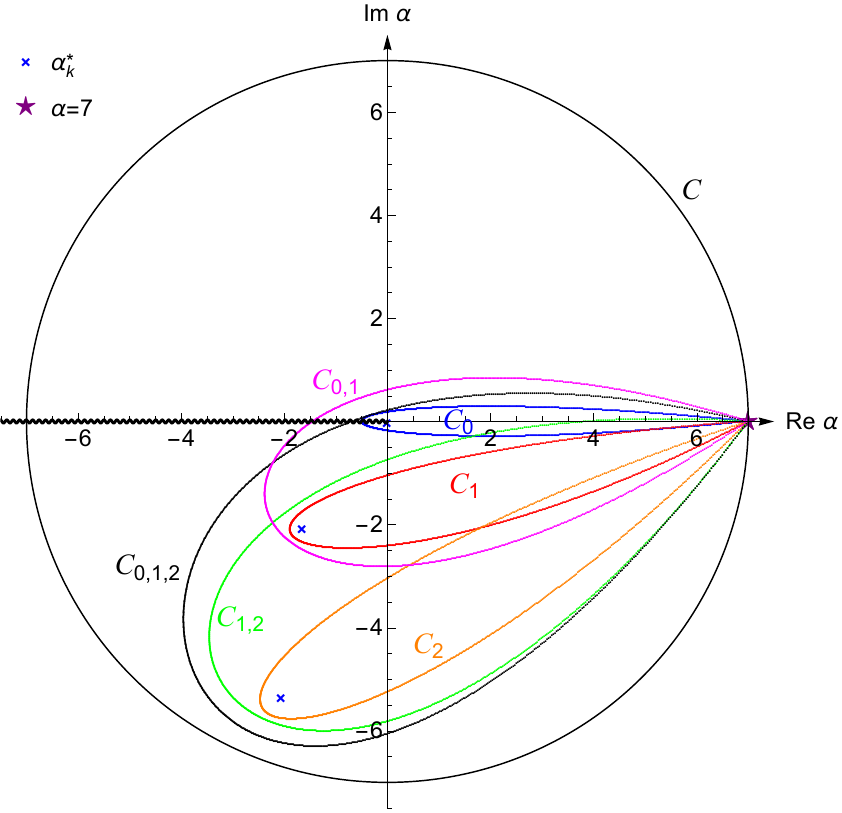}
    \caption{Various contours for analytical continuation into the complex region $\alpha$, starting at $\alpha=7$.}
    \label{fig:contours}
\end{figure}

\begin{figure}[h!]
    \centering
    \includegraphics[width=0.49\linewidth]{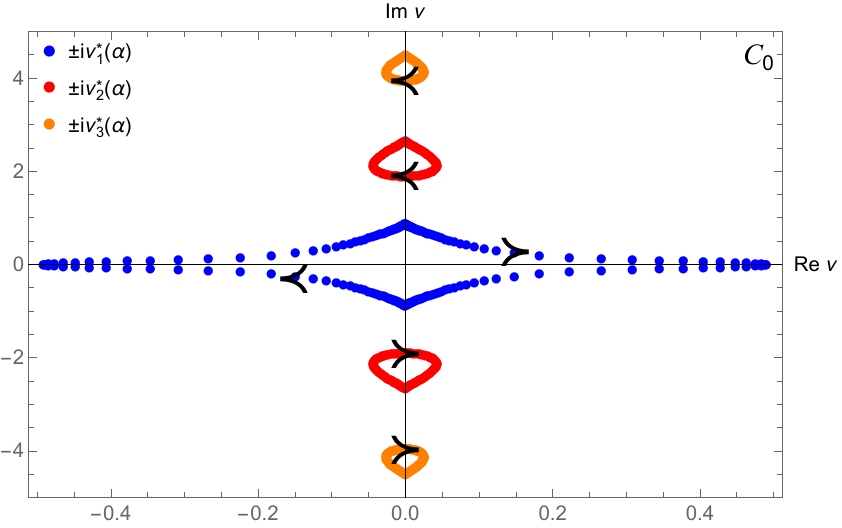}
    \includegraphics[width=0.49\linewidth]{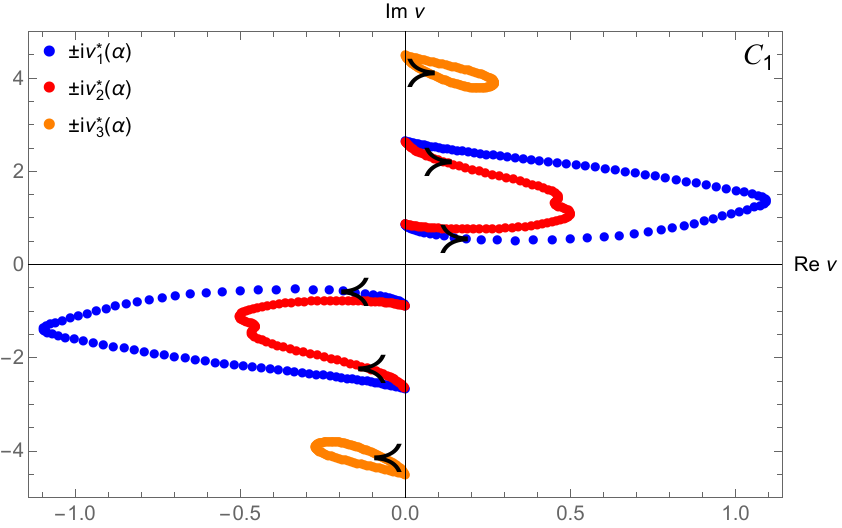}
    \includegraphics[width=0.49\linewidth]{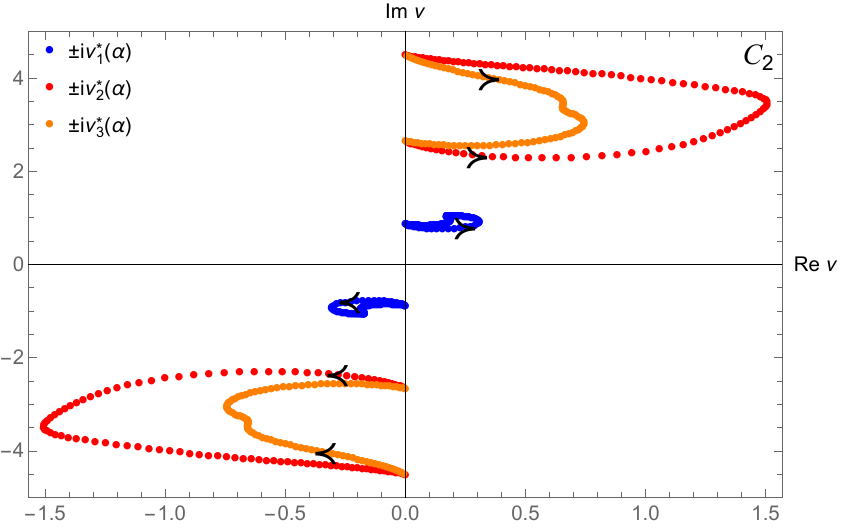}
    \includegraphics[width=0.49\linewidth]{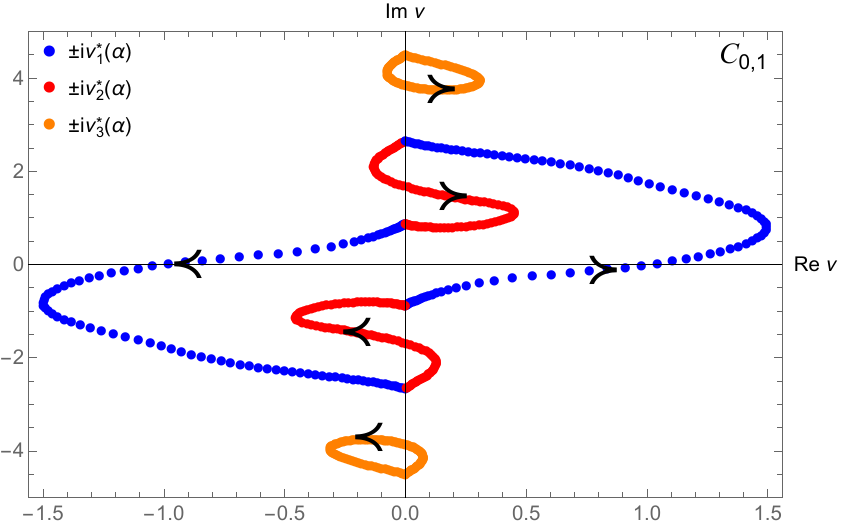}
    \includegraphics[width=0.49\linewidth]{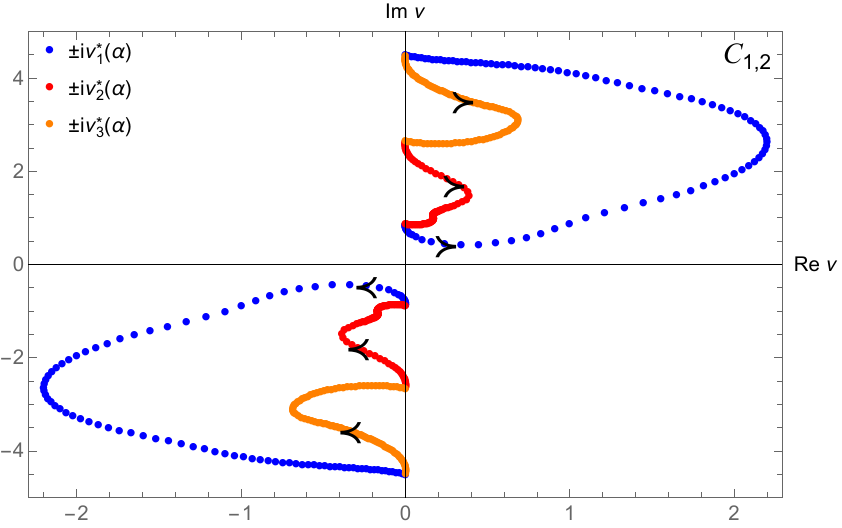}
    \includegraphics[width=0.49\linewidth]{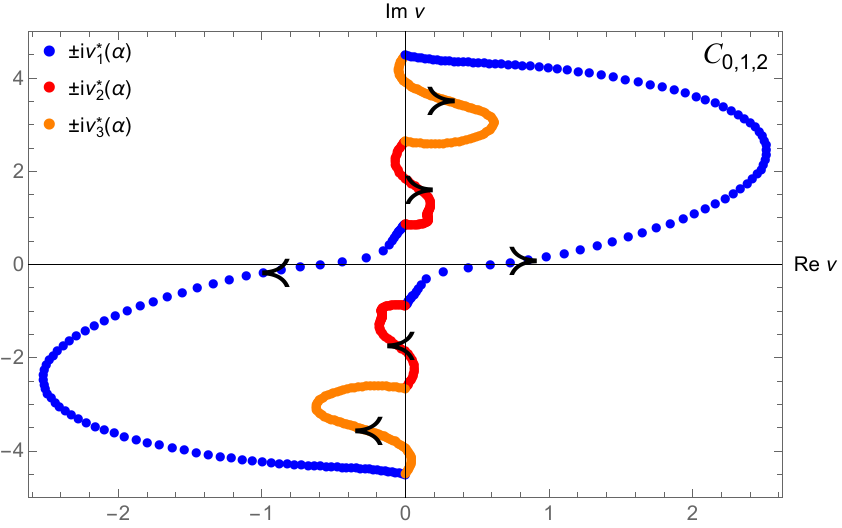}
    \caption{Evolution of the poles $\pm i\nu_k^*(\alpha)$, $k=1,2,3$, of the function $\Psi(\nu)$ under analytic continuation along the contours shown in Figure \ref{fig:contours}. Arrows indicate the direction of evolution. The pole evolution plots along contours that do not encircle $\alpha^*_0$ (not intersect branch-cut) serve as evidence that all other singularities reside on the second sheet.}
    \label{fig:poles-evolution-2}
\end{figure}

First, it is clear that more complicated contours encircling multiple critical points $\alpha_k^*$ can be decomposed into a sum of elementary loops, each encircling only a single critical point. In Figure \ref{fig:contours}, we depict the standard contour $\mathcal{C}$ -- a circle of radius $7$ centered at the origin -- as well as the elementary contours $\mathcal{C}_i$ (for $i = 0, 1, 2$), each enclosing one of the critical points $\alpha_i^*$ individually. In addition, the contour $\mathcal{C}_{0,1}$ encloses $\alpha_0^*$ and $\alpha_1^*$, $\mathcal{C}_{1,2}$ encloses $\alpha_1^*$ and $\alpha_2^*$, while $\mathcal{C}_{0,1,2}$ encloses all three: $\alpha_0^*$, $\alpha_1^*$, and $\alpha_2^*$. Figure \ref{fig:poles-evolution-2} illustrates the evolution of the first three poles in the upper and lower half-planes under analytic continuation along the contours $\mathcal{C}_i$, $\mathcal{C}_{0,1}$, $\mathcal{C}_{1,2}$, $\mathcal{C}_{0,1,2}$. The behavior of the poles under the standard contour $\mathcal{C}$ (a circle of radius $7$) was discussed earlier. We observe that continuation along $\mathcal{C}_0$ causes the first pair of poles to cross the real axis and exchange positions, while the second and third pairs return to their original locations. This behavior is not unexpected and is consistent with the pattern observed previously for $|\alpha|=2$ (see Figure \ref{fig:poles-evolution}).
In contrast, continuations along the remaining elementary contours do not lead to real-axis crossings, and therefore the Bethe–Salpeter equation \eqref{BS-eq} remains unchanged under these continuations. Instead, they induce specific permutations of the poles: the contour $\mathcal{C}_1$ swaps the first and second poles (in both the upper and lower half-planes), $\mathcal{C}_2$ exchanges the second and third. The detailed action of each contour on the poles is summarized in Table \ref{tab:action-on-poles}.
\begin{table}[h!]
    \centering
    \begin{tabular}{|c|c|c|c|c|c|c|c|c|}
    \hline
    \diagbox{poles}{contours} & $\mathcal{C}$ & $\mathcal{C}_0$ & $\mathcal{C}_1$ & $\mathcal{C}_2$ & $\mathcal{C}_{0,1}$ & $\mathcal{C}_{1,2}$ & $\mathcal{C}_{0,1,2}$ \\
    \hline
    $1$   & $1$  & $-1$ & $2$  & $1$  & $-2$ & $3$  & $-3$ \\
    \hline
    $-1$  & $-1$ & $1$  & $-2$ & $-1$ & $2$  & $-3$ & $3$ \\
    \hline
    $2$   & $2$  & $2$  & $1$  & $3$  & $1$  & $1$  & $1$ \\
    \hline
    $-2$  & $-2$ & $-2$ & $-1$ & $-3$ & $-1$ & $-1$ & $-1$ \\
    \hline
    $3$   & $-3$ & $3$  & $3$  & $2$  & $3$  & $2$  & $2$ \\
    \hline
    $-3$  & $3$  & $-3$ & $-3$ & $-2$ & $-3$ & $-2$ & $-2$ \\
    \hline
    \end{tabular}
    \caption{Effect of analytic continuation along the contours shown in Figure \ref{fig:contours} on the poles $\pm i\nu_k^*(\alpha)$, $k=1,2,3$, of the wave function $\Psi(\nu)$. For brevity, each pole is labeled by its index, with negative signs indicating poles in the lower half-plane.}
    \label{tab:action-on-poles}
\end{table}

This pattern of pole evolution confirms our statement about homotopic contours: the elementary contour $\mathcal{C}_0$ causes the first pole to cross the real axis and end up at $-i\nu^*_3(7)$ (with the second and third reaching their initial positions), while $\mathcal{C}_{1,2}$ cyclically permutes the poles ($1\to3\to2\to1$). The order of traversing the contours (from smaller to larger) is important.

This example illustrates several general patterns. We effectively work on a Riemann sphere with a countable set of isolated punctures $\alpha_k^*$ accumulating at infinity. Among them, $\alpha_0^*=0$ plays a special role: if a contour winds around it (and no others) an odd number of times, it modifies the Bethe-Salpeter equation by introducing additional terms from the residues at the first poles $\pm i\nu_1^*(\alpha)$ \eqref{BS_second_worldsheet}, which cross the real axis and interchange. All other poles $\pm i\nu_k^*(\alpha)$ return to their original positions\footnote{The behavior of all other poles follows from integer shifts \eqref{Psi-all-poles}.}. In contrast, contours that wind an odd number of times around a single critical point $\alpha_k^*$ with $k\ne0$ (and avoid all others) act as transpositions of the corresponding poles: $k\leftrightarrow k+1$. More generally, if a contour winds an odd number of times around a sequence of adjacent points $\alpha_k^*,\dots,\alpha_{k+l}^*$, it results in a cyclic permutation of the corresponding poles: $k \to k+l \to\ldots\to k+1 \to k$ (and analogously for the negative ones). If the sequence includes $\alpha_0^*$, then the first poles cross the real axis, and the transformation takes the form $1 \to -l \to -l+1 \to\ldots\to -1 \to l \to\ldots\to 2 \to 1$. All other cases reduce to combinations of these scenarios.

\paragraph{Physical interpretation of the critical points.} The precise physical interpretation of the critical points located on the second sheet of the complex plane remains an open question for both: IFT and 't Hooft model. A rigorous understanding of the physical mechanisms underlying these critical points---whether in the 't Hooft model or IFT---demands a systematic treatment beyond the strict asymptotic limits, specifically incorporating finite-$N_c$ corrections in two-dimensional QCD and multiparticle effects in IFT. To date, such a comprehensive analysis has not yet been accomplished for the 't Hooft model, and as mentioned in Introduction and Section \ref{BS-and-Integr} the corresponding developments for IFT remain at an early stage.

Nevertheless, one instructive example for IFT is well-established. The first singularity in the IFT spectrum, corresponding to the vanishing of the lightest meson mass, is identified in the full theory with the Yang-Lee edge singularity occurring at purely imaginary magnetic field $h_{\text{YL}}=\pm i(0.18935\ldots)m^{15/8}$. The corresponding scaling parameters for this field configuration are given by \eqref{scaling-parameters}, with $\xi_{\text{YL}}=\pm i(0.18935\ldots)$ and $\eta_{\text{YL}}=2.42917\ldots e^{\pm\frac{11\pi i}{15}}$. 

Since the definition of the parameter $\alpha$ in \eqref{FZ-our-notations}, which we have used throughout, involves the string tension $f_0$, which is nontrivially related to the magnetic field $h$ (see the discussion in Section \ref{BS-and-Integr}), an exact determination of the location of the Yang-Lee point $\alpha_{\text{YL}}$ in the full Ising Field Theory is quite challenging. To estimate the position of this point, we follow the discussion in Section \ref{BS-and-Integr} and neglect all multiquark effects responsible for dressing the bare mass $m$ and the string tension $f_0$, and we use the definition of string tension given in \eqref{f-series}.\footnote{Strictly speaking, the statement in \cite{Fonseca:2006au} -- that for sufficiently large real $\eta \gtrsim 0.5$, the standard definition for string tension \eqref{f0-def} becomes applicable -- was established only for real $\eta$. In our estimate of the Yang–Lee point, however, we use an alternative definition of $f_0$ \eqref{f0-def}, which is expected to remain sufficiently accurate for complex values of $h$ satisfying $|\eta|<|\eta_{\text{YL}}|$. For the purpose of estimation, we simply evaluate the expansion \eqref{f-series} at $\eta=\eta_{\text{YL}}$.} Consequently, this critical point corresponds to $\alpha_{\text{YL}}\approx-0.63-5.95i$ and is marked in Fig. \ref{fig:crit-points}. This estimate is based on the first $11$ terms of the expansion \eqref{f-series}, calculated from Table $3$ in \cite{Fonseca:2001dc}.

Importantly, the critical exponent governing this singularity differs from that predicted by the weak-field approximation \eqref{lambda-odd-critical}, taking in the full theory the form 
  \cite{Xu:2022mmw,Xu:2023nke} 
\begin{equation}
    \lambda_1(\xi)\overset{\eqref{FZ-our-notations}}{\sim}M^2_1(\xi)\sim(\xi^2-\xi^2_{\text{YL}})^{5/6}\overset{}{\sim}(\alpha-\alpha_{\text{YL}})^{5/6}.   
\end{equation}
This critical point is associated with the non-unitary Conformal Field Theory $\mathcal{M}_{2,5}$ with central charge $c=-\frac{22}{5}$, also known as Yang-Lee minimal model \cite{Cardy:1985yy}.

Extending this line of inquiry, it is natural to conjecture (this was first mention in \cite{Fateev:2009jf}) that the critical points residing on the second sheet in IFT and the 't Hooft model may likewise correspond to particular non-unitary Conformal Field Theories that govern the infrared behavior near these singularities. Identifying such CFTs remains an important open problem for future research, with potential implications for understanding non-perturbative dynamics in confining Quantum Field Theories.
\section{Conclusion}\label{Conclusion}
In this work, we explored the analytical structure of mesons in the Ising Field Theory using the two-particle Bethe-Salpeter approximation. We revealed that the Bethe-Salpeter equation admits a reformulation as a Baxter TQ equation -- a connection that opens the door to powerful integrability based methods. Motivated by the similarity between the integral equations of IFT and the large-$N_c$ two-dimensional QCD, we extended the elegant non-perturbative approach developed by Fateev, Lukyanov, and Zamolodchikov \cite{Fateev:2009jf} for the 't Hooft model, further building on its generalizations in \cite{Litvinov:2024riz,Artemev:2025cev}.

A central analytical achievement of this work is the discovery of a set of non-trivial relations: \eqref{DD-integral} (but modified using formula \eqref{newDet-oldDet}), \eqref{DD-QQ-relation-new}, and \eqref{2.17-new} -- that enable the direct extraction of spectral data from solutions to the TQ equation. These results yield explicit expressions for the spectral sums $\mathcal{G}^{(s)}_\pm(\alpha)$ and the large-$n$ WKB expansion \eqref{WKB-normal-alpha-odd}-\eqref{WKB-normal-alpha-even}, both of which recover known results in limiting cases with good accuracy. Perhaps most remarkably, the derived formulas unveil a rich analytic structure in the complex $\alpha$-plane, predicting the existence of mesons that vanish at a discrete set of critical points and providing a detailed description of their local behavior in the vicinity of this points.

Continuing the analogy with the 't Hooft model, a natural next step is to explore a generalization of our results to the case of unequal quark masses, $\alpha_1 \ne \alpha_2$. While such a scenario has no direct meaning within the standard Ising Field Theory---where the action consists of equal massive Majorana fermions---it becomes physically relevant in more elaborate setups. For example, when two transverse-field Ising chains (TFICs) with magnetic order are coupled as
\begin{equation}
    \mathcal{A}_{\text{TFIC}}=\mathcal{A}^{(1)}_{\text{FF}}+\mathcal{A}^{(2)}_{\text{FF}}+\mu\int \sigma^{(1)}(x)\sigma^{(2)}(x)\; d^2x,
\end{equation}
where $\mathcal{A}^{(i)}_{\text{FF}}$ denotes the free fermionic action with mass $m_i$ \eqref{free-fermion-action}, and $\mu\overset{\text{def}}{=}\frac{f_0}{2\bar{\sigma}^2}$ (see definition of $\bar{\sigma}$ in \eqref{spontaneous-magnetiation}) controls the interchain coupling strength, the interpretation of distinct “quark” masses becomes meaningful and opens a promising direction for future investigation. For example, the Bethe–Salpeter equation describing interchain mesons in the coupled TFIC model can be derived as a natural generalization of equation (30) in \cite{Gao:2025mcg}:
\begin{equation}
    \left(\frac{M^2}{4\cosh^2{\theta}}-\frac{m_1^2+m_2^2}{2}-\frac{m_1^2-m_2^2}{2}\tanh{\theta}\right)\psi(\theta)=-f_0\int_{-\infty}^{\infty}\limits\frac{d\theta'}{2\pi}\left(\frac{\cosh{(\theta-\theta')}}{\sinh^2(\theta-\theta')}-\frac{1}{2\cosh{\theta}\cosh{\theta'}}\right)\psi(\theta'),
\end{equation}
where the second term in r.h.s. differs from original BS equation for IFT  \eqref{BS-start} and in $\nu$-space it takes the form $\sim\frac{1}{\cosh{\frac{\pi\nu}{2}}}\int_{-\infty}^{\infty}\limits d\nu'\frac{1}{\cosh{\frac{\pi\nu'}{2}}} \Psi(\nu')$. This equation exhibits a striking resemblance to the ’t Hooft model with two flavors, which in $\theta$-space $\theta=\frac{1}{2}\log{\frac{x}{1-x}}$ takes the form:
\begin{equation}
\begin{gathered}
    \left(\frac{\pi^2\lambda}{2\cosh^2{\theta}}-\frac{\alpha_1+\alpha_2}{2}-\frac{\alpha_1-\alpha_2}{2}\tanh{\theta}\right)\phi(\theta)=-2\int_{-\infty}^{\infty}\limits\frac{d\theta'}{2\pi}\frac{\phi(\theta')}{\sinh^2(\theta-\theta')},\\
    \alpha_i=\frac{\pi m_i^2}{g^2}-1,\quad M^2=2\pi g^2\lambda,
\end{gathered}
\end{equation}
and was recently analyzed in details in \cite{Artemev:2025cev}.

Another promising direction for future investigation is a deeper analysis of the analytic continuation in the complex scaling parameter $\alpha$-plane. This involves uncovering how the meson spectrum reorganizes itself under continuation---which one meson states disappear near critical points. It also raises intriguing questions about the behavior of the Bethe–Salpeter wave function $\Psi(\nu)$ as a function of the scaling parameter $\alpha$. What insights do the critical points of the Bethe–Salpeter equation offer about the full Ising Field Theory beyond the two-particle approximation? We leave these questions open for future exploration.
\subsection*{Supplemental material}
Together with this submission  we provide three accompanying Wolfram Mathematica notebooks: \texttt{Spectral-sums-Ising.nb}, \texttt{Phi.nb}, and \texttt{WKB.nb}. The first notebook contains analytic expressions for the first five spectral sums $\mathcal{G}^{(s)}_\pm$ and the corresponding matrix elements $\langle p | \hat{K}^n | p \rangle$ for $n = 1,\ldots,5$. The second notebook includes the phase functions $\Phi_\pm^{(k)}(l)$ for $k = 0,\ldots,5$. The third file presents higher-order terms in the large-$n$ WKB expansion \eqref{WKB-normal-alpha-odd}–\eqref{WKB-normal-alpha-even}, including corrections up to $\frac{1}{\mathfrak{n}^5}$. All three notebooks make use of the fundamental integrals $\mathtt{v}(\alpha)$, $\mathtt{u}_{2k-1}(\alpha)$, $\mathtt{i}_{2k}(\alpha)$, and $\mathtt{i}_{2k-1}(\alpha)$, defined in \eqref{Proector-def}, \eqref{u-def}, \eqref{i2k_def} and \eqref{i2k-1_def}, respectively.
\section*{Acknowledgments}
We acknowledge discussions with Alexander Artemev, Sergei Lukyanov, Nikita Slavnov and Alexander Zamolodchikov. The work of P.M.  performed in Landau Institute has been supported by the Russian Science Foundation under the grant 23-12-00333. A.L. and E.S. were supported by Basis foundation.
\appendix

\section{Matrix elements of spectral operator \texorpdfstring{$\hat{K}$}{}}\label{matrix-el-of-K}
Here, we describe how to calculate the matrix elements $\bra{p}\hat{K}^n\ket{p}$ and derive equation \eqref{g-for-appendix}.
Consider the Liouville -- Neumann series for $\Psi_-(\nu|\lambda)$
\begin{equation}\label{Liouville-Neumann_for_Psi_m}
    f(\nu)\Psi_-(\nu|\lambda)=\frac{\nu}{\cosh{\frac{\pi \nu}{2}}}+\sum\limits_{k=1}^\infty\lambda^k\int_{-\infty}^\infty\limits\frac{\nu_k}{\cosh{\frac{\pi \nu_k}{2}}}\prod\limits_{j=1}^k\frac{d\nu_j}{f(\nu_j)}S(\nu_j-\nu_{j-1}),\quad \nu_0\equiv \nu,
\end{equation}
where as always 
 $f(\nu)=\frac{2\alpha}{\pi}+\nu\tanh{\frac{\pi\nu}{2}},\quad S(\nu)=\frac{\pi\nu}{2\sinh{\frac{\pi\nu}{2}}}$. By definition $Q_-(\nu|\lambda)=\cosh{\frac{\pi \nu}{2}} f(\nu) \Psi(\nu|\lambda)$ so one can rewrite \eqref{Liouville-Neumann_for_Psi_m} as 
\begin{equation}\label{Liouville-Neumann_for_Qp}
    \frac{Q_-(\nu)}{\cosh{\frac{\pi\nu}{2}}}=\frac{\nu}{\cosh{\frac{\pi\nu}{2}}}+\sum\limits_{k=1}^\infty\lambda^k\int_{-\infty}^\infty\limits\frac{\nu_k}{\cosh{\frac{\pi\nu_k}{2}}}\prod\limits_{j=1}^k \frac{d\nu_j}{f(\nu_j)}S(\nu_j-\nu_{j-1}).
\end{equation}

Let's consider the explicit form of the matrix element
\begin{equation}
    \bra{p}\hat{K}^n\ket{p}=\frac{1}{16 \mathtt{v}(\alpha)} \int_{-\infty}^\infty\limits \frac{d \nu_0}{f(\nu_0)} \prod_{j=1}^n \frac{d\nu_j}{f(\nu_j)} S(\nu_j-\nu_{j-1}) \frac{\nu_0}{\cosh{\frac{\pi \nu_0}{2}}} \cdot \frac{\nu_n}{\cosh{\frac{\pi \nu_n}{2}}}.
\end{equation}
From this it is easy to see that one can obtain $\bra{p}\hat{K}^n\ket{p}$ by expanding the left part of the equation \eqref{Liouville-Neumann_for_Qp} at point $i$ by $\lambda$
\begin{equation}\label{matrix-elements-series}
    \resizebox{\textwidth}{!}{$
    \begin{aligned}
        -\frac{i}{8\pi \mathtt{v}(\alpha)}\frac{Q_-(\nu)-\nu}{\cosh{\frac{\pi\nu}{2}}}\Biggl|_{\nu=i}=-\frac{Q'_-(i)-1}{4\pi^2\mathtt{v}(\alpha)}=\lambda+\sum\limits_{k=1}^\infty \lambda^{k+1}\bra{p}\hat{K}^k\ket{p}\quad\Rightarrow\quad -1-\frac{Q'_-(i)-1}{4\pi^2\mathtt{v}(\alpha)\lambda}=\sum\limits_{k=1}^\infty \lambda^{k}\bra{p}\hat{K}^k\ket{p}.
    \end{aligned}
    $}
\end{equation}
For example
\begin{multline}
    \bra{p}\hat{K}\ket{p}=\frac{1}{12\pi^6\mathtt{v}(\alpha)}\Biggl[2\pi^4\alpha(3+2\alpha)-8\pi^2\alpha^2(3+\alpha)\zeta(3)+12\alpha^3\left(\zeta^2(3)+\zeta(5)\right)-2\pi^6+
    \\
    +6\pi^2\alpha^3\left(2\alpha\zeta(3)-\pi^2(2+\alpha)\right)\mathtt{u}_3(\alpha)+3\pi^4\alpha^5\mathtt{u}^2_3(\alpha)-6\pi^4\alpha^4\mathtt{u}_5(\alpha)\Biggl],
\end{multline}
\begin{equation}
    \begin{aligned}
        \bra{p}\hat{K}^2\ket{p}=&\;-\frac{1}{216\pi^2\mathtt{v}(\alpha)}\Biggl[\pi^2(81+22\alpha)-12\alpha(5+\alpha(19+3\alpha))+
        \\&
        +\frac{2\alpha}{\pi^6}\Biggl(27\alpha^3\left(2\zeta(3)+\pi^2\alpha\mathtt{u}_3(\alpha)\right)^3-18\pi^2\alpha^2(9+4\alpha\left(2\zeta(3)+\pi^2\alpha\mathtt{u}_3(\alpha)\right)^2+
        \\&
        +\pi^2\alpha\left(3\pi^2-4\alpha(18+5\alpha)\right)\left(6\zeta(5)-\pi^4\alpha\left(\mathtt{u}_3(\alpha)+3\mathtt{u}_5(\alpha)\right)\right)+
        \\&
        +\left(2\zeta(3)+\pi^2\alpha\mathtt{u}_3(\alpha)\right)\left(\pi^4(216+\alpha(216+23\alpha))\alpha-6\pi^6(6+\alpha)+216\alpha^3\zeta(5)-\right.
        \\&\left.-36\pi^4\alpha^4\left(\mathtt{u}_3(\alpha)+3\mathtt{u}_5(\alpha)\right)\right)+\alpha^3\left(90\zeta(7)+\pi^6\alpha\left(2\mathtt{u}_3(\alpha)+30\mathtt{u}_5(\alpha)+45\mathtt{u}_7(\alpha)\right)\right)\Biggl)\Biggl].
    \end{aligned}
\end{equation}
We have collected more matrix elements (first $5$) in \texttt{Spectral-sums-Ising.nb}.

Here we also give a proof that the quantization condition \eqref{Q_quantization_cond} coincide with \eqref{Quant_cond2}. Indeed, using \eqref{Q_quantization_cond} and TQ equation \eqref{TQ-equation} one can obtain 
\begin{equation}
    Q_-(i) = \frac{i}{16} \int_{-\infty}^\infty\limits d \nu' \frac{\nu'}{\cosh^2{\frac{\pi \nu'}{2}}} \frac{(Q_-(\nu'+2i) - Q_-(\nu')) + (Q_-(\nu'-2i) - Q_-(\nu'))}{- 4 \pi  \lambda_n}.
\end{equation}
Let us introduce
\begin{equation}
    G_1(\nu)\overset{\text{def}}{=}\frac{\nu Q_-(\nu)}{\cosh^2{\frac{\pi\nu}{2}}},\quad G_2(\nu)\overset{\text{def}}{=}\frac{Q_-(\nu)}{\cosh^2{\frac{\pi\nu}{2}}}.
\end{equation}
Then under analytical continuation we can write
\begin{multline}
    Q_-(i)=\frac{-i}{64\pi\lambda_n}\left[-\oint_{\mathcal{C}_i}\limits d\nu\;G_1(\nu)+\oint_{\mathcal{C}_{-i}}\limits d\nu\;G_1(\nu)+2i\oint_{\mathcal{C}_{i}}\limits d\nu\;G_2(\nu)+2i\oint_{\mathcal{C}_{-i}}\limits d\nu\;G_2(\nu)\right]=
    \\
     =\frac{Q_-(i)-iQ'_-(i)}{4\pi^2\lambda_n}=\frac{i-iQ'_-(i)}{4\pi^2\lambda_n}=i,
\end{multline}
where $\mathcal{C}_i$ and $\mathcal{C}_{-i}$ are the contours that surround points $i$ and $-i$, respectively. Here we use normalization conditions \eqref{Qpm-normalization-conditions}. So we obtain the condition \eqref{Quant_cond2}.
\section{Asymptotics of the integrals \texorpdfstring{$\mathtt{u}_{2k-1}(\alpha)$, $\mathtt{i}_{2k}(\alpha)$, $\mathtt{i}_{2k-1}(\alpha)$}{}}\label{u-asymptotics}
Here, we describe the asymptotic behavior of the integrals $\mathtt{u}_{2k-1}(\alpha)$, $\mathtt{i}_{2k}(\alpha)$, $\mathtt{i}_{2k}(\alpha)$ defined in \eqref{u-def}, \eqref{i2k_def}  and \eqref{i2k-1_def}.
\subsection{Chiral limit: \texorpdfstring{$\alpha \rightarrow0$}{}}\label{u-asymptotics-chiral}
We examine the limit $\alpha\rightarrow0$, which corresponds to the vanishing quark mass $m\rightarrow0$. In this regime, the poles of the integrand $\mathtt{u}_{2k-1}(\alpha)$ located within the strip $[-2i,2i]$ i.e., the roots of the transcendental equation \eqref{main-transcendental-equation} can be expanded as follows:
\begin{equation}
    it^*_0=i\sqrt{\alpha}\left(1-\frac{1}{6}\alpha+\frac{11}{360}\alpha^2-\frac{17}{5040}\alpha^3-\frac{281}{604800}\alpha^4+\mathcal{O}(\alpha^5) \right).
\end{equation}
Since these poles move toward the real axis as $\alpha \rightarrow 0$, we deform the integration contour accordingly and write the integral, assuming the contour passes above the pole $it^*_0$
\begin{equation}
    \mathtt{u}_{2k-1}(\alpha)\Big|_{\alpha\rightarrow0}=\underbrace{2\pi i\underset{t=it^*}{\text{ Res }}\frac{\cosh^2{t}}{t\sinh^{2k-1}{t}\cdot(\alpha\cosh{t}+t\sinh{t})}}_{\mathtt{u}^{(1)}_{2k-1}(\alpha)}+\underbrace{\int^{\infty+i\epsilon}_{-\infty+i\epsilon}\limits dt\frac{\cosh^2{t}}{t\sinh^{2k-1}{t}\cdot(\alpha\cosh{t}+t\sinh{t})}}_{\mathtt{u}^{(2)}_{2k-1}(\alpha)}.
\end{equation}

The first term is a series expansion in half-integer powers of the parameter $\alpha$
\begin{equation}
    \mathtt{u}^{(1)}_{2k-1}(\alpha)\Big|_{\alpha\to0}=(-1)^k\frac{\pi}{\alpha^{k}}\Biggl[\frac{1}{\sqrt{\alpha}}+\frac{4k-7}{6} \sqrt{\alpha}+\frac{80k^2-328k+303}{360}\alpha^{3/2}+\mathcal{O}\left(\alpha^{5/2}\right)\Biggl],
\end{equation}
and the second one is a series over integer powers
\begin{equation}
    \mathtt{u}^{(2)}_{2k-1}(\alpha)\Big|_{\alpha\to0}=\sum_{l=0}^{\infty}c^{(l)}_{2k-1}\alpha^l,\quad c^{(l)}_{2k-1}=(-1)^l\int_{-\infty+i\epsilon}^{\infty+i\epsilon}\limits dt\frac{\cosh^{2+l}{t}}{t^{2+l}\sinh^{2k+l}{t}},
\end{equation}
where the coefficients $c^{(l)}_{2k-1}$ can be computed numerically.

The integrals $\mathtt{i}_k(\alpha)$ are calculated in a similar way (note that even integrals can be obtained alternatively via \eqref{u-i2k-relation})
\begin{equation}
    \begin{aligned}
        &\mathtt{i}^{(1)}_{2k-1}(\alpha)\Big|_{\alpha\to0}=\frac{4\pi}{\sqrt{\alpha}}+\frac{2\pi}{3}(6k-5)\sqrt{\alpha}+\frac{\pi}{90}(180k^2-360k+179)\alpha^{3/2}+\mathcal{O}(\alpha^{5/2}),\\
        &\mathtt{i}^{(2)}_{2k-1}(\alpha)\Big|_{\alpha\to0}=\sum_{l=0}^{\infty}C^{(l)}_{2k-1}\alpha^l,\quad C^{(l)}_{2k-1}=(-1)^l\int_{-\infty+i\epsilon}^{\infty+i\epsilon}\limits dt\;\frac{\sinh{2t}+2t}{t^{2+l}\cosh^{2k-1-l}{t}\sinh^{1+l}{t}},
    \end{aligned}    
\end{equation}
\begin{equation}
    \begin{aligned}
        &\mathtt{i}^{(1)}_{2k}(\alpha)\Big|_{\alpha\to0}=(-1)^k\frac{\pi}{\alpha^{k}}\left[\frac{4}{\sqrt{\alpha}}+\frac{2(4k-5)}{3}\sqrt{\alpha}+\frac{80k^2-248k+179}{90}\alpha^{3/2}+\mathcal{O}\left(\alpha^{5/2}\right)\right],\\
        &\mathtt{i}^{(2)}_{2k}(\alpha)\Big|_{\alpha\to0}=\sum_{l=0}^{\infty}C^{(l)}_{2k}\alpha^l,\quad C^{(l)}_{2k}=(-1)^l\int_{-\infty+i\epsilon}^{\infty+i\epsilon}\limits dt\;\frac{\cosh^{1+l}{t}(\sinh{2t}+2t)}{t^{2+l}\sinh^{2k+1+l}{t}}.
    \end{aligned}
\end{equation}

\subsection{Heavy quark limit: \texorpdfstring{$\alpha\to\infty$}{}}\label{u-asymptotics-heavy}
The heavy quark mass limit corresponds to $\alpha \to \infty$. In this regime, for $k > 1$, one can expand in inverse powers of $\alpha$ (note that for $k = 1$, the leading term diverges)
\begin{equation}
    \mathtt{u}_{2k-1}(\alpha)\Big|_{\alpha\rightarrow\infty}=\frac{1}{\alpha}\fint_{-\infty}^{\infty}\limits dt\frac{\cosh{t}}{t\sinh^{2k-1}{t}}\sum_{l=0}^{\infty}\limits\left(-\frac{t\tanh{t}}{\alpha}\right)^l
\end{equation}
The integrals appearing in the summands are evaluated analytically. For example,
\begin{equation}
    \resizebox{\textwidth}{!}{$
    \begin{aligned}
        \mathtt{u}_{3}(\alpha)=&\;-\frac{2\zeta(3)}{\pi^2\alpha}+\frac{2}{\alpha^2}+\frac{\pi^2}{4\alpha^3}-\frac{\pi^2}{6\alpha^4}+\frac{\pi^2}{4\alpha^5}-\frac{\pi^2(120+7\pi^2)}{360\alpha^6}+\frac{\pi^2(30+7\pi^2)}{72\alpha^7}+\mathcal{O}\left(\frac{1}{\alpha^8}\right),\\
        \mathtt{u}_{5}(\alpha)=&\;\frac{2\zeta(3)}{3\pi^2\alpha}-\frac{4}{3\alpha^2}-\frac{4+\pi^2}{4\alpha^3}-\frac{\pi^2}{6\alpha^4}+\frac{\pi^2(\pi^2-8)}{32\alpha^5}+\frac{\pi^2(60-7\pi^2)}{180\alpha^6}-\frac{\pi^2(60-7\pi^2)}{144\alpha^7}+\mathcal{O}\left(\frac{1}{\alpha^8}\right),\\
        \mathtt{u}_{7}(\alpha)=&\;-\frac{16\zeta(3)}{45\pi^2\alpha}+\frac{16}{15\alpha^2}+\frac{16+3\pi^2}{12\alpha^3}+\frac{12+7\pi^2}{18\alpha^4}+\frac{\pi^2(12-\pi^2)}{16\alpha^5}-\frac{\pi^2(120-11\pi^2)}{360\alpha^6}+\frac{\pi^2(240-112\pi^2+9\pi^4)}{576\alpha^7}+\mathcal{O}\left(\frac{1}{\alpha^8}\right).
    \end{aligned}
    $}
\end{equation} 
For $k=1$ we use
\begin{equation}
    \mathtt{u}_{1}(\alpha)=2\fint_{0}^{\infty}\limits\frac{dt}{t}\frac{\Theta(t-1)}{t+\alpha}+2\fint_{0}^{\infty}\limits\frac{dt}{t}\left\{\frac{1}{\tanh{t}(\alpha+t\tanh{t})}-\frac{\Theta(t-1)}{t+\alpha}\right\},
\end{equation}
where $\Theta$ is the Heaviside step function. The first term is straightforward and equals $\frac{2\log(1+\alpha)}{\alpha}$, while the second term admits a regular expansion in the limit $\alpha \to \infty$
\begin{equation}
    \fint_{0}^{\infty}\limits\frac{dt}{t}\left\{\frac{1}{\tanh{t}(\alpha+t\tanh{t})}-\frac{\Theta(t-1)}{t+\alpha}\right\}=\frac{1}{\alpha}\sum_{l=0}^{\infty}\frac{(-1)^l}{\alpha^l}\fint_{0}^{\infty}\limits dt\left\{\tanh^{l-1}{t}-\Theta(t-1)\right\}t^{l-1}.
\end{equation}
As a result, we have the following asymptotic form for integrals $\mathtt{u}_{1}(\alpha)$
\begin{equation}
    \resizebox{\textwidth}{!}{$
    \begin{aligned}
        \mathtt{u}_{1}(\alpha)\Big|_{\alpha\to\infty}=&\;\frac{2\log\alpha}{\alpha}+\frac{\log\left(\frac{e^{\gamma_E}}{\pi}\right)}{\alpha}-\frac{\pi^2}{12\alpha^3}+\frac{8-\pi^2}{6\alpha^4}-\frac{\pi^2(120+7\pi^2)}{480\alpha^5}+\frac{\pi^2(30+7\pi^2)}{90\alpha^6}-\frac{\pi^2(1680+980\pi^2+31\pi^4)}{4032\alpha^7}+\mathcal{O}\left(\frac{1}{\alpha^8}\right).
    \end{aligned}
    $}
\end{equation}
Expressions for expansion  $\mathtt{i}_{2k}(\alpha)$,  we can instantly find from relation \eqref{u-i2k-relation}. For example,
\begin{equation}
    \resizebox{\textwidth}{!}{$
    \begin{aligned}
        \mathtt{i}_{2}(\alpha)\Big|_{\alpha\rightarrow\infty}=&\;\frac{4\log\left(\frac{e^{\gamma_E-1}}{\pi}\alpha\right)}{\alpha}-\frac{\pi^2}{2\alpha^2}+\frac{\pi^2}{6\alpha^3}+\frac{16-5\pi^2}{6\alpha^4}+\frac{\pi^2(120+7\pi^2)}{720\alpha^5}-\frac{\pi^2(30+7\pi^2)}{180\alpha^6}+\frac{\pi^2(1680+980\pi^2+31\pi^4)}{10080\alpha^7}+\mathcal{O}\left(\frac{1}{\alpha^8}\right).
    \end{aligned}
    $}
\end{equation}
The evaluation of the integrals $\mathtt{i}_{2k-1}$ is largely similar to that of $\mathtt{u}_{2k-1}$. For $\mathtt{i}_1$, one needs to use a trick involving the addition and subtraction of the integrand’s asymptotic behavior. Higher-order integrals can be computed by expanding the integrand, with the resulting terms evaluated analytically.

\subsection{Critical points: \texorpdfstring{$\alpha\to\alpha^*_k\ne0$}{}}\label{u-asymptotics-critical-points}
Let's find the expansion of the function $\mathtt{u}_{2n-1}(\alpha)$ in the neighborhood of the critical point $\alpha^*_k\ne0$. Under analytic continuation of $\alpha$, as described in detail in Section \ref{BS-analyt-Section}, the zeros of the equation \eqref{main-transcendental-equation} cross the real axis, generating additional terms associated with residues (similar to the analytic continuation of the BS equation \eqref{BS_second_worldsheet}). By slightly lifting the contour above the real axis, we obtain:
\begin{equation}
\resizebox{\textwidth}{!}{$
    \begin{aligned}
        \mathtt{u}_{2n-1}(\alpha)=\underbrace{2\pi i\left(\underset{t=it^*}{\text{Res}}+\frac{1}{2}\underset{t=0}{\text{Res}}-\underset{t=-it^*}{\text{Res}}\right)\frac{\cosh^2{t}}{t\sinh^{2n-1}{t}\cdot(\alpha\cosh{t}+t\sinh{t})}}_{\mathtt{u}^{(1)}_{2n-1}(\alpha)}+\underbrace{\int^{\infty+i\epsilon}_{-\infty+i\epsilon}\limits dt\frac{\cosh^2{t}}{t\sinh^{2n-1}{t}\cdot(\alpha\cosh{t}+t\sinh{t})}}_{\mathtt{u}^{(2)}_{2n-1}(\alpha)}.
    \end{aligned}
    $}
\end{equation}
It is easy to see that $\underset{t=0}{\text{Res}}(...)=0,\;\underset{t=it^*}{\text{Res}}(...)=-\underset{t=-it^*}{\text{Res}}(...)$ (because integrands in $\mathtt{u}_{2n-1}(\alpha)$ are even functions). As in the chiral limit, the first term is a series expansion in half-integer powers of the parameter $\alpha-\alpha_k^*$ 
\begin{equation}
    \begin{aligned}
        \mathtt{u}^{(1)}_{2n-1}(\alpha)=&-\frac{2\pi}{(\alpha^*_k)^{n}}\frac{1}{\sqrt{\alpha-\alpha^*_k}}-\frac{4i\pi(3(1+\alpha^*_k)n-4\alpha^*_k-3)}{3(\alpha^*_k)^{n+\frac{1}{2}}\sqrt{1+\alpha^*_k}}+
        \\&+\frac{\pi\left(12+13\alpha^*_k(3+2\alpha^*_k)+12(1+\alpha^*_k)^2n^2-18(1+2\alpha^*_k)(1+\alpha^*_k)n\right)}{3(\alpha^*_k)^{n+1}(1+\alpha^*_k)}\sqrt{\alpha-\alpha^*_k}+\mathcal{O}(\alpha-\alpha^*_k),
    \end{aligned}
\end{equation}
and the second term is an expansion in integer powers
\begin{equation}
    \mathtt{u}^{(2)}_{2n-1}(\alpha)=\sum_{l=0}^{\infty}c_{2n-1}^{(l)}(\alpha^*_k)(\alpha-\alpha^*_k)^l,\quad  c_{2n-1}^{(l)}(\alpha^*_k)=(-1)^l\int^{\infty+i\epsilon}_{-\infty+i\epsilon}\limits dt\frac{\cosh^{2+l}t}{t\sinh^{2n-1}t(\alpha^*_k\cosh{t}+t\sinh{t})^{l+1}},
\end{equation}
where the coefficients $c^{(l)}_{2n-1}$ are accessible through numerical computation.

The integrals $\mathtt{i}_k(\alpha)$ are calculated in a similar way 
\begin{equation}
    \begin{aligned}
        &\mathtt{i}^{(1)}_{2n-1}(\alpha)=-\frac{8\pi i}{(\alpha_k^*)^{n}}+\frac{16\pi}{(\alpha_k^*)^{n}}\sqrt{\frac{1+\alpha_k^*}{\alpha_k^*}}(n-1)\sqrt{\alpha-\alpha_k^*}+\mathcal{O}(\alpha-\alpha_k^*),\\
        &\mathtt{i}^{(2)}_{2n-1}(\alpha)=\sum_{l=0}^{\infty}C^{(l)}_{2n-1}(\alpha-\alpha_k^*)^l,\quad C^{(l)}_{2n-1}=(-1)^l\int_{-\infty+i\epsilon}^{\infty+i\epsilon}\limits dt\;\frac{\sinh{2t}+2t}{t\cosh^{2n-1-l}{t}(\alpha_k^*\cosh{t}+\sinh{t})^{l+1}},
    \end{aligned}
\end{equation}
\begin{equation}
    \begin{aligned}
        &\mathtt{i}^{(1)}_{2n}(\alpha)=\frac{8\pi i}{(\alpha^*_k)^n}\sqrt{\frac{1+\alpha_k^*}{\alpha^*_k}}+\frac{8\pi}{(\alpha_k^*)^{n+1}}\left(1+2\alpha_k^*-2(\alpha_k^*+1)n\right)\sqrt{\alpha-\alpha_k^*}+\mathcal{O}(\alpha-\alpha_k^*),\\
        &\mathtt{i}^{(2)}_{2n}(\alpha)=\sum_{l=0}^{\infty}C^{(l)}_{2n}(\alpha-\alpha_k^*)^l,\quad C^{(l)}_{2n}=(-1)^l\int_{-\infty+i\epsilon}^{\infty+i\epsilon}\limits dt\;\frac{\cosh^{1+l}{t}(\sinh{2t}+2t)}{t\sinh^{2n}{t}(\alpha_k^*\cosh{t}+t\sinh{t})^{l+1}}.
    \end{aligned}
\end{equation}

Here we also present explicit expressions for the expansions of the first three spectral sums in the vicinity of critical points $\alpha^*_k$, whose leading terms were used in Section \ref{BS-analyt-Section}
\begin{equation}
    \begin{aligned}
        &\mathcal{G}^{(1)}_{+}(\alpha)\Big|_{\alpha\to\alpha^*_k}=\log2\pi-3-2\pi i\sqrt{\frac{1+\alpha^*_k}{\alpha^*_k}}-\frac{\alpha^*_k}{2}\left(c^{(0)}_{1}-\alpha^*_kc^{(0)}_{3}\right)-\frac{\alpha^*_k\mathtt{c}_2}{2}+\mathcal{O}\left(\sqrt{\alpha-\alpha^*_k}\right),
        \\
        &\mathcal{G}^{(1)}_{-}(\alpha)\Big|_{\alpha\to\alpha^*_k}=\log2\pi-5-2\pi i\sqrt{\frac{1+\alpha^*_k}{\alpha^*_k}}-\frac{15\pi^2}{4\alpha^*_k}-\frac{\alpha^*_k}{2}\left(c^{(0)}_{1}-\alpha^*_kc^{(0)}_{3}\right)-\frac{\alpha^*_k(8+\mathtt{c}_2)}{4}+\mathcal{O}\left(\sqrt{\alpha-\alpha^*_k}\right),
    \end{aligned}
\end{equation}
\begin{equation}
    \begin{aligned}
        \mathcal{G}^{(2)}_{+}(\alpha)\Big|_{\alpha\to\alpha^*_k}=&\;-\frac{4\pi(1+\alpha^*_k)}{\sqrt{\alpha-\alpha^*_k}}+2+\frac{8(\alpha^*_k)^2\zeta(3)}{3\pi^2}-\frac{4(\alpha^*_k)^2\zeta(5)}{\pi^4}-\frac{4\alpha^*_k}{3}+\frac{\pi^2}{\alpha^*_k}+
        \\&+2(\alpha^*_k)^3\left(c^{(0)}_{3}+c^{(0)}_{5}-\frac{8\pi i\sqrt{1+\alpha^*_k}(3+\alpha^*_k)}{3(\alpha^*_k)^{7/2}}\right)+\mathcal{O}\left(\sqrt{\alpha-\alpha^*_k}\right),
        \\
        \mathcal{G}^{(2)}_{-}(\alpha)\Big|_{\alpha\to\alpha^*_k}=&\;\frac{675\pi^4+32(\alpha^*_k)^2(1+\alpha^*_k)^2+16\pi^2\alpha^*_k(3+2\alpha^*_k)}{48(\alpha^*_k)^2}+\mathcal{O}\left(\sqrt{\alpha-\alpha^*_k}\right),
    \end{aligned}
\end{equation}
\begin{equation}
    \begin{aligned}
        \mathcal{G}^{(3)}_{+}(\alpha)\Big|_{\alpha\to\alpha^*_k}=&\;\frac{2\pi(1+\alpha^*_k)\left(\pi^2-4\alpha^*_k(1+\alpha^*_k)\right)}{\alpha^*_k\sqrt{\alpha-\alpha^*_k}}+\mathcal{O}\left(1\right),
        \\
        \mathcal{G}^{(3)}_{-}(\alpha)\Big|_{\alpha\to\alpha^*_k}=&\;\frac{1}{15}\left(65+\alpha^*_k\left(45+\alpha^*_k-5(\alpha^*_k)^2\right)\right)-\frac{\pi^2(105+\alpha^*_k(274+147\alpha^*_k))}{36\alpha^*_k}-\frac{3375\pi^6}{64(\alpha^*_k)^3}-
        \\&-(\alpha^*_k)^2c^{(0)}_{3}+2(\alpha^*_k)^3c^{(0)}_{5}-(\alpha^*_k)^4c^{(0)}_{7}-\frac{2\left(45+30\alpha^*_k+8(\alpha^*_k)^2\right)\alpha^*_k\zeta(3)}{45\pi^2}-
        \\&
        -\frac{4(3+\alpha^*_k)(\alpha^*_k)^2\zeta(5)}{3\pi^4}-\frac{2(\alpha^*_k)^3\zeta(7)}{\pi^6}+\mathcal{O}\left(\sqrt{\alpha-\alpha^*_k}\right).
    \end{aligned}
\end{equation}
\bibliographystyle{MyStyle}
\bibliography{MyBib}

\providecommand{\href}[2]{#2}\begingroup\raggedright\begin{thebibliography}{10}

\bibitem{Belavin:1984vu}
A.~A. Belavin, A.~M. Polyakov and A.~B. Zamolodchikov, \emph{{Infinite Conformal Symmetry in Two-Dimensional Quantum Field Theory}}, \href{https://doi.org/10.1016/0550-3213(84)90052-X}{\emph{Nucl. Phys. B} {\bfseries 241} (1984) 333}.

\bibitem{McCoy:1978ta}
B.~M. McCoy and T.~T. Wu, \emph{{Two-dimensional Ising Field Theory in a Magnetic Field: Breakup of the Cut in the Two Point Function}}, \href{https://doi.org/10.1103/PhysRevD.18.1259}{\emph{Phys. Rev. D} {\bfseries 18} (1978) 1259}.

\bibitem{Onsager:1943jn}
L.~Onsager, \emph{{Crystal statistics. 1. A Two-dimensional model with an order disorder transition}}, \href{https://doi.org/10.1103/PhysRev.65.117}{\emph{Phys. Rev.} {\bfseries 65} (1944) 117}.

\bibitem{Zamolodchikov:1989fp}
A.~B. Zamolodchikov, \emph{{Integrals of Motion and S Matrix of the (Scaled) T=T(c) Ising Model with Magnetic Field}}, \href{https://doi.org/10.1142/S0217751X8900176X}{\emph{Int. J. Mod. Phys. A} {\bfseries 4} (1989) 4235}.

\bibitem{Zamolodchikov:1989zs}
A.~Zamolodchikov, \emph{{Integrable field theory from conformal field theory}}, \href{https://doi.org/doi:10.2969/aspm/01910641}{\emph{Adv. Stud. Pure Math.} {\bfseries 19} (1989) 641}.

\bibitem{Yang:1952be}
C.-N. Yang and T.~D. Lee, \emph{{Statistical theory of equations of state and phase transitions. 1. Theory of condensation}}, \href{https://doi.org/10.1103/PhysRev.87.404}{\emph{Phys. Rev.} {\bfseries 87} (1952) 404}.

\bibitem{Lee:1952ig}
T.~D. Lee and C.-N. Yang, \emph{{Statistical theory of equations of state and phase transitions. 2. Lattice gas and Ising model}}, \href{https://doi.org/10.1103/PhysRev.87.410}{\emph{Phys. Rev.} {\bfseries 87} (1952) 410}.

\bibitem{Fonseca:2001dc}
P.~Fonseca and A.~Zamolodchikov, \emph{{Ising field theory in a magnetic field: Analytic properties of the free energy}},  \href{https://arxiv.org/abs/hep-th/0112167}{{\ttfamily hep-th/0112167}}.

\bibitem{Xu:2022mmw}
H.-L. Xu and A.~Zamolodchikov, \emph{{2D Ising Field Theory in a magnetic field: the Yang-Lee singularity}}, \href{https://doi.org/10.1007/JHEP08(2022)057}{\emph{JHEP} {\bfseries 08} (2022) 057} [\href{https://arxiv.org/abs/2203.11262}{{\ttfamily 2203.11262}}].

\bibitem{Mangazeev:2023bnt}
V.~V. Mangazeev, B.~Hagan and V.~V. Bazhanov, \emph{{Corner transfer matrix approach to the Yang-Lee singularity in the two-dimensional Ising model in a magnetic field}}, \href{https://doi.org/10.1103/PhysRevE.108.064136}{\emph{Phys. Rev. E} {\bfseries 108} (2023) 064136} [\href{https://arxiv.org/abs/2308.15113}{{\ttfamily 2308.15113}}].

\bibitem{Fisher:1978pf}
M.~E. Fisher, \emph{{Yang-Lee Edge Singularity and $\phi^3$ Field Theory}}, \href{https://doi.org/10.1103/PhysRevLett.40.1610}{\emph{Phys. Rev. Lett.} {\bfseries 40} (1978) 1610}.

\bibitem{Cardy:1985yy}
J.~L. Cardy, \emph{{Conformal Invariance and the Yang-lee Edge Singularity in Two-dimensions}}, \href{https://doi.org/10.1103/PhysRevLett.54.1354}{\emph{Phys. Rev. Lett.} {\bfseries 54} (1985) 1354}.

\bibitem{Cardy:1989fw}
J.~L. Cardy and G.~Mussardo, \emph{{S Matrix of the Yang-Lee Edge Singularity in Two-Dimensions}}, \href{https://doi.org/10.1016/0370-2693(89)90818-6}{\emph{Phys. Lett. B} {\bfseries 225} (1989) 275}.

\bibitem{Xu:2023nke}
H.-L. Xu and A.~Zamolodchikov, \emph{{Ising field theory in a magnetic field: $\varphi^3$ coupling at $T > T_{c}$}}, \href{https://doi.org/10.1007/JHEP08(2023)161}{\emph{JHEP} {\bfseries 08} (2023) 161} [\href{https://arxiv.org/abs/2304.07886}{{\ttfamily 2304.07886}}].

\bibitem{Fonseca:2006au}
P.~Fonseca and A.~Zamolodchikov, \emph{{Ising spectroscopy. I. Mesons at T $<$ T$_c$}},  \href{https://arxiv.org/abs/hep-th/0612304}{{\ttfamily hep-th/0612304}}.

\bibitem{Zamolodchikov:2013ama}
A.~Zamolodchikov, \emph{{Ising Spectroscopy II: Particles and poles at $T > T_c$}},  \href{https://arxiv.org/abs/1310.4821}{{\ttfamily 1310.4821}}.

\bibitem{Delfino:1996xp}
G.~Delfino, G.~Mussardo and P.~Simonetti, \emph{{Nonintegrable quantum field theories as perturbations of certain integrable models}}, \href{https://doi.org/10.1016/0550-3213(96)00265-9}{\emph{Nucl. Phys. B} {\bfseries 473} (1996) 469} [\href{https://arxiv.org/abs/hep-th/9603011}{{\ttfamily hep-th/9603011}}].

\bibitem{Delfino:2005bh}
G.~Delfino, P.~Grinza and G.~Mussardo, \emph{{Decay of particles above threshold in the Ising field theory with magnetic field}}, \href{https://doi.org/10.1016/j.nuclphysb.2005.12.024}{\emph{Nucl. Phys. B} {\bfseries 737} (2006) 291} [\href{https://arxiv.org/abs/hep-th/0507133}{{\ttfamily hep-th/0507133}}].

\bibitem{Rutkevich:2005ai}
S.~B. Rutkevich, \emph{{Large-n excitations in the ferromagnetic Ising field theory in a small magnetic field: Mass spectrum and decay widths}}, \href{https://doi.org/10.1103/PhysRevLett.95.250601}{\emph{Phys. Rev. Lett.} {\bfseries 95} (2005) 250601} [\href{https://arxiv.org/abs/hep-th/0509149}{{\ttfamily hep-th/0509149}}].

\bibitem{Rutkevich:2009zz}
S.~B. Rutkevich, \emph{{Formfactor perturbation expansions and confinement in the Ising field theory}}, \href{https://doi.org/10.1088/1751-8113/42/30/304025}{\emph{J. Phys. A} {\bfseries 42} (2009) 304025} [\href{https://arxiv.org/abs/0901.1571}{{\ttfamily 0901.1571}}].

\bibitem{Rutkevich:2017qdw}
S.~B. Rutkevich, \emph{{Radiative corrections to the quark masses in the ferromagnetic Ising and Potts field theories}}, \href{https://doi.org/10.1016/j.nuclphysb.2017.08.009}{\emph{Nucl. Phys. B} {\bfseries 923} (2017) 508} [\href{https://arxiv.org/abs/1706.05281}{{\ttfamily 1706.05281}}].

\bibitem{Lencses:2015bpa}
M.~Lencses and G.~Takacs, \emph{{Confinement in the q-state Potts model: an RG-TCSA study}}, \href{https://doi.org/10.1007/JHEP09(2015)146}{\emph{JHEP} {\bfseries 09} (2015) 146} [\href{https://arxiv.org/abs/1506.06477}{{\ttfamily 1506.06477}}].

\bibitem{Gabai:2019ryw}
B.~Gabai and X.~Yin, \emph{{On the S-matrix of Ising field theory in two dimensions}}, \href{https://doi.org/10.1007/JHEP10(2022)168}{\emph{JHEP} {\bfseries 10} (2022) 168} [\href{https://arxiv.org/abs/1905.00710}{{\ttfamily 1905.00710}}].

\bibitem{Fitzpatrick:2023aqm}
A.~L. Fitzpatrick, E.~Katz and Y.~Xin, \emph{{{Lightcone Hamiltonian for Ising Field Theory I: $T< T_c$}}}, \href{https://doi.org/10.1103/9dxz-k5wb}{\emph{SciPost Phys.} {\bfseries 18} (2025) 179} [\href{https://arxiv.org/abs/2311.16290}{{\ttfamily 2311.16290}}].

\bibitem{Jha:2024jan}
R.~G. Jha, A.~Milsted, D.~Neuenfeld, J.~Preskill and P.~Vieira, \emph{{Real-time scattering in Ising field theory using matrix product states}}, \href{https://doi.org/10.1103/9dxz-k5wb}{\emph{Phys. Rev. Res.} {\bfseries 7} (2025) 023266} [\href{https://arxiv.org/abs/2411.13645}{{\ttfamily 2411.13645}}].

\bibitem{THOOFT1974461}
G.~'t~Hooft, \emph{{A Two-Dimensional Model for Mesons}}, \href{https://doi.org/10.1016/0550-3213(74)90088-1}{\emph{Nucl. Phys. B} {\bfseries 75} (1974) 461}.

\bibitem{Fateev:2009jf}
V.~A. Fateev, S.~L. Lukyanov and A.~B. Zamolodchikov, \emph{{On mass spectrum in 't Hooft's 2D model of mesons}}, \href{https://doi.org/10.1088/1751-8113/42/30/304012}{\emph{J. Phys. A} {\bfseries 42} (2009) 304012} [\href{https://arxiv.org/abs/0905.2280}{{\ttfamily 0905.2280}}].

\bibitem{Litvinov:2024riz}
A.~Litvinov and P.~Meshcheriakov, \emph{{Meson mass spectrum in QCD2 't Hooft's model}}, \href{https://doi.org/10.1016/j.nuclphysb.2024.116766}{\emph{Nucl. Phys. B} {\bfseries 1010} (2025) 116766} [\href{https://arxiv.org/abs/2409.11324}{{\ttfamily 2409.11324}}].

\bibitem{Artemev:2025cev}
A.~Artemev, A.~Litvinov and P.~Meshcheriakov, \emph{{QCD2 \textquoteright{}t Hooft model: Two-flavor mesons spectrum}}, \href{https://doi.org/10.1103/7311-kdbj}{\emph{Phys. Rev. D} {\bfseries 111} (2025) 125001} [\href{https://arxiv.org/abs/2504.12081}{{\ttfamily 2504.12081}}].

\bibitem{Wu:1975mw}
T.~T. Wu, B.~M. McCoy, C.~A. Tracy and E.~Barouch, \emph{{Spin spin correlation functions for the two-dimensional Ising model: Exact theory in the scaling region}}, \href{https://doi.org/10.1103/PhysRevB.13.316}{\emph{Phys. Rev. B} {\bfseries 13} (1976) 316}.

\bibitem{Fonseca:2003ee}
P.~Fonseca and A.~Zamolodchikov, \emph{{Ward identities and integrable differential equations in the Ising field theory}},  \href{https://arxiv.org/abs/hep-th/0309228}{{\ttfamily hep-th/0309228}}.

\bibitem{Voloshin:1985id}
M.~B. Voloshin, \emph{{Decay of false vacuum in $(1+1)$-dimensions}}, {\emph{Yad. Fiz.} {\bfseries 42} (1985) 1017}.

\bibitem{Gao:2025mcg}
Y.~Gao, Y.~Jiang and J.~Wu, \emph{{Mesons in a quantum Ising ladder}},  \href{https://arxiv.org/abs/2502.15463}{{\ttfamily 2502.15463}}.

\bibitem{Baxter:1972hz}
R.~J. Baxter, \emph{{Partition function of the eight vertex lattice model}}, \href{https://doi.org/10.1016/0003-4916(72)90335-1}{\emph{Annals Phys.} {\bfseries 70} (1972) 193}.

\bibitem{Its:1980}
A.~R. Its, A.~G. Izergin and V.~E. Korepin, \emph{{Temperature correlators of the impenetrable Bose gas as an integrable system}}, \href{https://doi.org/10.1007/BF02096786}{\emph{Commun. Math. Phys.} {\bfseries 129} (1990) 205 }.

\bibitem{Its:1990MPhysB}
A.~Its, A.~Izergin, V.~Korepin and N.~Slavnov, \emph{Differential equations for quantum correlation functions}, \href{https://doi.org/10.1142/s0217979290000504}{\emph{Int. J. Mod. Phys. B} {\bfseries 04} (1990) 1003–1037}.

\bibitem{zbMATH01284258}
P.~Deift, \emph{Integrable operators},  in \emph{Differential operators and spectral theory. M. Sh. Birman's 70th anniversary collection}, pp.~69--84, Providence, RI: American Mathematical Society, (1999).

\bibitem{Fateev:1993av}
V.~A. Fateev, \emph{{The exact relations between the coupling constants and the masses of particles for the integrable perturbed conformal field theories}}, \href{https://doi.org/10.1016/0370-2693(94)00078-6}{\emph{Phys. Lett.} {\bfseries B324} (1994) 45}.

\bibitem{ZIYATDINOV:2010ModPhysA}
I.~Ziyatdinov, \emph{{Asymptotic properties of mass spectrum in 't Hooft's model of mesons}}, \href{https://doi.org/10.1142/S0217751X10050287}{\emph{Int. J. Mod. Phys. A} {\bfseries 25} (2010) 3899} [\href{https://arxiv.org/abs/1003.4304}{{\ttfamily 1003.4304}}].

\bibitem{Zamolodchikov:2009pres}
A.~Zamolodchikov, \emph{{On Confining Interactions in $1+1$. Talk at Conference in the Memory of Aliosha Zamolodchikov, Saclay}}, {\emph{\url{https://indico.in2p3.fr/event/1886/sessions/3945/attachments/17798/21781/Zamolodchikov.pdf}} (2009) }.

\end{thebibliography}\endgroup
\end{document}